\documentclass[preprint,12pt]{elsarticle}
\usepackage[margin=1.5cm,includefoot]{geometry}
\usepackage{setspace}

\usepackage{bibentry}
\usepackage{graphicx}
\usepackage{amssymb}
\usepackage{gensymb}
\usepackage[utf8]{inputenc}
\usepackage{lineno}
\usepackage{mathtools}
\usepackage[separate-uncertainty=true]{siunitx}
\usepackage[colorlinks]{hyperref}
\PassOptionsToPackage{hyphens}{url}\usepackage{hyperref} %allow URLs to break across lines
\usepackage[nameinlink,capitalise]{cleveref} %needs to appear after hyperref, https://tex.stackexchange.com/questions/396728/my-equations-referencing-not-working
\Crefname{figure}{Figure}{Figures} %needs to appear after hyperref and cleveref
\usepackage{mathrsfs}
\usepackage{chemformula} %\usepackage{mhchem} %\usepackage[version=4]{mhchem}
\usepackage{booktabs}
\usepackage{longtable}
\usepackage{url}
\usepackage{enumitem}
\usepackage{tabulary}
\usepackage{makecell}
\usepackage{multirow}
\usepackage{graphics}
\graphicspath{{lit-figs/}{figures/}} %Setting the graphicspath
\usepackage [english]{babel}
\usepackage [autostyle, english = american]{csquotes}
\MakeOuterQuote{"}
\usepackage[shortcuts,abbreviations,stylemods={mcols},style=mcolindex]{glossaries-extra} %page 29 of http://tug.ctan.org/macros/latex/contrib/glossaries/glossariesbegin.pdf for mcolindex example
\usepackage{glossary-mcols}
\glssetcategoryattribute{abbreviation}{indexonlyfirst}{true}
\glssetcategoryattribute{abbreviation}{nohyper}{true}
\makeglossaries
\robustify{\gls}% Make \gls not fragile

\newabbreviation{ml}{ML}{machine learning}
\newabbreviation{svm}{SVM}{support vector machine}
\newabbreviation{mse}{MSE}{mean square error}
\newabbreviation{mae}{MAE}{mean absolute error}
\newabbreviation{rmse}{RMSE}{root mean square error}
\newabbreviation{nlp}{NLP}{natural language processing}
\newabbreviation{bo}{BO}{Bayesian optimization}
\newabbreviation{gpr}{GPR}{Gaussian process regression}
\newabbreviation{mbo}{MBO}{Bayesian model-based optimization}
\newabbreviation{dft}{DFT}{density functional theory}
\newabbreviation{tddft}{TD-DFT}{time-dependent density functional theory}
\newabbreviation{ann}{ANN}{artificial neural network}
\newabbreviation{ea}{EA}{evolutionary algorithm}
\newabbreviation{crfs}{CR-FS}{cluster resolution feature selection}
\newabbreviation{sr}{SR}{symbolic regression}
\newabbreviation{ad}{AD}{adaptive design}
\newabbreviation{cgcnn}{CGCNN}{crystal graph convolutional neural network}
\newabbreviation{icgcnn}{iCGCNN}{improved crystal graph convolutional neural network}
\newabbreviation{fem}{FEM}{finite element method}
\newabbreviation{rf}{RF}{random forest}
\newabbreviation{rfe}{RFE}{recursive feature elimination}
\newabbreviation{doe}{DoE}{design of experiments}
\newabbreviation{dt}{DT}{decision tree}
\newabbreviation{fs}{FS}{feature selection}
\newabbreviation{cr-fs}{CR-FS}{cluster resolution feature selection}
\newabbreviation{lr}{LR}{linear regression}
\newabbreviation{pr}{PR}{polynomial regression}
\newabbreviation{gbr}{GBR}{gradient boosting regression}
\newabbreviation{krr}{KRR}{kernel ridge regression}
\newabbreviation{ert}{ERT}{extremely randomized tree}
\newabbreviation{plsda}{PLS-DA}{partial least-squares discriminant analysis}
\newabbreviation{knn}{kNN}{k-nearest neighbor}
\newabbreviation{pls}{PLS}{partial least squares}
\newabbreviation{smc}{SMC}{sequential Monte Carlo}
\newabbreviation{combo}{COMBO}{COMmon Bayesian Optimization}
\newabbreviation{smiles}{SMILES}{simplified molecular-input line-entry system}
\newabbreviation{max}{MAX}{$M_{n+1}AX_n$}
\newabbreviation{rrr}{RRR}{residual resistivity ratio}
\newabbreviation{bma}{BMA}{Bayesian model averaging}
\newabbreviation{mbe}{MBE}{molecular beam epitaxy}
\newabbreviation{ei}{EI}{expected improvement}
\newabbreviation{fabhmes}{FAB-HMEs}{factorized asymptotic Bayesian inference hierarchical mixture of experts}
\newabbreviation{ste}{STE}{spin-driven thermoelectric}
\newabbreviation{xrd}{XRD}{X-ray diffraction}
\newabbreviation{xrf}{XRF}{X-ray fluorescence}
\newabbreviation{oer}{OER}{oxygen evolution reaction}
\newabbreviation{ego}{EGO}{efficient global optimization}
\newabbreviation{kg}{KG}{Knowledge Gradient}
\newabbreviation{hea}{HEA}{high-entropy alloy}
\newabbreviation{gbdt}{GBDT}{gradient boosting decision tree}
\newabbreviation{smote}{SMOTE}{synthetic minority oversampling technique}
\newabbreviation{rbf}{RBF}{radial basis function}
\newabbreviation{lasso}{LASSO}{least absolute shrinkage and selection operator}
\newabbreviation{ga}{GA}{genetic algorithm}
\newabbreviation{mof}{MOF}{metal-organic framework}
\newabbreviation{gcmc}{GCMC}{grand canonical Monte Carlo}
\newabbreviation{mpb}{MPB}{morphotropic phase boundary}
\newabbreviation{cif}{CIF}{crystallographic information file}
\newabbreviation{cac}{CAC}{cation/anion contribution}
\newabbreviation{rmsecv}{RMSECV}{cross-validated root mean square error}
\newabbreviation{maecv}{MAECV}{cross-validated mean absolute error}
\newabbreviation{cvmr}{CVMR}{cross-validated misclassification rate}
\newabbreviation{tmr}{TMR}{training misclassification rate}
\newabbreviation{dac}{DAC}{diamond anvil cell}
\newabbreviation{ca}{CA}{correlation analysis}
\newabbreviation{bpnn}{BPNN}{back propagated neural network}
\newabbreviation{nmf}{NMF}{non-negative matrix factorization}
\newabbreviation{svd}{SVD}{singular value decomposition}
\newabbreviation{cpd}{CPD}{canonical polyadic decomposition}
\newabbreviation{crc}{CRC}{chemically relevant composition}
\newabbreviation{cms}{CMS}{combinatorial magnetron sputtering}
\newabbreviation{cv}{CV}{cross-validation}
\newabbreviation{gcv}{G-CV}{grouping cross-validation}
\newabbreviation{lococv}{LOCO-CV}{leave-one-cluster-out cross-validation}
\newabbreviation{hitp}{HiTp}{high-throughput}
\newabbreviation{oled}{OLED}{organic light-emitting diode}
\newabbreviation{tadf}{TADF}{thermally activated delayed fluorescence}
\newabbreviation{eqe}{EQE}{external quantum efficiency}
\newabbreviation{hoip}{HOIP}{hybrid organic-inorganic perovskite}
\newabbreviation{crr}{CRR}{capacity retention rate}
\newabbreviation{cnt}{CNT}{carbon nanotube}
\newabbreviation{ares}{ARES}{Autonomous Research System}
\newabbreviation{rr}{RR}{ridge regression}

\newcommand{\nvalid}{50} %number of experimentally and computationally validated articles

\newcommand*{\glsentrytitlecasefull}[1]{\glsentrytitlecase{#1}{long} (\glsentryshort{#1}) }

\newcommand{\ignore}[1]{}
\newcommand{\nobibentry}[1]{{\let\nocite\ignore\bibentry{#1}}}

\biboptions{sort&compress}
\title{Data-Driven Materials Discovery and Synthesis using Machine Learning Methods}

\date{September 2020}

\begin{document}

\author[myu]{Sterling G. Baird}
\author[myu,west]{Marianne Liu} %I think no spaces in optional arguments
\author[myu]{Hasan M. Sayeed}
\author[myu]{Taylor D. Sparks\corref{cor1}}%
\ead{sparks@eng.utah.edu}
\cortext[cor1]{Corresponding author.}

\address[myu]{Department of Materials Science and Engineering, University of Utah, Salt Lake City, Utah 84112, USA}

\address[west]{West High School, Salt Lake City, Utah 84112, USA}

\begin{abstract}
    {\centering
    \includegraphics[width=89mm]{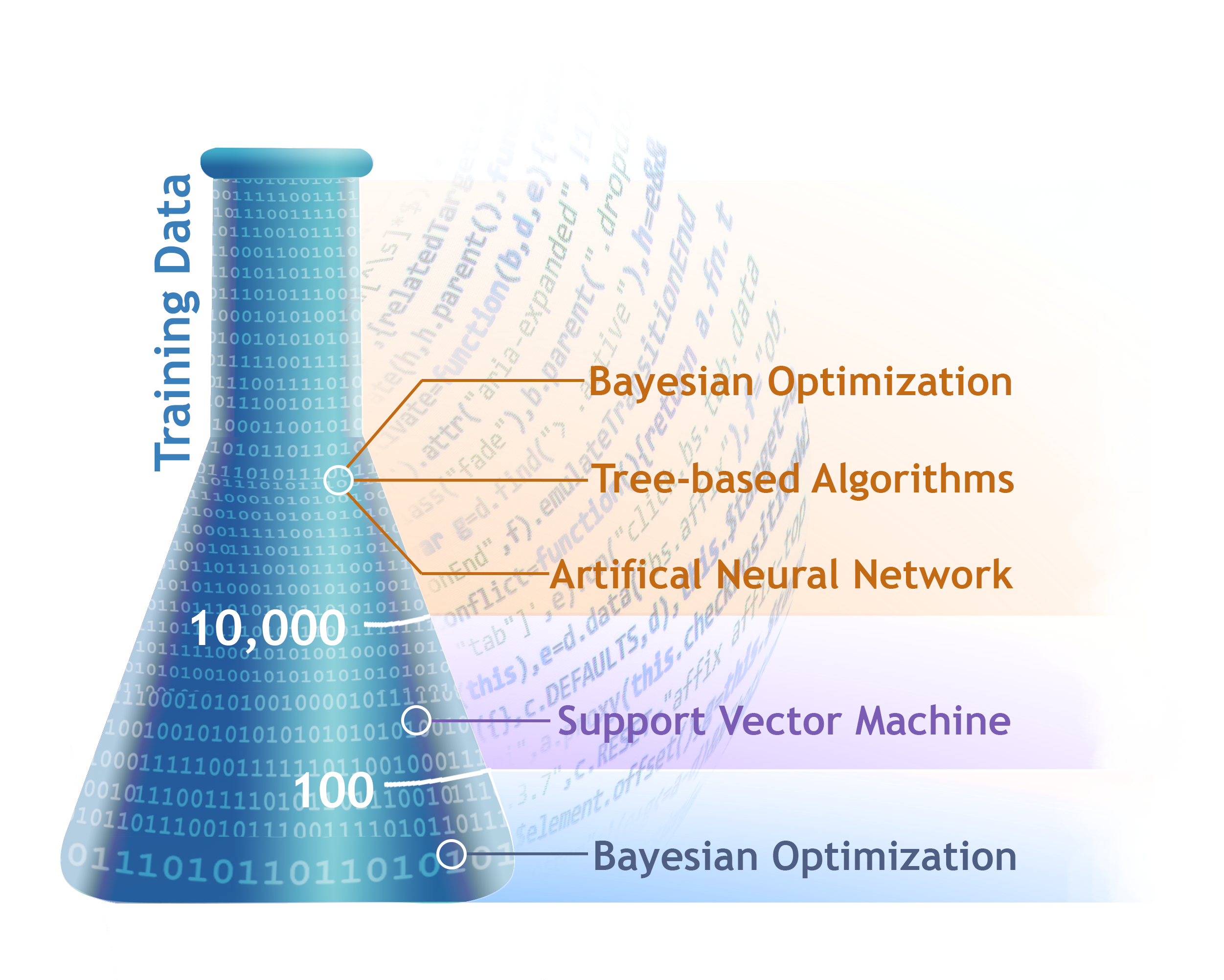}
    \par
    } %alternative:
    Experimentally \cite{balachandranExperimentalSearchHightemperature2018,buciorEnergybasedDescriptorsRapidly2019,caoHowOptimizeMaterials2018,chenMachineLearningAssisted2020,gaultoisPerspectiveWebbasedMachine2016,gomez-bombarelliDesignEfficientMolecular2016,gzylHalfHeuslerStructuresFullHeusler2020,gzylSolvingColoringProblem2019,hommaOptimizationHeterogeneousTernary2020,houMachineLearningAssistedDevelopmentTheoretical2019,iwasakiIdentificationAdvancedSpindriven2019,kauweMachineLearningPrediction2018,kimDeeplearningbasedInverseDesign2018,liEfficientOptimizationPerformance2018,menonMolecularEngineeringSuperplasticizers2019,minMachineLearningAssisted2018,nikolaevAutonomyMaterialsResearch2016,oliynykClassifyingCrystalStructures2016,oliynykDisentanglingStructuralConfusion2017,oliynykHighThroughputMachineLearningDrivenSynthesis2016,raccugliaMachinelearningassistedMaterialsDiscovery2016,renAcceleratedDiscoveryMetallic2018,rickmanMaterialsInformaticsScreening2019,sakuraiUltranarrowBandWavelengthSelectiveThermal2019,shampDecompositionProductsPhosphine2016,tehraniMachineLearningDirected2018,wahabMachinelearningassistedFabricationBayesian2020,wakabayashiMachinelearningassistedThinfilmGrowth2019,wengSimpleDescriptorDerived2020,wenMachineLearningAssisted2019,wuMachinelearningassistedDiscoveryPolymers2019,xueAcceleratedSearchBaTiO3based2016,xueInformaticsApproachTransformation2017,yuanAcceleratedDiscoveryLarge2018,zhangFindingNextSuperhard2020,zhangNotJustPar2018,zhuoEvaluatingThermalQuenching2020,zhuoIdentifyingEfficientThermally2018} and computationally \cite{balachandranAdaptiveStrategiesMaterials2016,balachandranDatadrivenDesignB202020,balachandranLearningDataDesign2017,balachandranPredictingDisplacementsOctahedral2017,juDesigningNanostructuresPhonon2017,luAcceleratedDiscoveryStable2018,mannodi-kanakkithodiMachineLearningStrategy2016,meredigCombinatorialScreeningNew2014,parkDevelopingImprovedCrystal2020,sekoMatrixTensorbasedRecommender2018,sendekHolisticComputationalStructure2017,talapatraAutonomousEfficientExperiment2018} validated \gls{ml} articles are sorted based on size of the training data: \numrange{1}{100}, \numrange{101}{10000}, and \num{10000}+ in a comprehensive set summarizing legacy and recent advances in the field. The review emphasizes the interrelated fields of synthesis, characterization, and prediction. Size range \numrange{1}{100} consists mostly of \gls{bo} articles, whereas \numrange{101}{10000} consists mostly of \gls{svm} articles. The articles often use combinations of \gls{ml}, \gls{fs}, \gls{ad}, \gls{hitp} techniques, and domain knowledge to enhance predictive performance and/or model interpretability. \Gls{gcv} techniques curb overly optimistic extrapolative predictive performance. % and extraordinary predictions are frequently found for small, interpolative datasets and less frequently for large, extrapolative datasets.
    Smaller datasets relying on \gls{ad} are typically able to identify new materials with desired properties but do so in a constrained design space. In larger datasets, the low-hanging fruit of materials optimization are typically already discovered, and the models are generally less successful at extrapolating to new materials, especially when the model training data favors a particular type of material.
    The large increase of \gls{ml} materials science articles that perform experimental or computational validation on the predicted results demonstrates the interpenetration of materials informatics with the materials science discipline and an accelerating materials discovery for real-world applications.
\end{abstract}


\glsresetall %reset the abbreviations

\maketitle

\section{Introduction to Experimental and Computational Machine Learning Validation} \label{sec:intro}

Data-driven materials science is plagued by sparse, noisy, multi-scale, heterogeneous, small datasets in contrast to many traditional \gls{ml} fields \cite{meredigFiveHighImpactResearch2019}. The budding field brings together experts from both materials science and \gls{ml} disciplines; a great challenge is to incorporate domain knowledge with the appropriate \gls{ml} tools to discover new materials with better properties \cite{murdockDomainKnowledgeNecessary2020}. When predictions of new materials are made, experimental or computational validation of those results is less common in the sea of \gls{ml} articles. This perhaps stems from a requirement to mesh deep expertise from two topics (e.g. \gls{dft} and \glspl{ann}) and the difficulty in publishing if validation results do not align with the proposed model or do not produce exemplary results \cite{raccugliaMachinelearningassistedMaterialsDiscovery2016}. 

Some have addressed the former issue of interdisciplinary expertise requirements by providing user-friendly web apps \cite{gaultoisPerspectiveWebbasedMachine2016} or clearly documented install and use instructions for code hosted on sites such as GitHub \cite{wangMachineLearningMaterials2020}. An example of this was the work by \citet{zhangNotJustPar2018}, which used a previously constructed \gls{ml} web app\cite{gaultoisPerspectiveWebbasedMachine2016} (\url{http://thermoelectrics.citrination.com/}) which takes only chemical formulas as inputs and went on to validate these predictions of low thermal conductivity for novel quaternary germanides.

The expertise issue is aided by advances in flexible code packages in e.g. Python (\texttt{PyTorch} \cite{paszkePyTorchImperativeStyle2019}, \texttt{scikit-learn} \cite{pedregosaScikitlearnMachineLearning2011}, \texttt{COMBO} \cite{uenoCOMBOEfficientBayesian2016}, \texttt{pymatgen} \cite{ongPythonMaterialsGenomics2013}, \texttt{Magpie} \cite{wardGeneralpurposeMachineLearning2016}, \texttt{JARVIS} \cite{choudharyJointAutomatedRepository2020}), MATLAB (\texttt{Statistics and Machine Learning Toolbox} \cite{themathworksStatisticsMachineLearning2020}, \texttt{Deep Learning Toolbox} \cite{themathworksDeepLearningToolbox2020}), and R (\texttt{caret} \cite{kuhnBuildingPredictiveModels2008}, \texttt{e1071} \cite{meyerE1071MiscFunctions2020}, \texttt{nnet} \cite{venablesModernAppliedStatistics2002}) (see also Table 2 of \citet{butlerMachineLearningMolecular2018}), which shifts some of the burden of computational optimization, speed, and flexibility away from materials scientists and engineers. Additionally, experimental (e.g. arc melting \cite{gaultoisDatadrivenReviewThermoelectric2013,gaultoisPerspectiveWebbasedMachine2016,houMachineLearningAssistedDevelopmentTheoretical2019,tehraniMachineLearningDirected2018,wenMachineLearningAssisted2019,xueInformaticsApproachTransformation2017,zhangFindingNextSuperhard2020} and \gls{cms} \cite{iwasakiIdentificationAdvancedSpindriven2019,renAcceleratedDiscoveryMetallic2018}) and computational (e.g. \gls{dft} \cite{balachandranAdaptiveStrategiesMaterials2016,balachandranDatadrivenDesignB202020,balachandranLearningDataDesign2017,balachandranPredictingDisplacementsOctahedral2017,luAcceleratedDiscoveryStable2018,mannodi-kanakkithodiMachineLearningStrategy2016,meredigCombinatorialScreeningNew2014,parkDevelopingImprovedCrystal2020,sekoMatrixTensorbasedRecommender2018,sendekHolisticComputationalStructure2017} and \gls{fem} \cite{hoarMachineLearningEnabledExplorationMorphology2020,yanOptimizationThermalConductivity2020}) high throughput techniques and materials databases/tools such as the Materials Project \cite{jainMaterialsProjectMaterials2013}, Open Quantum Materials Database \cite{kirklinOpenQuantumMaterials2015}, Pearson's Crystal Database \cite{villarsPearsonCrystalData2014}, Matminer \cite{wardMatminerOpenSource2018}, \href{https://darkreactions.haverford.edu/}{Dark Reactions Project} \cite{raccugliaMachinelearningassistedMaterialsDiscovery2016}, \href{http://www.pdb.nmse-lab.ru/}{2D Perovskites Database}, \href{http://www.mrl.ucsb.edu:8080/datamine/thermoelectric.jsp}{Energy Materials Datamining}, and a \href{https://doi.org/10.6084/m9.figshare.11888115.v2}{battery materials database} (see also Table 3 of \citet{butlerMachineLearningMolecular2018}) are available. These techniques, databases, and tools allow for consistent, curated datasets to be more easily produced, accessed, and added to. Thus, for experimental and computational scientists and engineers, an in-depth knowledge of \gls{ml} algorithms or experimental/computational data production methods may not be necessary to leverage data-driven materials predictions. However, it is likely that when datasets are used for materials discovery, an understanding of the strengths and weaknesses of various algorithms, effect of parameters, and database entry details will improve prediction results. Some publications may also give recommendations of potential, promising compounds for the materials community which are then open for other groups to test \cite{balachandranDatadrivenDesignB202020}.

\citet{meredigFiveHighImpactResearch2019} brought up five high impact research areas for materials science \gls{ml}, namely: validation by experiment or physics-based simulation, \gls{ml} approaches tailored for materials data and applications, \gls{hitp} data acquisition capabilities, \gls{ml} that makes us better scientists, and integration of physics within \gls{ml} and \gls{ml} with physics-informed simulations. \citet{oliynykVirtualIssueMachineLearning2019} describe \num{26} articles validated by either experiment or \gls{dft} simulation, and \citet{saalMachineLearningMaterials2020} give a summary of information from \num{23} validation articles (all of which are included in the \num{26} references of \cite{oliynykVirtualIssueMachineLearning2019}) and discuss the five topics in \cite{meredigFiveHighImpactResearch2019}. They point out case studies of appropriately matching an algorithm to a training set for a given prediction type and mention the influence of dataset size on choice of algorithm.

In this work, we 
%discuss common \gls{ml} algorithms (\cref{sec:common-ml}) and common experimental and computational validation techniques (\cref{sec:common-validation}). We compile a list of \nvalid{} experimentally and computationally validated articles, and
sort experimentally and computationally validated articles into three categories based on training dataset size --- \numrange{1}{100} (\cref{sec:1-100}), \numrange{101}{10000} (\cref{sec:101-10000}), \num{10000}+ (\cref{sec:10000+}) --- % as summarized in \cref{tab:summary},
and discuss trends and unique examples within each. We then discuss \gls{cv} approaches geared towards materials discovery (\cref{sec:cv}) and the pursuit of extraordinary materials predictions (\cref{sec:extraordinary}).

We will assume that the reader is familiar with the basic \gls{ml} algorithms discussed in this work. For a treatment of these algorithms, we refer the reader to \citet{butlerMachineLearningMolecular2018}.

% What have other people looked at?
% Why is it interesting to look at size?
% What is our approach in organizing by size?
% What can we learn from the trends in size and technique?

% \input{tab-comp-exp.tex}

% \input{tab-ml-type.tex}

% \input{common-ml} common machine learning algorithms

\section{Training Dataset Size Organization of Validation Articles} \label{sec:size}

To our knowledge, no work before has organized and analyzed the corpus of materials informatics literature as a function of dataset size. However, this could be an appropriate way to organize the literature. After all, different algorithms are certainly better suited for different training data sizes. For example, \glspl{ann} are commonly referred to as data hungry, whereas others such as \gls{gpr} are well-suited to small datasets and generally require sparse approximations for large datasets. We take a rigorous approach by summarizing and comparing \nvalid{} validation articles for three training dataset size ranges, 1-100 (\cref{sec:1-100}), 101-10000 (\cref{sec:101-10000}), and \num{10000}+ (\cref{sec:10000+}), identifying the most common methods used for each, highlighting unique approaches, and commenting on general trends with respect to data.

Some articles \cite{wengSimpleDescriptorDerived2020,xueAcceleratedSearchBaTiO3based2016,sakuraiUltranarrowBandWavelengthSelectiveThermal2019} showed ambiguity with respect to interpreting training dataset size, which could potentially place the article into multiple size ranges for which we take a case-by-case approach. We assign \cite{wengSimpleDescriptorDerived2020,xueAcceleratedSearchBaTiO3based2016} to the \numrange{1}{100} size range and \cite{sakuraiUltranarrowBandWavelengthSelectiveThermal2019} to the \num{10000}+ size range.

% %% Master Table
% \begin{table}[]
%     \centering
% \begin{tabular}{@{}lllllll@{}}
% \toprule
% Ref. &
%   Comp/Exp &
%   \# Training Datapoints &
%   \gls{ml} Type &
%   Material &
%   Classification &
%   Regression \\* \midrule
%     \end{tabular}
%     \caption{Summary of computationally and experimentally validated articles.}
%     \label{tab:summary}
% \end{table}

% \input{tab-summary.tex} %tasks: include specific # training datapoints instead of only size categories (e.g. 78 instead of 1-100), sort Excel table by size

    \subsection{\numrange{1}{100} Training Datapoints} \label{sec:1-100}
    
        \gls{ml} articles that use less than 100 training datapoints \cite{balachandranDatadrivenDesignB202020,balachandranLearningDataDesign2017,balachandranPredictingDisplacementsOctahedral2017,beraIntegratedComputationalMaterials2014,chenMachineLearningAssisted2020,hommaOptimizationHeterogeneousTernary2020,houMachineLearningAssistedDevelopmentTheoretical2019,iwasakiIdentificationAdvancedSpindriven2019,liEfficientOptimizationPerformance2018,rickmanMaterialsInformaticsScreening2019,sendekHolisticComputationalStructure2017,shampDecompositionProductsPhosphine2016,talapatraAutonomousEfficientExperiment2018,wahabMachinelearningassistedFabricationBayesian2020,wakabayashiMachinelearningassistedThinfilmGrowth2019,wengSimpleDescriptorDerived2020,wuMachinelearningassistedDiscoveryPolymers2019,xueAcceleratedSearchBaTiO3based2016,xueInformaticsApproachTransformation2017} are typically \gls{bo} and \gls{bo}/\gls{ad} techniques \cite{hommaOptimizationHeterogeneousTernary2020,houMachineLearningAssistedDevelopmentTheoretical2019,iwasakiIdentificationAdvancedSpindriven2019,liEfficientOptimizationPerformance2018,talapatraAutonomousEfficientExperiment2018,wahabMachinelearningassistedFabricationBayesian2020,wakabayashiMachinelearningassistedThinfilmGrowth2019,wuMachinelearningassistedDiscoveryPolymers2019,xueAcceleratedSearchBaTiO3based2016}, with some \gls{svm} \cite{balachandranDatadrivenDesignB202020,balachandranPredictingDisplacementsOctahedral2017,chenMachineLearningAssisted2020,xueInformaticsApproachTransformation2017} among others (e.g. \gls{sr} \cite{wengSimpleDescriptorDerived2020} and \gls{rf} \cite{wahabMachinelearningassistedFabricationBayesian2020}). This is to be expected, as \gls{bo} and \gls{ad} techniques can allow fewer experiments to be performed while maximizing the exploratory (probing high uncertainty regions) and exploitative (probing favorable prediction regions) gains of optimization. \Gls{bo} techniques benefit from the inherent availability of uncertainty quantification in addition to property predictions. This can be used for uncertainty quantification through models and offer better explanation of results that deviate from predictions or confirmation of results in areas with low uncertainty and high predictive accuracy. Uncertainty can also be quantified with varying degrees of success for other methods (e.g. bootstrapping \gls{svm} results \cite{wahabMachinelearningassistedFabricationBayesian2020,balachandranExperimentalSearchHightemperature2018,wenMachineLearningAssisted2019,xueInformaticsApproachTransformation2017}). We now share examples of experimental \cite{chenMachineLearningAssisted2020,hommaOptimizationHeterogeneousTernary2020,houMachineLearningAssistedDevelopmentTheoretical2019,iwasakiIdentificationAdvancedSpindriven2019,liEfficientOptimizationPerformance2018,rickmanMaterialsInformaticsScreening2019,shampDecompositionProductsPhosphine2016,wahabMachinelearningassistedFabricationBayesian2020,wakabayashiMachinelearningassistedThinfilmGrowth2019,wengSimpleDescriptorDerived2020,wuMachinelearningassistedDiscoveryPolymers2019,xueAcceleratedSearchBaTiO3based2016,xueInformaticsApproachTransformation2017} and computational \cite{balachandranDatadrivenDesignB202020,balachandranLearningDataDesign2017,balachandranPredictingDisplacementsOctahedral2017,sendekHolisticComputationalStructure2017,talapatraAutonomousEfficientExperiment2018} validation articles, first addressing \gls{bo} and \gls{ad}  (\cref{sec:1-100:bo}) followed by those of other \gls{ml} types (\cref{sec:1-100:other}).
    
        \subsubsection{\glsentrytitlecasefull{bo} and \glsentrytitlecasefull{ad} Techniques}
        \label{sec:1-100:bo}
            
\begin{figure}
                \centering
                \includegraphics[width=\textwidth]{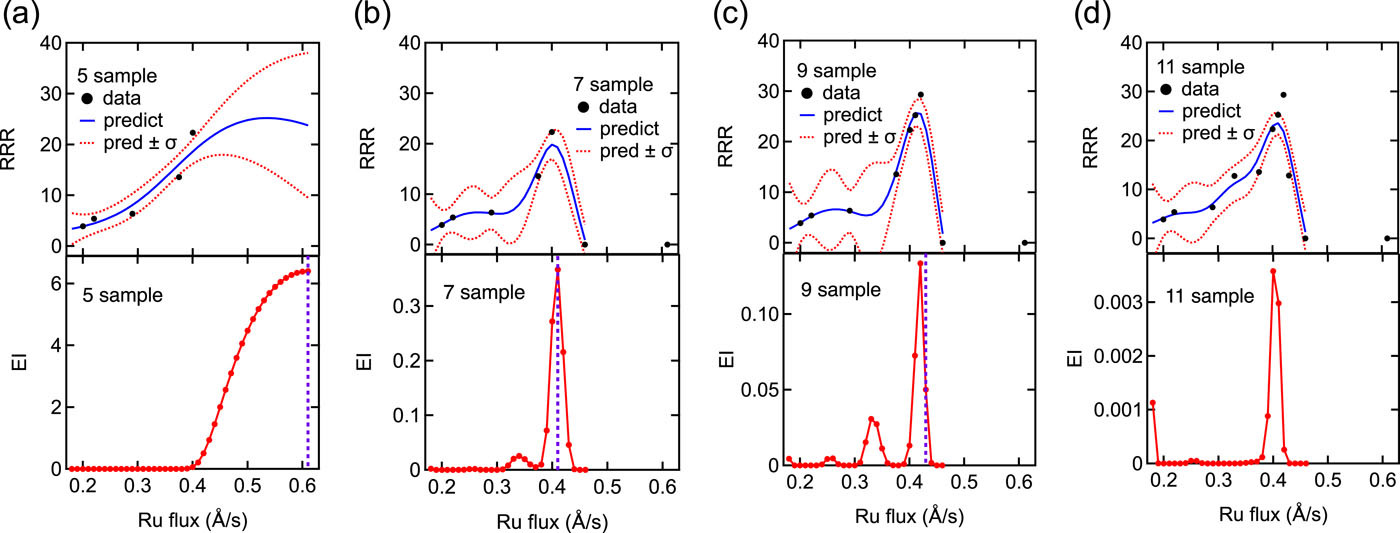}
                \caption{Sequential (i.e. one-variable-at-a-time) \acrfull{bo}/\acrfull{ad} results. Experimental and predicted \acrfull{rrr}, defined as the ratio of resistivity at 300 K to that at 4 K, for 5 random (a), 7 (b), 9 (c), and 11 (d) samples (\#6-11 via \gls{ad}) and \acrfull{ei} values for which the maximum gives the next experiment to perform in the \gls{bo}/\gls{ad} algorithm. Uncertainty tends to decrease in regions near new \gls{ad} datapoints. Reproduced from Wakabayashi, Y. K.; Otsuka, T.; Krockenberger, Y.; Sawada, H.; Taniyasu, Y.; Yamamoto, H. APL Materials 2019, 7 (10)\cite{wakabayashiMachinelearningassistedThinfilmGrowth2019}; licensed under a Creative Commons Attribution (CC BY) license (\url{http://creativecommons.org/licenses/by/4.0/}). }
                \label{fig:wakabayashi2019-bayesian-teaching}
            \end{figure}

            %\citet{liEfficientOptimizationPerformance2018} predicts the optimal doping ratio of \ch{Mn^2+} of \ch{CZTSSe} using 4 training datapoints, 1 validation datapoints, and two adaptive iterations, yielding a solar cell efficiency of 8.9\%.
            \citet{wakabayashiMachinelearningassistedThinfilmGrowth2019} seeks to improve the \gls{rrr} (ratio of resistivity at 300 K to that at 4 K), which is a good measure of the purity of a metallic system, of \gls{mbe} deposited single-crystalline \ch{SrRuO3} thin films. Eleven sequential runs per parameter for three parameters in a \gls{gpr}/\gls{ad} scheme over 33 total growth runs were used. Maximization of \gls{ei} gave the next experiment (\cref{fig:wakabayashi2019-bayesian-teaching}), as is common to many \gls{gpr} implementations. First, 11 runs were used to optimize the Ru flux rate, followed by 11 runs to optimize the growth temperature, and finally 11 runs to optimize the \ch{O_3}-nozzle-to-substrate distance. The highest \gls{rrr} of 51.79 was obtained relative to the highest value ever reported of 80. \citet{wakabayashiMachinelearningassistedThinfilmGrowth2019} comment that a \gls{gpr}/\gls{ad} optimization in 3-dimensional space can be used to further increase the \gls{rrr}. Naturally, the global optimum is constrained by the scope of the design space, as defined by the three parameters used, their upper and lower bounds, and the resolution used, with trade-offs in the complexity and costs associated with additional experiments.
            
            \citet{wahabMachinelearningassistedFabricationBayesian2020} performed 4-dimensional simultaneous optimization to increase the Raman G/D ratios (ratio of the height of the D peak, \SI{1350}{\per\centi\meter}, relative to the height of the G peak, \SI{1580}{\per\centi\meter}) of laser-induced graphene films. Higher G/D ratios indicate better crystallinity and therefore less laser ablation damage. Within 50 optimization iterations, a fourfold increase of Raman G/D ratios (indicating degree of graphitization) relative to common literature values was attained. Twenty initial training datapoints were used, totalling 70 experiments. Instrument precision, gas availability, and user-defined lower and upper limits defined the design space per \cref{tab:wahab-design-space}, which again, constrain the global optimum. While three of the four optimization parameters are technically non-negative continuous variables (i.e. all except gas type), this is a case where instrument resolution constraints dictate a finite number of testable combinations, which we calculate by the Cartesian product to be $554 \times \num{195000} \times 100 \times 3 = \num{32409000000}$. While the total possible number of combinations is large, this finite number only takes on meaning in the context of a minimum correlation length within the true property-design space; if subtle variations in the parameters cause large changes in Raman D/G ratios, this is indicative of a small correlation length and that many more parameter combinations would need to be tested in a brute force approach.
            
            The more likely scenario is that a slight change in e.g. irradiation power is unlikely to produce a significant change in Raman G/D ratios, as the relatively smooth trends exhibited in the partial dependence plots of Figure 6 of \cite{wahabMachinelearningassistedFabricationBayesian2020} suggest. Kernel scale or correlation length (also referred to as smoothness length) is often a hyperparameter of \gls{bo} methods, for which a proper choice can greatly affect the rate at which a sequential optimization improves property predictions and approximates the true property-design space. This is an important case where domain knowledge can play an important role, such as by imposing initial conditions or constraints on the kernel scale or other hyperparameters such as property standard deviation. Even in non-\gls{bo} algorithms, estimations of the local smoothness of the true function being predicted gives context to large combinatoric metrics given in some property-design \gls{ml} articles; a large number of possible parameter combinations (especially of arbitrarily discretized variables that would otherwise be continuous) does not necessarily correlate with high model complexity if the design space has large correlation lengths.
            
            % \usepackage{booktabs}
            % \usepackage{graphicx}
            \begin{table}
            \centering
            \caption{Parameter space limits for \glsentrytitlecasefull{mbo} of laser-induced graphene Raman G/D ratio maximization. Reproduced with permission from Wahab, H.; Jain, V.; Tyrrell, A. S.; Seas, M. A.; Kotthoff, L.; Johnson, P. A. Carbon 2020, 167, 609–619. \cite{wahabMachinelearningassistedFabricationBayesian2020}.}
            \label{tab:wahab-design-space}
            \resizebox{\textwidth}{!}{%
            \begin{tabular}{@{}lllll@{}}
            \toprule
            Parameters                       & Lower Limit         & Upper Limit        & Instrument Precision & Number possible values \\ \midrule
            CW-laser power (\si{\watt})      & 0.01                & 5.55               & 0.01                 & 554                    \\
            Irradiation time (\si{\second})  & 0.500               & 20.000             & 0.001                & \num{195000}           \\
            Gas pressure (\si{\kilo\pascal}) & 0                   & 6894.76            & 68.9476              & 100                    \\
            Gas type                         & \multicolumn{2}{l}{Argon  Nitrogen  Air} & -                    & 3                      \\ \bottomrule
            \end{tabular}%
            }
            \end{table}
            
            \citet{hommaOptimizationHeterogeneousTernary2020} give another effective and straightforward application of \gls{bo} in pursuit of enhanced Li-ion conductivities in hetereogenous ternary \ch{Li_3PO_4{-}Li_3BO_3{-}Li_2SO_4} solid electrolytes. The ternary mixture is adaptively tuned, beginning with 15 gridded training data, followed by 10 \gls{ad} iterations and yielding a compound 3x higher than any binary composition. Such \gls{bo}/\gls{ad} approaches are becoming increasingly accessible by experimentalists due to the increasing number of powerful, easy-to-use code packages such as \gls{combo} \cite{uenoCOMBOEfficientBayesian2016} as used in \cite{hommaOptimizationHeterogeneousTernary2020} and the similarity with \gls{doe}, a familiar technique to many experimentalists.
            
            \citet{liEfficientOptimizationPerformance2018} used \gls{gpr} to predict the optimal doping ratio of \ch{Mn^{2+}} ions in \ch{CZTSSe} solar cells, experimentally achieving a highest solar cell efficiency of 8.9\%. Four training datapoints and two \gls{ad} iterations were used where all training data were a multiple of 5\%. It appears that the solar cell exhibits a single peak as a function of dopant ratio, suggesting a smooth and simple underlying function which is predicted.
            
            \sloppy \citet{houMachineLearningAssistedDevelopmentTheoretical2019} used the \gls{gpr} implementation in \gls{combo} to maximize the power factor of \ch{Al_{23.5+x}Fe_{36.5}Si_{40-x}} thermoelectrics by 40\% at 510\,K relative to their starting sample ($x=0$) via tuning the Al-Si ratio ($x$). Forty-eight training datapoints were used across two variables, namely temperature (measured at approximately fixed spacing between \SI{300}{\kelvin} and \SI{850}{\kelvin}) and Al/Si ratio ($x$).
            
\begin{figure}
                \centering
                \includegraphics[width=0.8\textwidth]{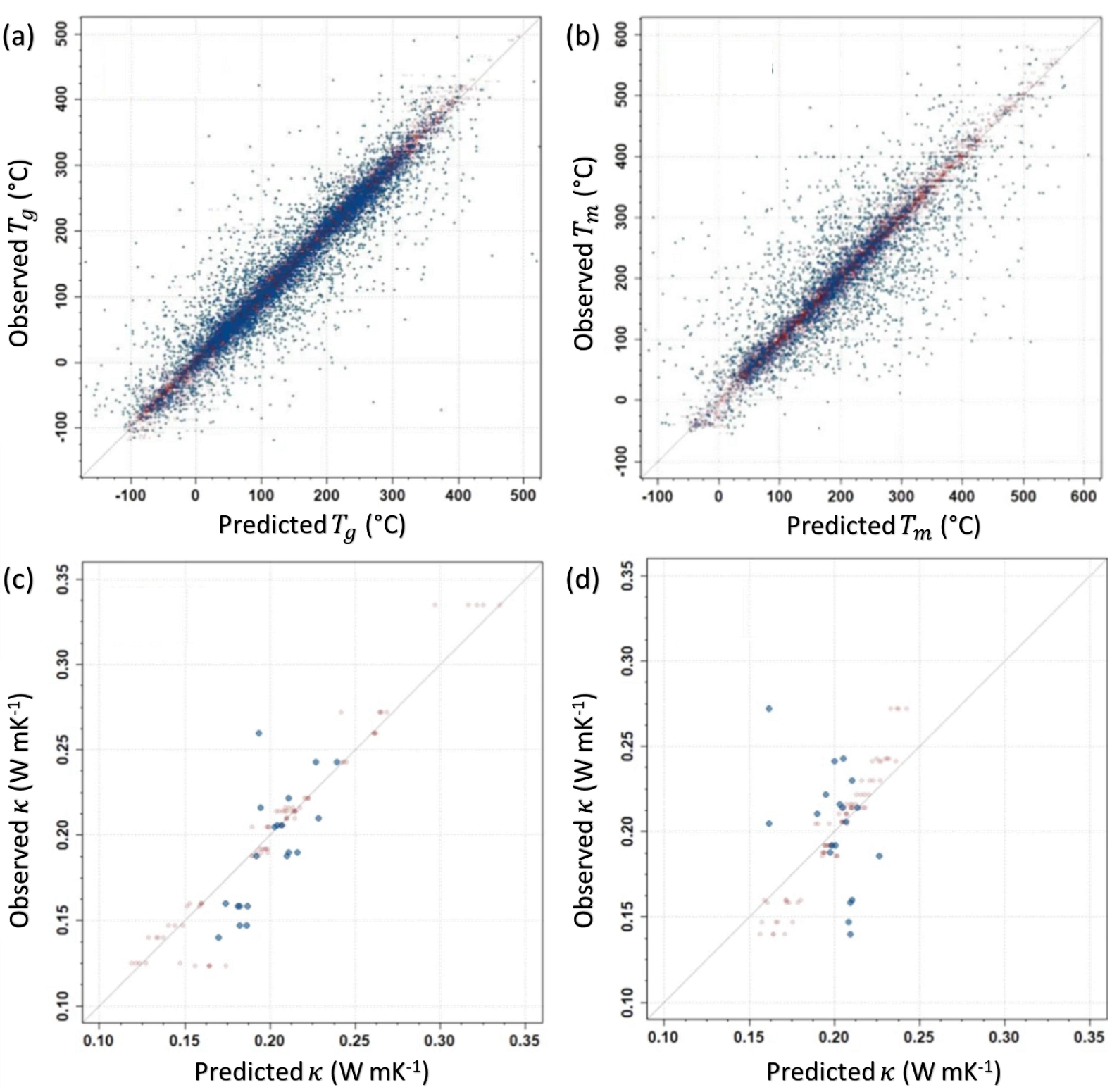}
                \caption{Transfer learning results for a Bayesian molecular design of polymer structures. Glass transition temperature and melting temperature act as proxy models for thermal conductivity and for which parity plots are shown in (a) and (b), respectively. Use of transfer learning enhances prediction accuracy relative to a direct learning approach for which parity plots are shown in (c) and (d), respectively. Adapted from Wu, S.; Kondo, Y.; Kakimoto, M.; Yang, B.; Yamada, H.; Kuwajima, I.; Lambard, G.; Hongo, K.; Xu, Y.; Shiomi, J.; Schick, C.; Morikawa, J.; Yoshida, R. npj Comput Mater 2019, 5 (1), 66\cite{wuMachinelearningassistedDiscoveryPolymers2019}; licensed under a Creative Commons Attribution (CC BY) license (\url{http://creativecommons.org/licenses/by/4.0/}).}
                \label{fig:wu2019-transfer-results}
            \end{figure}

            \citet{wuMachinelearningassistedDiscoveryPolymers2019} employed Bayesian molecular design paired with transfer learning towards discovering high thermal conductivity polymers. The Bayesian molecular design strategy generated a library of potential polymer structures by representing polymer structures digitally via a \gls{smiles} string. For example, phenol (\ch{C_6H_6O}) would be represented as $\texttt{C1=CC=C(C=C1)O}$, encoding double bonds as \texttt{=}, start and terminal of ring closures by common digits such as \texttt{1}, and side chains via parentheses enclosures. They imposed prior information that reduced sampling probability of chemically unfavorable or unrealistic structures and sampled the updated distribution by a \gls{smc} scheme. Twenty-eight training structures with thermal conductivity data were used (total 322 observations), and \num{5917} and \num{3234} structures were used for the surrogate properties of glass transition temperature and melting temperature, respectively. The transfer learning approach improves \gls{mae} from \SI{0.0327}{\watt\per\milli\kelvin} to \SI{0.0204}{\watt\per\milli\kelvin} as shown in parity plots (\cref{fig:wu2019-transfer-results}c, \cref{fig:wu2019-transfer-results}d), and surrogate model parity plots are also shown (\cref{fig:wu2019-transfer-results}a, \cref{fig:wu2019-transfer-results}b). Additionally, they synthesized three predicted polymers and demonstrated experimental thermal conductivities similar to state-of-the-art polymers in non-composite thermoplastics.
        
            \citet{talapatraAutonomousEfficientExperiment2018} used an extension of the typical \gls{gpr} scheme in a \gls{bma} approach. Rather than select a single model for a small training dataset, a weighted average of \gls{gpr} models with different parameters was used. The weights were assigned based on the prior probability and likelihood of the observed data for each model, and the weights were updated as more data was iteratively added (i.e. the likelihood of the observed data for each model was updated). As the number of observations increases, it is expected that better predictive models progressively are weighted more heavily and that the \gls{bma} model predictions improve. Because their \gls{bma} implementation depends on many individual \gls{gpr} models, without sparse approximations, such an approach may be limited to small datasets for which many \gls{gpr} models can be fitted efficiently. The \gls{bma} approach was applied to polycrystalline nanolaminates ternary layered carbides/nitrides. These are also called \gls{max} phases, where M is a transition metal, A is an A group element, X is C and/or N, and n = \numrange{1}{3}  \cite{ghoshConsolidationSynthesisMAX2012}. They performed single-objective optimization of maximal polycrystalline bulk modulus and minimal shear modulus structures (\gls{dft}-based values) and multi-objective of maximum bulk and minimum shear modulus. Six feature sets were used as the candidate models for averaging based on domain knowledge. The authors used 10 initial training \gls{dft} simulations and a budget of 70 \gls{dft} optimization iterations where a second-order \gls{bma} approximation is used. Both maximizing bulk modulus and minimizing shear modulus revealed that the best model almost always had the highest weight, indicating to us that hyperparameter optimization (e.g. choice of feature set) via an appropriate acquisition function (e.g. \gls{ei}) may be sufficient and yield faster convergence than \gls{bma}. However, it remains to be seen if \gls{bma} effectively safeguards against poor models better than a (simpler) single hyperparameter optimization step near the beginning of an \gls{ad} process; poor models rarely have high weight coefficients for the considered dataset, especially when at least 10 training datapoints are available.
            
            \citet{xueAcceleratedSearchBaTiO3based2016} incorporated domain knowledge in the form of a quadratic equation describing a phase boundary of interest based on Landau-Devonshire theory into a \gls{bo} scheme in pursuit of more vertical \glspl{mpb} in Pb-free \ch{BaTiO3}-based piezoelectrics. State-of-the-art Pb-free \ch{BaTiO3}-based piezoelectrics exhibit large electromechanical responses; however, they also exhibit high temperature sensitivity. More vertical \glspl{mpb} are correlated with less temperature sensitivity, providing motivation for the work in \citet{xueAcceleratedSearchBaTiO3based2016}. They used 19 training phase diagrams based on 231 experiments, 83 of which were used as inputs and served the sole purpose of obtaining a fit to the quadratic equation for each phase diagram. Six features based on atomic, crystal chemistry, and electronic structure properties were considered. They successfully predicted and synthesized a piezoelectric with \SI{15}{\percent} less curvature in the \gls{mpb} and lower temperature sensitivity. An important measure of electromechanical response is the longitudinal piezoelectric strain coefficient ($d_{33}$) for which higher values are more favorable. While $d_{33}$ values around \SIrange{500}{600}{\pico\coulomb\per\newton} have been achieved and are present in the initial training data (the performance being a partial motivator for Pb-free \ch{BaTiO3}-based piezoelectrics), the synthesized material exhibits a $d_{33}$ value of \SI{85}{\pico\coulomb\per\newton}, highlighting an opportunity to use multi-objective optimization to create a material with both large $d_{33}$ and low temperature sensitivity.
            
\begin{figure}
                \centering
                \includegraphics[width=350pt]{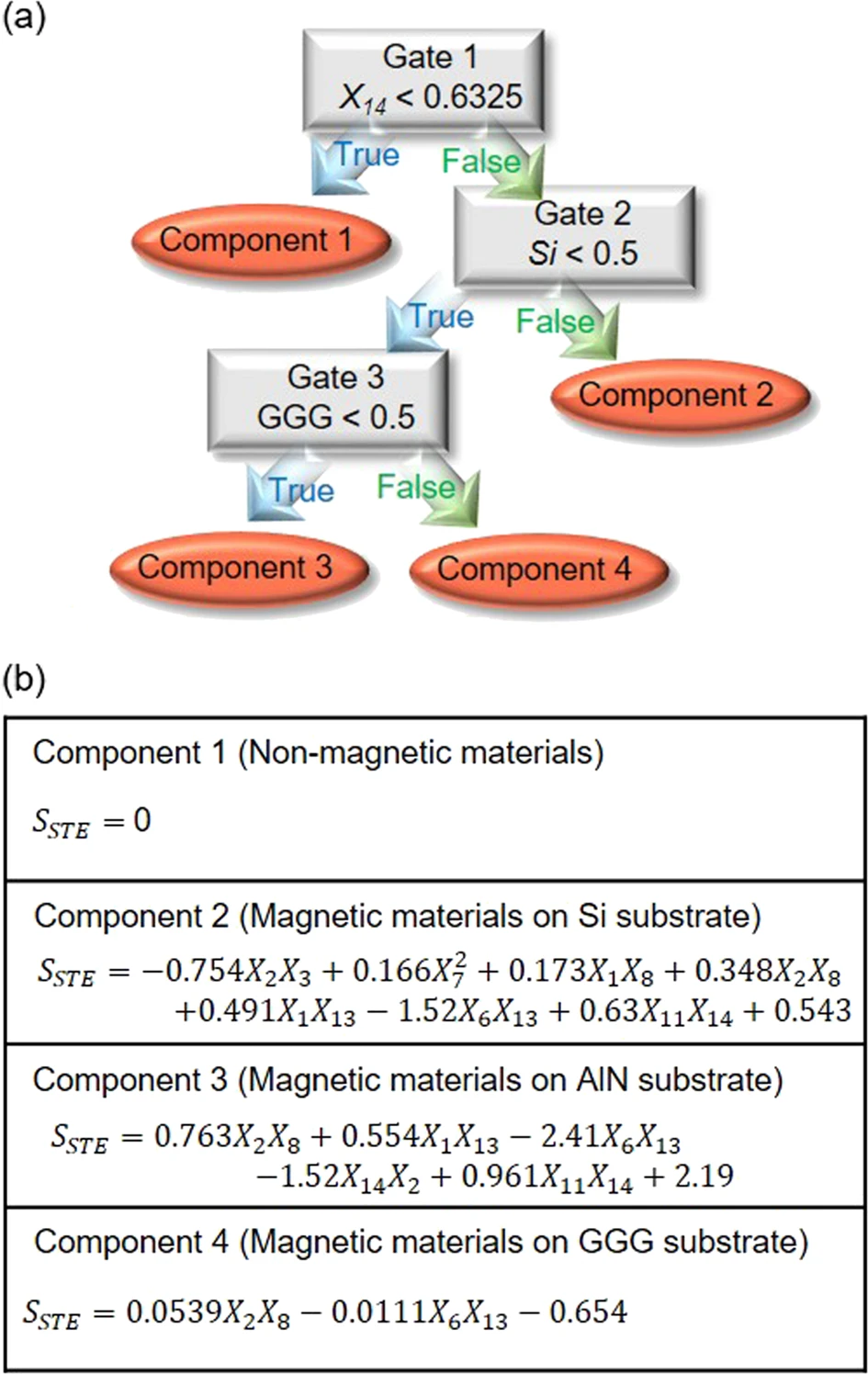}
                \caption{An interpretable model produced by a state-of-the-art \gls{ml} method, \acrfull{fabhmes}, which can be summarized/visualized via a tree structure with components (i.e. regression models) (a) that are accessed by gates (a). Regression models for the four components selected via a Bayesian approach (b). Reproduced from Iwasaki, Y.; Sawada, R.; Stanev, V.; Ishida, M.; Kirihara, A.; Omori, Y.; Someya, H.; Takeuchi, I.; Saitoh, E.; Yorozu, S. npj Computational Materials 2019, 5 (1), 6–11.\cite{iwasakiIdentificationAdvancedSpindriven2019}; licensed under a Creative Commons Attribution (CC BY) license (\url{http://creativecommons.org/licenses/by/4.0/}).}
                \label{fig:iwasaki2019-fab-hmes}
            \end{figure}

            \citet{iwasakiIdentificationAdvancedSpindriven2019} employ a state-of-the-art, accurate, interpretable \gls{ml} method called \gls{fabhmes}, which "constructs a piecewise sparse linear model that assigns sparse linear experts to individual partitions in feature space and expresses whole models as patches of local experts" \cite{iwasakiIdentificationAdvancedSpindriven2019}. They use 21 training datapoints and 17 predictors to identify and synthesize a \gls{ste} material with the largest spin-driven thermopower measured to date and provide possible insights into new domain knowledge. Thermopower, or the Seebeck coefficient, gives a measure of the voltage induced by a thermal gradient and higher thermopower leads to better performance of thermoelectric generators and coolers. While the first 14 features come from \gls{dft} calculations, it is important to realize that the \gls{dft} parameters were set up based on experimental composition information from \gls{xrf} and experimental crystal structure information from \gls{xrd}. They took \gls{xrf} and \gls{xrd} data at different points along a "[compositional] spread thin film" made via a \gls{cms} technique (\gls{hitp}). "For instance, fcc, bcc, and L1\_0 structures are the possible crystal structures in \ch{FePt} binary alloy, which were determined by the combinatorial XRD experiments" (from Supporting Information of \cite{iwasakiIdentificationAdvancedSpindriven2019}). Features 15-17 are experimental; they cut the sample into small sections and measured thermopower. Their approach is reminiscent of a digital twin, where an object goes through complementary simulation and experimental testing. Their validation was experimental, yielding a material with a thermopower of approximately \SI{13}{\micro\volt\per\kelvin} compared to to typical state of the art \glspl{ste} thermopowers below \SI{10}{\micro\volt\per\kelvin}. The authors argue that the interpretable and visualizable \gls{fabhmes} model they generated (\cref{fig:iwasaki2019-fab-hmes}) allowed them to discover new insight that thermopower ($S_{\mathrm{STE}})$ and the product term ($X_2X_8$) of Pt content ($X_2$) and Pt spin polarization ($X_8$) are positively correlated. They suggest that \gls{ml} could be useful in observing previously unexplained phenomena.
            
        % \subsubsection{Inverse Design}
        %             Experimentally validated inverse design article \cite{talapatraAutonomousEfficientExperiment2018}
        %             Other inverse design articles \cite{jainInverseMethodsMaterial2014,kimMaterialsSynthesisInsights2017,malikMaterialsGraphTransformer2020,pellegrinoMachineLearningApproach2020,recatala-gomezAcceleratedThermoelectricMaterials2020}
        
        \subsubsection{Non- \glsentrytitlecasefull{bo}}
            \label{sec:1-100:other}

            Other \gls{ml} methods used in the 1-100 training dataset size include \gls{sr} \cite{wengSimpleDescriptorDerived2020}, \gls{svm} \cite{balachandranDatadrivenDesignB202020,balachandranPredictingDisplacementsOctahedral2017,chenMachineLearningAssisted2020,xueInformaticsApproachTransformation2017}, \gls{pr} \cite{xueInformaticsApproachTransformation2017}, and \gls{rf} \cite{wahabMachinelearningassistedFabricationBayesian2020}.
            
            In a \gls{sr} scheme, \citet{wengSimpleDescriptorDerived2020} randomly generated \num{43000000} symbolic equations and used these to predict and synthesize 13 new perovskites based on lowest ratio of octahedral factor ($\mu$) to tolerance factor ($t$), a new descriptor ($\mu/t$) they identified by visually analyzing equations on the Pareto front of \gls{mae} vs. equation complexity. Five of the thirteen synthesized perovskites turned out to be pure, and four out of those five are among the highest \gls{oer} perovskites, where high \gls{oer} correlates with better catalytic performance of perovskites in e.g. water-splitting into hydrogen or metal-air batteries. Training data consisted of 90 datapoints across 18 in-house synthesized, well-studied, oxide perovskite catalysts (18 perovskites × 4 samples × 3 measurements × 5 current densities = 1080 measurements). Because \gls{mae} was used as the metric in the approach, from a model perspective, using a set of repeated measurements of a given perovskite and current density as training data is identical to using the average of the set. Naturally, using repeated measurements across multiple samples to decrease observed noise in the average measured property likely improved the final results of their model and is certainly a wise practice when feasible. Their implementation of \gls{sr} involved a genetic algorithm approach according to Figure 2b of \cite{wengSimpleDescriptorDerived2020}. With this global optimization approach, a Pareto front of \gls{mae} vs. complexity for 8460 mathematical formulas was generated from which they identified and studied the recurring $\mu/t$ descriptor and generated a list of promising perovskite compounds based on minimizing $\mu/t$.
            
            \citet{balachandranDatadrivenDesignB202020} applied \gls{svm} using 18 training datapoints and a single test datapoint from experimental literature to enhance helical transition temperature of known B20 compounds for spintronics applications via elemental substitution. \Gls{dft} validated the prediction that Sn can enhance the transition temperature of Fe(Ge,Sn) compounds and they suggest certain experiments for other researchers to perform. \citet{balachandranPredictingDisplacementsOctahedral2017} employed \gls{svm} to predict breaks in spatial inversion symmetry due to displacement of cations using 14 published \gls{dft} training data and made 10 predictions for materials without existing \gls{dft} data which they then validated by \gls{dft}. This is useful for identifying promising ferroelectrics because of a correlation between ionic displacement magnitude and Curie temperature, where a high Curie temperature is desired for applications such as ferroelectric capacitor-based computer RAM and heat sensors.
            
            \citet{chenMachineLearningAssisted2020} performs a multi-objective, \gls{ad} optimization to increase the strength and ductility of an as-cast ZE62 (Mg6 wt.\% Zn-2 wt.\% RE (Y, Gd, Ce, Nd)) Mg alloy, which is of interest for aerospace, vehicle, electronic, and biomedical applications due to low density, high stiffness, and high biocompatibility. Ten initial training datapoints selected by orthogonal design are used to train a \gls{svm} model, followed by iterative recommendations of next parameters for a four-parameter experiment via either a Pareto front vector or scalarization approach. In the Pareto front vector approach, the angle between two vectors $w^t$ and $w^p$ is minimized, where $w^t$ and $w^p$ are vectors from the origin to the target and the virtual (i.e. \gls{svm}-based) Pareto front, respectively. The target point used in their work was \SI{15.6}{\percent} strain and \SI{157.2}{\mega\pascal} yield strength, as obtained via Figure 2c of \cite{chenMachineLearningAssisted2020} and DataThief III \cite{tummersDataThiefIIISoftware2015}. In the scalarization approach, a point in the virtual space with minimum distance to the target is found. In either approach, when minimization is complete, the minimized point in the virtual space defines the set of parameters for the next experiment. Both approaches performed similarly, and the latter gave a material with strength and ductility improved by 27\% and 13.5\%, respectively, relative to the initial training dataset via 4 iterations of experiments.
            
            While \citet{wahabMachinelearningassistedFabricationBayesian2020} falls primarily into the category of \gls{bo} and was discussed in \cref{sec:1-100:bo}, a \gls{rf} surrogate model with 500 trees is used due to the presence of both continuous numerical and discrete categorical variables; however, it is worth noting that \gls{gpr} and other methods can handle both types simultaneously via dummy variables \cite{MathWorksHelpCenter2020}.
            
            \citet{sendekHolisticComputationalStructure2017} demonstrated a new large-scale computational screening method capable of identifying promising candidate materials for solid state electrolytes for lithium-ion batteries. First, \num{12831} lithium containing crystalline solids were screened for high structural and chemical stabilities, low electronic conductivity, and low cost down to 300 potential candidates. A training set of 40 crystal structures and experimentally reported ionic conductivity values from literature were used to train a superionic classification model using logistic regression to identify which of those candidate structures are most likely to exhibit fast lithium conduction. They identify a 5-feature model, selected from 20 potential atomic and chemical property features, that resulted in the lowest \gls{cvmr} and \gls{tmr} of \SI{10}{\percent} (in other words, 4 of the 40 training points are misclassified). From the 300 potential candidates, the model narrowed that down to 21 crystal structures that showed promise as electrolytes. \citet{sendekHolisticComputationalStructure2017} concluded that a multi-descriptor model exhibits the highest degree of predictive power, compared to stand alone simple atomistic descriptor functions, and it also served as a first step towards a robust data-driven model to screen for promising solid electrolyte structures.

            \citet{xueInformaticsApproachTransformation2017} trained five different iterative statistical learning models to make rapid predictions of the transformation temperature of NiTi-based alloys from a training set of 53 synthesized alloys and three features (Pauling electronegativity, metallic radius, and Waber Cromer’s pseudopotential radii). A bootstrap resampling method was applied to the dataset with 53 points and used to train a \gls{lr}, \gls{pr}, \gls{svm} with a \gls{rbf} kernel, \gls{svm} with a linear kernel, and \gls{svm} with a polynomial kernel. Using validation from a high precision testing dataset with 23 points on the transformation temperatures of \ch{NiTi}-based shape-memory alloys, the \gls{pr} model had the lowest error out of the 5 with a \gls{mse} of about \SI{40}{\celsius}. Next, an adaptive design loop used a trade-off between exploration and exploitation to find the highest transformation temperature alloy in the virtual dataset consisting of \num{1652417} unexplored alloys. Three different selectors (max, \gls{ego}, and \gls{kg}) were employed for two iterations to improve the virtual dataset by suggesting the next candidate material for experiment. Experimental validation found that the \gls{pr} model significantly improves after the virtual dataset is improved (the \gls{mse} decreases from \SI{15.7}{\celsius} to \SI{15.0}{\celsius}). \citet{xueInformaticsApproachTransformation2017} demonstrated a systematic learning and adaptive design framework that can guide future synthesis and discovery of new materials with certain desired properties.
            
\begin{figure}
                \centering 
                \includegraphics[width=15cm]{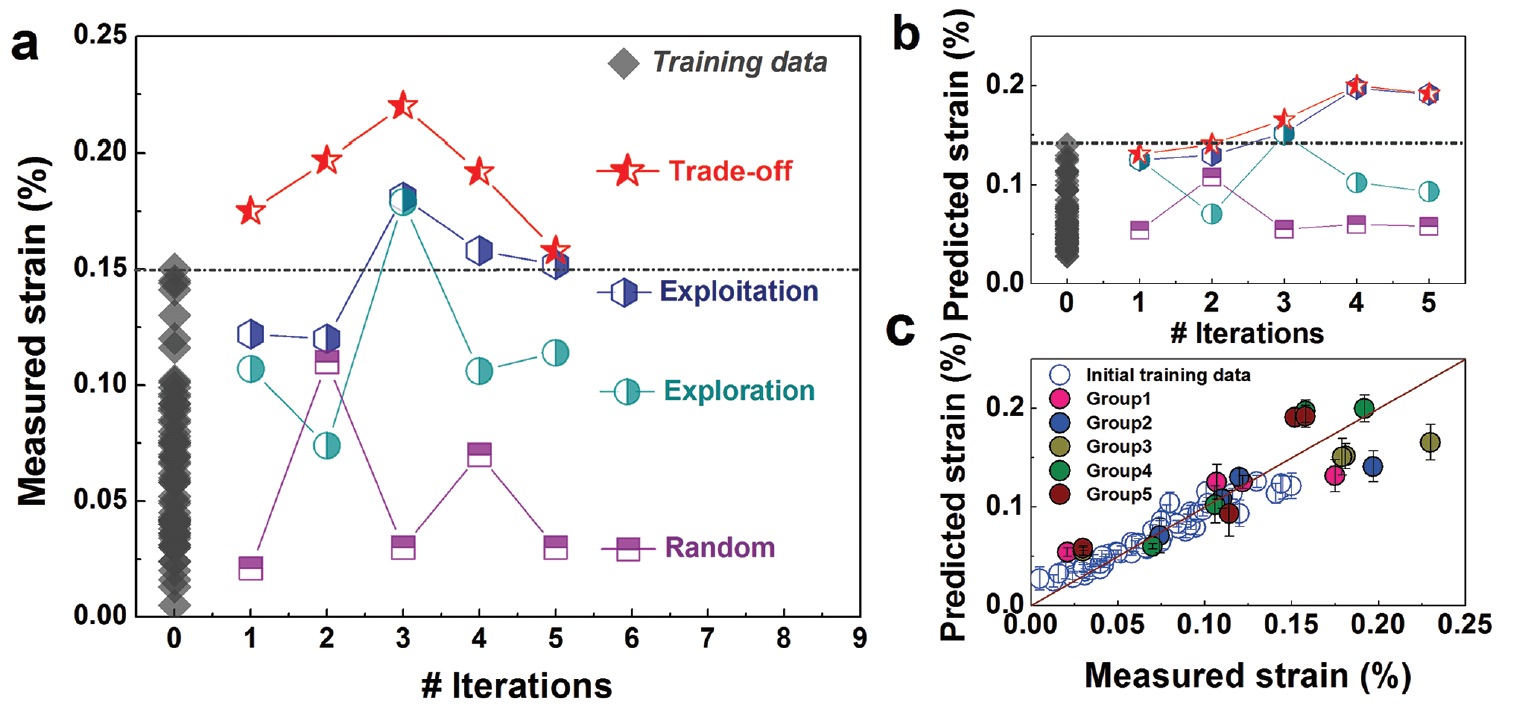}
                \caption{Overall performance of the trade-off between exploration (probing high uncertainty) and exploitation (probing high performance) design methodology. (a) The trade-off between exploration and exploitation methodology gives higher measured electrostrain (\SI{}{\percent}) in comparison with the other four design methodologies for an increasing number of iterations. (b) Predictions made from the model using the trade-off strategy. (c) Parity plot showing the accuracy of the trade-off model's predicted strains \SI{}{\percent} in comparison to new synthesized compounds. Reproduced with permission from Yuan, R.; Liu, Z.; Balachandran, P. V.; Xue, D. D.; Zhou, Y.; Ding, X.; Sun, J.; Xue, D. D.; Lookman, T. Advanced Materials 2018, 30 (7).\cite{yuanAcceleratedDiscoveryLarge2018} }
                \label{fig:yuan2018-exploration-exploitation}
            \end{figure}

            \citet{yuanAcceleratedDiscoveryLarge2018} used an \gls{svm} model with a \gls{rbf} kernel and 61 experimental training datapoints to discover new Pb-free \ch{BaTiO3} (BTO) based piezoelectrics with large electrostrain. The model screened \num{605000} unexplored compositions and performed five \gls{ad} iterations in sets of four experiments. Validation compounds were experimentally synthesized following predictions from four strategies: exploitation, exploration, trade-off between the former two, and random selection (\cref{fig:yuan2018-exploration-exploitation}). An optimized trade-off between exploration (high uncertainty regions) and exploitation (best predicted performance regions), was achieved by experimentally comparing multiple design strategies. Thus, they were able to produce an optimal criterion for the synthesis of the piezoelectric \ch{(Ba_{0.84}Ca_{0.16})(Ti_{0.90}Zr_{0.07}Sn_{0.03})O3}, for which the largest electrostrain was \SI{0.23}{\percent} in the BTO family. The trade-off between exploration and exploitation is especially significant because it provides a good precedent in guiding experiments in materials design.
            
    \subsection{\numrange{101}{10000} Training Datapoints} \label{sec:101-10000}
    
        Many of the \gls{ml} validation articles that have 101-10000 training datapoints \cite{balachandranAdaptiveStrategiesMaterials2016,balachandranExperimentalSearchHightemperature2018,buciorEnergybasedDescriptorsRapidly2019,caoHowOptimizeMaterials2018,gzylHalfHeuslerStructuresFullHeusler2020,gzylSolvingColoringProblem2019,kauweMachineLearningPrediction2018,luAcceleratedDiscoveryStable2018,mannodi-kanakkithodiMachineLearningStrategy2016,minMachineLearningAssisted2018,nikolaevAutonomyMaterialsResearch2016,oliynykClassifyingCrystalStructures2016,oliynykDisentanglingStructuralConfusion2017,oliynykHighThroughputMachineLearningDrivenSynthesis2016,raccugliaMachinelearningassistedMaterialsDiscovery2016,renAcceleratedDiscoveryMetallic2018,sekoMatrixTensorbasedRecommender2018,tehraniMachineLearningDirected2018,wenMachineLearningAssisted2019,yanOptimizationThermalConductivity2020,zhangFindingNextSuperhard2020,zhuoEvaluatingThermalQuenching2020,zhuoIdentifyingEfficientThermally2018} use \gls{svm} \cite{balachandranAdaptiveStrategiesMaterials2016,balachandranExperimentalSearchHightemperature2018,caoHowOptimizeMaterials2018,gzylHalfHeuslerStructuresFullHeusler2020,gzylSolvingColoringProblem2019,kauweMachineLearningPrediction2018,luAcceleratedDiscoveryStable2018,minMachineLearningAssisted2018,oliynykClassifyingCrystalStructures2016,oliynykDisentanglingStructuralConfusion2017,raccugliaMachinelearningassistedMaterialsDiscovery2016,tehraniMachineLearningDirected2018,wenMachineLearningAssisted2019,yanOptimizationThermalConductivity2020,zhuoEvaluatingThermalQuenching2020,zhuoIdentifyingEfficientThermally2018}. There are also other examples such as ensemble \cite{gzylHalfHeuslerStructuresFullHeusler2020}, \gls{ann} \cite{wenMachineLearningAssisted2019}, \gls{rf} \cite{kauweMachineLearningPrediction2018}, \gls{dt} \cite{raccugliaMachinelearningassistedMaterialsDiscovery2016,wenMachineLearningAssisted2019}, \gls{rfe} \cite{zhuoEvaluatingThermalQuenching2020}, \gls{lasso} \cite{buciorEnergybasedDescriptorsRapidly2019}, \gls{crfs} \cite{oliynykDisentanglingStructuralConfusion2017,gzylSolvingColoringProblem2019,gzylHalfHeuslerStructuresFullHeusler2020}, \gls{doe} \cite{caoHowOptimizeMaterials2018}, \gls{lr} \cite{kauweMachineLearningPrediction2018,raccugliaMachinelearningassistedMaterialsDiscovery2016,wenMachineLearningAssisted2019}, \gls{pr} \cite{wenMachineLearningAssisted2019}, \gls{pls} \cite{gzylHalfHeuslerStructuresFullHeusler2020}, matrix-based recommender \cite{sekoMatrixTensorbasedRecommender2018}, \gls{smote} \cite{gzylHalfHeuslerStructuresFullHeusler2020}, \gls{knn} \cite{gzylHalfHeuslerStructuresFullHeusler2020,raccugliaMachinelearningassistedMaterialsDiscovery2016,wenMachineLearningAssisted2019}, and \gls{krr} \cite{luAcceleratedDiscoveryStable2018} approaches. Of the "other" \gls{ml} articles, only \cite{buciorEnergybasedDescriptorsRapidly2019,mannodi-kanakkithodiMachineLearningStrategy2016,nikolaevAutonomyMaterialsResearch2016,oliynykHighThroughputMachineLearningDrivenSynthesis2016,renAcceleratedDiscoveryMetallic2018,sekoMatrixTensorbasedRecommender2018,zhangFindingNextSuperhard2020} are not already included in the \gls{svm} group, indicating that \gls{svm} is often combined or compared with other methods. Most of the \gls{svm} articles described here employ a \gls{rbf} kernel, imposing smooth, Gaussian behavior on the predicted properties. %Some benefits of \gls{svm} are... Some limitations of \gls{svm} are... 
        We now share examples of experimental \cite{balachandranExperimentalSearchHightemperature2018,buciorEnergybasedDescriptorsRapidly2019,caoHowOptimizeMaterials2018,gzylHalfHeuslerStructuresFullHeusler2020,gzylSolvingColoringProblem2019,kauweMachineLearningPrediction2018,minMachineLearningAssisted2018,oliynykClassifyingCrystalStructures2016,oliynykDisentanglingStructuralConfusion2017,oliynykHighThroughputMachineLearningDrivenSynthesis2016,raccugliaMachinelearningassistedMaterialsDiscovery2016,renAcceleratedDiscoveryMetallic2018,tehraniMachineLearningDirected2018,wenMachineLearningAssisted2019,yanOptimizationThermalConductivity2020,zhangFindingNextSuperhard2020,zhuoEvaluatingThermalQuenching2020,zhuoIdentifyingEfficientThermally2018} and computational \cite{balachandranAdaptiveStrategiesMaterials2016,luAcceleratedDiscoveryStable2018,mannodi-kanakkithodiMachineLearningStrategy2016,sekoMatrixTensorbasedRecommender2018} validation articles, addressing \gls{svm}/\gls{ad} (\cref{sec:101-10000:svm-ad}), \gls{svm}/\gls{crfs} (\cref{sec:101-10000:svm-crfs}), general \gls{svm} (\cref{sec:101-10000:svm-general}), and non-\gls{svm} (\cref{sec:101-10000:non-svm}).
        
        \subsubsection{\glsentrytitlecasefull{svm} and \glsentrytitlecasefull{ad} }
        \label{sec:101-10000:svm-ad}
        
\begin{figure}
                \centering
                \includegraphics[width=\textwidth]{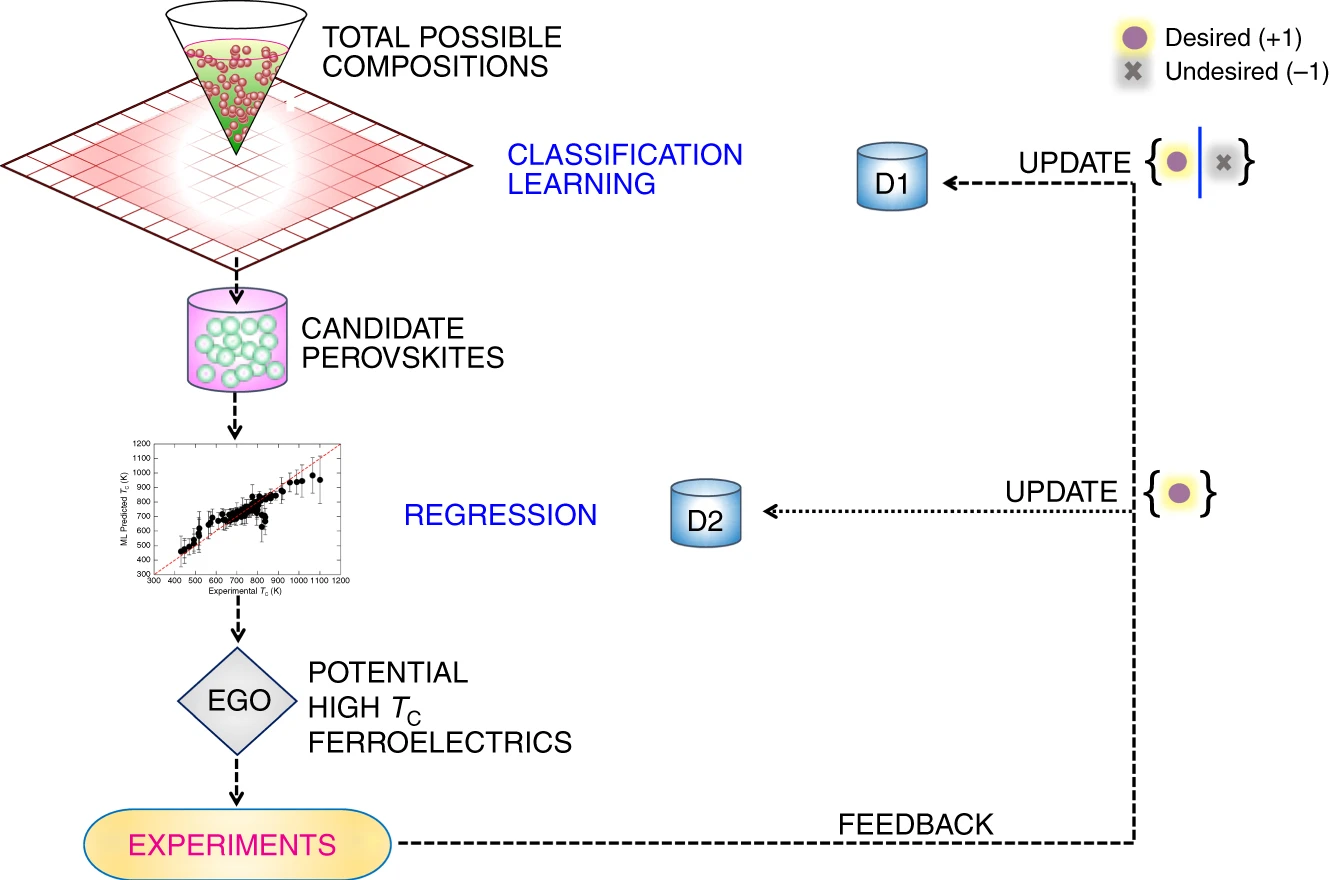}
                \caption{Two-step machine learning algorithm involving \acrfull{ad}. Step 1: Screening by classification algorithm to identify perovskite compositions that can be made without impure phases. Step 2: Predict Curie temperature via \acrfull{svm} regression and identify promising candidates using \acrfull{ego}. Both successful and failed experiments train the classification model via \gls{ad}, for which only successful experiments are passed on to the regression model. Reproduced from Balachandran, P. V.; Kowalski, B.; Sehirlioglu, A.; Lookman, T. Nature Communications 2018, 9 (1)\cite{balachandranExperimentalSearchHightemperature2018}; licensed under a Creative Commons Attribution (CC BY) license.}
                \label{fig:balachandran2018-feedback}
            \end{figure}

            \citet{balachandranExperimentalSearchHightemperature2018} used \gls{svm} and a two-step classification then regression approach with 167 and 117 initial training datapoints, respectively, to predict new high Curie temperature ($T_C$) \ch{xBi[Me_y'Me_y'']O3}--(1-x)\ch{PbTiO3} perovskite compounds through 5 iterations of \gls{ad}. Of the 10 compounds they experimentally synthesized, 6 were perovskites. With an initial approach using only regression and no classification, a perovskite was predicted and synthesized, but discovered to be non-pure. The classification algorithm includes training data from non-pure perovskites and is aimed at identifying promising regions in the four-parameter design space ($x$, $y$, \ch{Me_y'}, and \ch{Me_y''} in \ch{xBi[Me_y'Me_y'']O3}--(1-x)\ch{PbTiO3}) that are more likely to produce pure perovskite phases. The regression step is then aimed at identifying specific compositions with high $T_C$ for ferroelectric applications. In the \gls{ad} scheme, only compositions which are classified as perovskites are updated in the regression model, and a \gls{ego} scheme \cite{jonesEfficientGlobalOptimization1998} is used to identify new compositions for synthesis (\cref{fig:balachandran2018-feedback}). Since only a single iteration was used for the regression-only approach before switching to a two-step approach, it is unclear to what extent the classification algorithm affected the regression model and subsequent success of choosing high $T_C$ candidates. However, of the six discovered perovskites, \ch{0.2 Bi(Fe_{0.12} Co_{0.88})O3 – 0.8 PT} had the highest experimental $T_C$ of \SI{898}{\kelvin}, and three were novel {\ch{Me_y'}\ch{Me_y''}} pairs: {\ch{FeCo}}, {\ch{CoAl}}, and {\ch{NiSn}}. For comparison, the highest and median $T_C$ perovskites in the training data are approximately \SI{1100}{\kelvin} and \SI{750}{\kelvin}, respectively.
    
\begin{figure}
                \centering
                \includegraphics[scale=1]{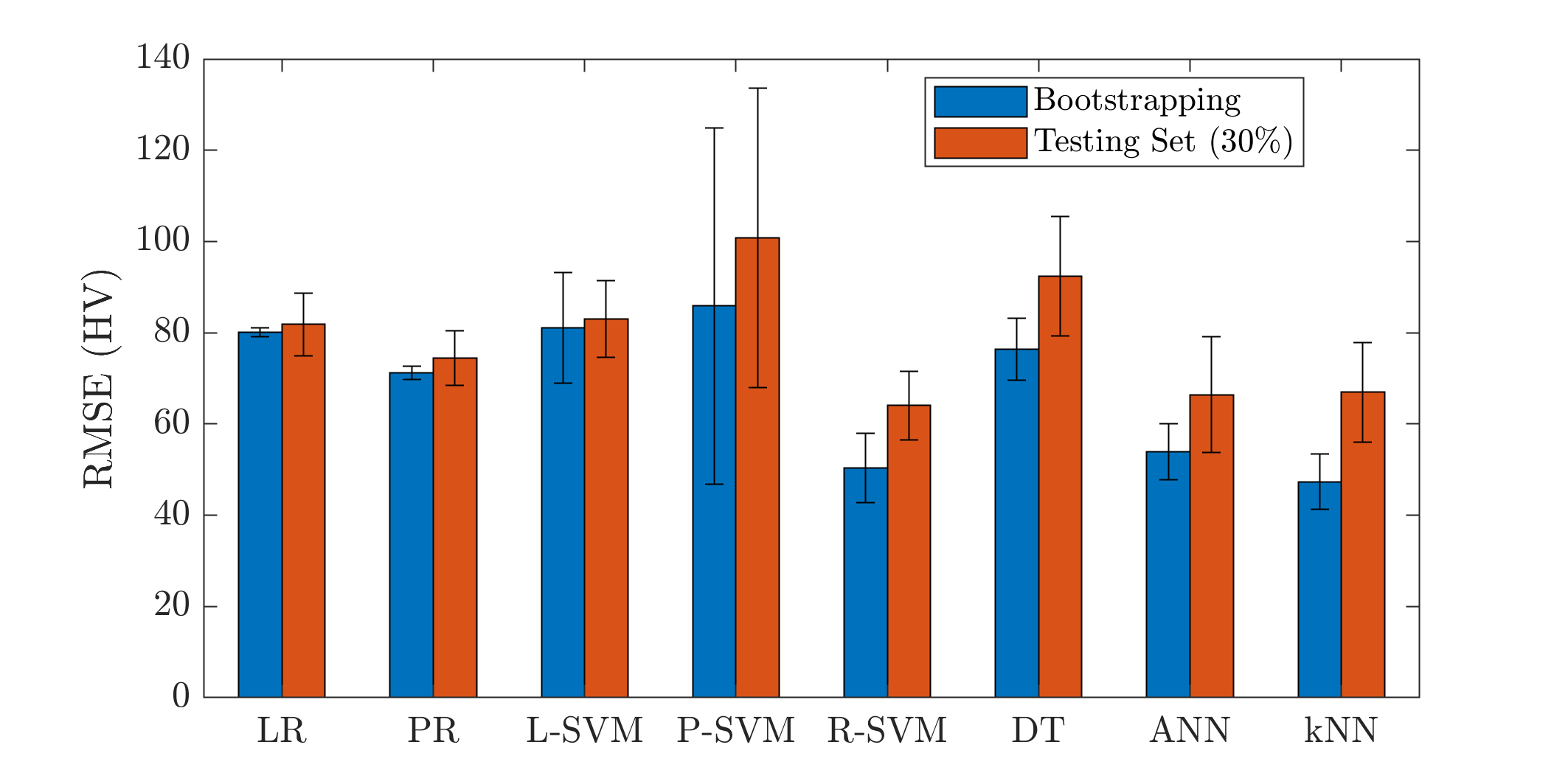}
                \caption{\Acrfull{rmse} and uncertainty standard deviation for bootstrapped and test set predictions for various models: \acrfull{lr}, \acrfull{pr}, linear \acrfull{svm} (L-SVM), polynomial \acrshort{svm} (P-SVM), \acrfull{rbf} \acrshort{svm} (R-SVM), \acrfull{dt}, \acrfull{ann}, and \acrfull{knn}. \gls{rbf} \gls{svm} had the lowest test dataset error and was used as a surrogate model. Reproduced with permission from Wen, C.; Zhang, Y.; Wang, C.; Xue, D.; Bai, Y.; Antonov, S.; Dai, L.; Lookman, T.; Su, Y. Acta Materialia 2019, 170, 109–117.\cite{wenMachineLearningAssisted2019} Bar chart data was extracted via \url{https://apps.automeris.io/wpd/} and replotted using MATLAB. }
                \label{fig:wen2019-model-compare-b}
            \end{figure}

            \citet{wenMachineLearningAssisted2019} searched for \glspl{hea} having high hardness using 135 training data samples (18 experimentally from their lab) and demonstrated that learning from composition and descriptors exploiting \gls{hea} domain knowledge outperformed \gls{ml} models that use only compositional descriptors. They compared performance across several different models (\gls{lr}, \gls{pr}, \gls{svm}, \gls{dt}, \gls{ann}, and \gls{knn}), for which \gls{svm} with a \gls{rbf} kernel had the best performance on test data (\cref{fig:wen2019-model-compare-b}). The \gls{svm} surrogate model was used in a \gls{doe}-based \gls{ad} scheme and \gls{fs} was performed via a hybrid \gls{ca}/wrapper. Using arc melting, they synthesized 42 alloys, 35 of them having higher hardness than the hardest candidates of the training set, 17 of them having $\sim$\SI{10}{\percent} higher hardness, and the highest with $\sim$\SI{14}{\percent} higher hardness (\num{883(47)} HV relative to \num{775} HV). They suggested extending this framework to bulk metallic glasses and superalloys.
            
            \citet{caoHowOptimizeMaterials2018} optimized power conversion efficiency of \ch{PDCTBT:PC71BM} organic photovoltaics via \gls{svm}, \gls{doe}, and 16 \gls{ad} iterations using a total of 150 experimental devices to achieve a maximum power conversion efficiency of approximately \SI{7.7}{\percent}.
        
            \citet{balachandranAdaptiveStrategiesMaterials2016} used a dataset of 223 \ch{M2AX} family of compounds containing information about bulk, shear, and Young’s modulus that were calculated using \gls{dft} and used it on an iterative \gls{ml} design strategy composed of two main steps: 1) \gls{ml} trained a regressor that predicts elastic properties by elementary orbital radii of the individual components of the materials, and 2) a selector used these predictions and their uncertainties to choose the next material to investigate. Additionally, \gls{dft} calculations were used to measure the desirability of the properties of a potential materials candidate. Three different regressors, \gls{gpr}, \gls{svm} with a \gls{rbf} kernel, and \gls{svm} with a linear kernel, were compared along with two different selectors, \gls{ego} and \gls{kg}. Ideally, the resulting model should provide a balance between exploration and exploitation and obtain a material with the desired elastic properties in as few iterations as possible. The performance of each model was measured in terms of “opportunity cost” and the number of iterations used to find a material. They found that selectors that use information about the prediction uncertainty perform better than by themselves. %This new design strategy illustrated how adaptive design tools can be used to lead future searches for new materials.
        
        \subsubsection{\glsentrytitlecasefull{svm} and \glsentrytitlecasefull{crfs}}
        \label{sec:101-10000:svm-crfs}
        
\begin{figure}
            \centering
            \includegraphics[width=400pt]{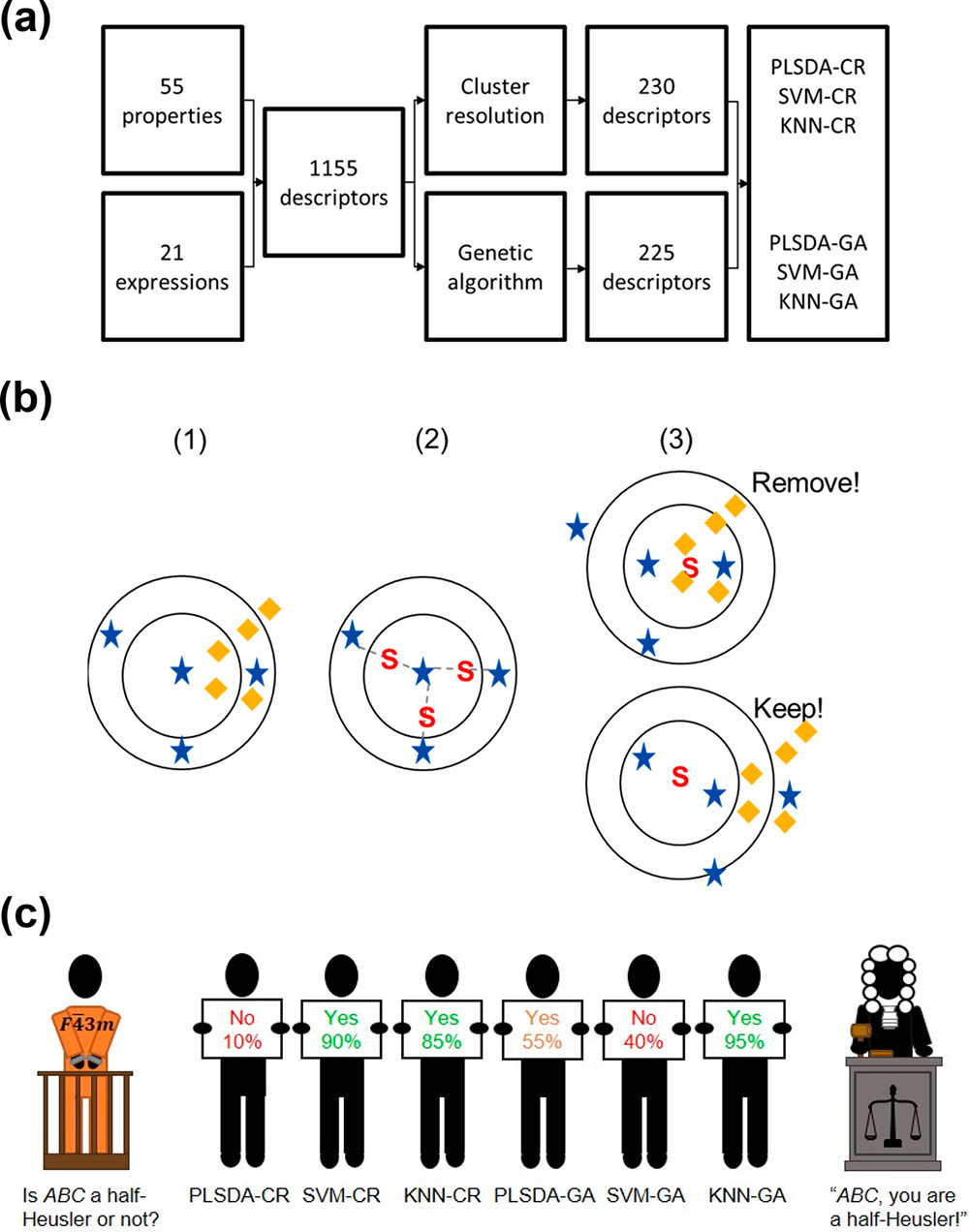}
            \caption{\acrfull{crfs} and \acrfull{ga} approaches selected \num{230} and \num{225} descriptors from a set of 1155 descriptors, respectively, and each approach was paired with \acrfull{pls}, \acrfull{svm}, and \acrfull{knn}, resulting in 6 models (a). \acrfull{smote} is used to address the issue of imbalanced data, where synthetic samples (S) are generated between pairs of minority (i.e. less frequently occurring) samples ($\bigstar$). If most nearest neighbors to S are minority samples, S is kept, otherwise if most nearest neighbors to S are majority samples (\protect\rotatebox[origin=c]{45}{$\blacksquare$}), S is removed. Finally, a soft-voting ensemble of the 6 models is used to classify whether a material is half-Heusler or not (yes if $>$\SI{50}{\percent}, no if $<$\SI{50}{\percent}). Reproduced with permission from Gzyl, A. S.; Oliynyk, A. O.; Mar, A. Crystal Growth \& Design 2020.\cite{gzylHalfHeuslerStructuresFullHeusler2020}}
            \label{fig:gzyl2020-fs-smote-ensemble}
        \end{figure}

                \citet{gzylHalfHeuslerStructuresFullHeusler2020} predicted half-Heusler structures,  compounds with equiatomic proportions $ABC$ (important for thermoelectrics, spintronics, and topological insulators), with a sensitivity, selectivity, and accuracy of \SI{88.3}{\percent}, \SI{98.2}{\percent}, and \SI{97.6}{\percent}, respectively. They used \num{2818} experimental training data points and an ensemble of \gls{pls}, \gls{svm}, and \gls{knn} \gls{ml} models. Each of the three \gls{ml} techniques was combined with a \gls{crfs} and \gls{ga} (also referred to as \gls{ea}) descriptor selection model, giving in total six models (\cref{fig:gzyl2020-fs-smote-ensemble}a). Additionally, the ensemble classification scheme was combined with \gls{smote} to address issues of unbalanced datasets and overfitting (\cref{fig:gzyl2020-fs-smote-ensemble}b). The ensemble classification schemes used soft-voting where predicted probabilities of being half-Heusler were averaged among the six models, and compounds with averaged probabilities above \SI{50}{\percent} were classified as half-Heusler (\cref{fig:gzyl2020-fs-smote-ensemble}c). Six of seven and 7/7 predicted half-Heusler and non-half-Heusler compounds, respectively, were successfully synthesized and confirmed. Once \gls{smote} had been applied, use of an ensemble approach increased the validation set sensitivity (rate of true positives) from \SI{83.3}{\percent} (best individual model, \gls{svm} \gls{crfs}) to \SI{88.3}{\percent} while maintaining near identical validation specificity and accuracy.
        
\begin{figure}
            \centering
            \includegraphics[width=89mm]{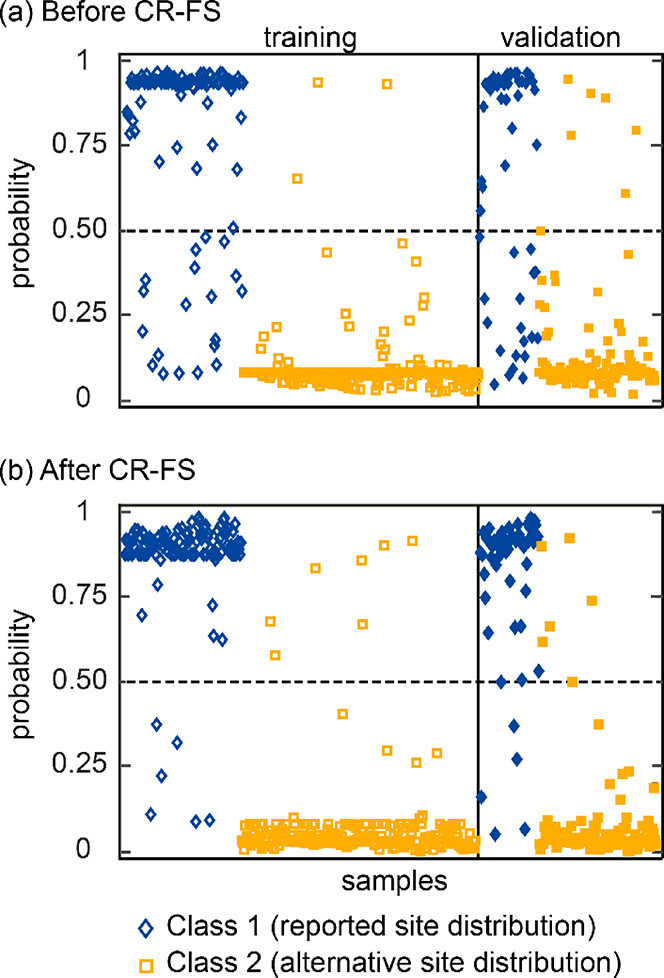}
            \caption{Applying \acrfull{crfs} improves \acrfull{svm} classification of reported (class 1) vs. alternative (class 2) site distributions for preferential site distributions in half-Heusler compounds (before \gls{crfs} (a), after \gls{crfs} (b)). A perfect classification accuracy would show all blue diamonds with a probability of 1 and all orange squares with a probability of 0. The algorithm was trained on 119 class 1 datapoints and 239 class 2 datapoints and validated on 60 class 1 datapoints and 119 class 2 datapoints. Reproduced with permission from Gzyl, A. S.; Oliynyk, A. O.; Adutwum, L. A.; Mar, A. Inorg. Chem. 2019, 58 (14), 9280–9289.\cite{gzylSolvingColoringProblem2019}}
            \label{fig:gzyl2019-crfs}
        \end{figure}

        \citet{gzylSolvingColoringProblem2019} used 179 experimentally reported structures, 23 descriptors (selected via \gls{crfs} from 243 descriptors based on 43 elemental properties), and \gls{svm} to classify half-Heusler site preferences resulting in a sensitivity, selectivity, and accuracy of 93\%, 96\%, and 95\%, respectively. One goal of the work was to apply data sanitation by retesting classified candidates with various classification probabilities. Three compounds, \ch{MnIrGa}, \ch{MnPtSn}, and \ch{MnPdSb}, gave probabilities of \num{0.127}, \num{0.043}, and \num{0.069}, respectively, before \gls{crfs} and \num{0.881}, \num{0.881}, and \num{0.680}, respectively, after \gls{crfs}, of which the higher probabilities were more accurate. Thus, using a \gls{crfs} scheme had notable benefits as further demonstrated by better delineation between Heusler and non-Heusler in \cref{fig:gzyl2019-crfs}. Two compounds, \ch{GdPtSb} and \ch{HoPdBi}, which were considered misclassified based on existing input data, were resynthesized. The results were confirmed for both compounds by powder \gls{xrd}; additionally, a single-crystal \ch{HoPdBi} sample was available, for which a full structural determination and unambiguous proof was obtained. This characterization demonstrated that the model's classification was indeed correct while the original input data was not. Revised \glspl{cif} were then prepared and submitted to the appropriate database, highlighting a successful example of data sanitation validated by experiment as well as a caution about possible discrepancies in input data.
        
        \citet{oliynykDisentanglingStructuralConfusion2017} filtered 990 features down to 113 by \gls{crfs} and applied \gls{svm} to \num{1037} training datapoints of 1:1:1 ternary structures (\ch{TiNiSi}-, \ch{ZrNiAl}-, \ch{PbFCl}-, \ch{LiGaGe}-, \ch{YPtAs}-, \ch{UGeTe}-, and \ch{LaPtSi}-type). They validated on 19 experimental samples and found that in a "structurally confused" region ($0.3 <= \mathrm{probability} <= 0.7$), both phases can coexist. This indicates that the "confused" region of a properly trained, appropriate classification scheme can indicate more than just sparsity or noisiness of data; it can also point to physical phenomena where either classification type may exist or even coexist.
        
        \citet{oliynykClassifyingCrystalStructures2016} trained a \gls{plsda} and \gls{svm} to develop a crystal structure predictor for binary AB compounds from 706 AB compounds with the seven most common structure types (\ch{CsCl}, \ch{NaCl}, \ch{ZnS}, \ch{CuAu}, \ch{TlI}, \ch{$\beta${-}FeB}, and \ch{NiAs}) and 31 elemental property features. In predicting crystal structure, \gls{plsda} and \gls{svm} showed an accuracy of \SI{77.1}{\percent} and \SI{93.2}{\percent}, respectively, after validation. Both models made quantitative predictions of hypothetical compounds. For example, \gls{plsda} and \gls{svm} predicted \ch{RhCd} to have a \ch{CsCl}-type structure with \num{0.669} and \num{0.918} probability, respectively, which was then later confirmed after experimental synthesis. \citet{oliynykClassifyingCrystalStructures2016} concluded \gls{svm} is the superior classification method in crystallography that can make quick and accurate predictions on crystal structure and has potential to be applied to identify the structure of any unknown compounds.
        
        \subsubsection{General \glsentrytitlecasefull{svm}}
        \label{sec:101-10000:svm-general}
        
\begin{figure}
            \centering
            \includegraphics[width=\textwidth]{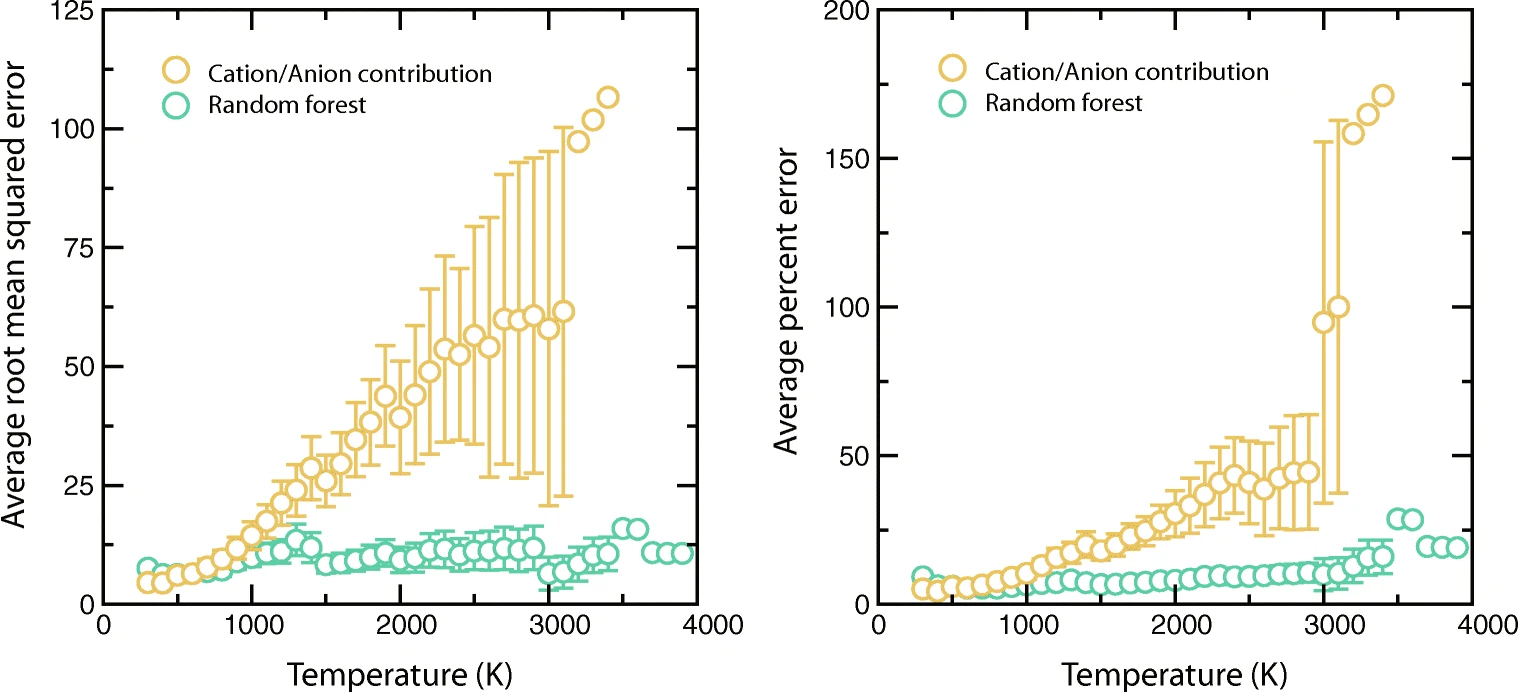}
            \caption{Average \gls{rmse} for heat capacity \gls{svm} predictions of inorganic solids (\SI{}{\joule\per\mole\per\kelvin}) vs. temperature (\SI{}{\kelvin}), where average \gls{rmse} was calculated by averaging temperatures in \SI{100}{\kelvin} increments or groups. \SI{95}{\percent} confidence intervals were calculated using \acrfull{cv} metrics. Reproduced from Kauwe, S. K.; Graser, J.; Vazquez, A.; Sparks, T. D. Integrating Materials and Manufacturing Innovation 2018, 7 (2), 43–51\cite{kauweMachineLearningPrediction2018}; licensed under a Creative Commons Attribution (CC BY) license (\url{http://creativecommons.org/licenses/by/4.0/}). }
            \label{fig:kauwe2018-rf-cac-temp}
        \end{figure}

        \citet{kauweMachineLearningPrediction2018} used \num{263} chemical formulae (e.g. \ch{Al2O3}) and temperatures from \SIrange{298.15}{3900}{\kelvin} obtained from NIST:JANAF tables (in total \num{3986} training datapoints) to predict heat capacity ($C_p$) of inorganic solids with \gls{svm}, \gls{lr}, and \gls{rf}. \Gls{gcv} was used to test extrapolative prediction (\cref{fig:kauwe2018-svm-lr-rf-cac}), giving $C_p$ \glspl{rmse} of \SIlist{21.07 \pm 3.6;19.22 \pm 2.4;15.15 \pm 2.5}{\joule\per\mole\per\kelvin} for \gls{svm}, \gls{lr}, and \gls{rf}, respectively. This showed significant improvement over conventional Neumann-Kopp (based on summing heat capacities of constituent elements in a chemical formula) and comparable performance to \gls{cac} (based on cation/anion pairs and a temperature dependent power series), the latter of which had 161/263 chemical formulae with available data. \citet{kauweMachineLearningPrediction2018} also noted that \gls{cac} likely used many of the same chemical formulae to obtain \gls{cac} fitting parameters which probably caused an overestimation of \gls{cac} performance. While the \gls{rmse} of \gls{cac} was on par with the \gls{ml} methods, the systematic errors and steep over- or under-estimation in some regions (in some cases even with a negative parity slope) highlights the need to consider more than a single metric in evaluating model performance and account for systemic error in the data. Indeed, \gls{rf} performed much better than \gls{cac} across the full temperature range (\cref{fig:kauwe2018-rf-cac-temp}).
        
        \citet{tehraniMachineLearningDirected2018} predicted two ultraincompressible, superhard materials, \ch{ReWC2} and \ch{Mo_{0.9}W_{1.1}BC}. The former was synthesized as \ch{ReWC_{0.8}} due to unreacted graphite in \ch{ReWC2} and is a brand-new ultraincompressible, high-hardness material. \ch{Mo_{0.9}W_{1.1}BC} had been previously studied in the literature and was further confirmed as an inexpensive, earth-abundant, ultraincompressible hard material. They used \gls{svm} of \num{2572} elastic moduli training datapoints from the Materials Project database \cite{jainMaterialsProjectMaterials2013} and \num{150} descriptors built from arithmetic operations on compositional and structural properties. Training data was curated from an original set of \num{3248} elastic moduli entries by eliminating inaccessible phases at ambient pressure and temperature and removing unreasonable entries with e.g. negative values, among other restrictions. While the full descriptor set was used for prediction, for perspective, descriptors were fed through a \gls{ga}-based \gls{fs} algorithm, indicating that \SIlist{52;44}{\percent} of the descriptors were essential for Bulk modulus ($B$) and shear modulus ($G$), respectively. Both $B$ and $G$ correlate positively with hardness and are used as proxies in the approach. Careful attention is paid to trends of errors relative to the compound type being predicted; for example, metallic/covalent bonding materials general exhibit lower error than highly ionic compounds. The \gls{svm} model predicts $B$ and $G$ for ~\num{120000} binary, ternary, and quaternary inorganic solids in Pearson's Crystal Database\cite{villarsPearsonCrystalData2014} with \glspl{rmsecv} of \SIlist{17.2; 16.5}{\giga\pascal}, respectively. \ch{ReWC2} and \ch{Mo_{0.9}W_{1.1}BC} are identified as potential high hardness candidates, having the highest predicted $B$ and $G$ out of all ternary (\ch{ReWC2}) and quaternary (\ch{Mo_{0.9}W_{1.1}BC}) candidates, and were amenable to synthesis via ambient pressure arc melting. Due to presence of unreacted graphite peaks in powder \gls{xrd} experiments of \ch{ReWC2}, eventually \ch{ReWC_{0.8}} was settled on for testing. High-pressure \gls{dac} experiments confirmed ultraincompressibility and Vicker's microhardness experiments confirmed superhardness at low loads, \SIlist{40 \pm 3; 42 \pm 2}{\giga\pascal} for \ch{ReWC_{0.8}} and \ch{Mo_{0.9}W_{1.1}BC}, respectively. %A generally accepted minimum threshold for superhardness is \SI{40}{\giga\pascal}, and \citet{tehraniMachineLearningDirected2018} discussed some ambiguity for defining a superhardness threshold.
        
        % \citet{oliynykHighThroughputMachineLearningDrivenSynthesis2016} why did I think this was SVM?
        
        \citet{raccugliaMachinelearningassistedMaterialsDiscovery2016} used in-house "dark" or failed experiments to enhance a \gls{svm} model, achieving 89\% accuracy relative to 79\% accuracy via human intuition. No comparison against a \gls{ml} model without failed experiments was reported. A web database (\url{https://darkreactions.haverford.edu/}) was made publicly accessible for failed chemical reaction experiments.
        
        %\citet{yanOptimizationThermalConductivity2020} optimized microstructure in search of high thermal conductivity ($\kappa$) \ch{UO2} pellets for nuclear energy applications using 1000 training \gls{fem} simulations and by testing \gls{gbdt}, \gls{svm}, \gls{gpr}, and Bayesian-regularized \gls{ann} models. The Bayesian-regularized \gls{ann} model was chosen for studying microstructural effect on $\kappa$, and they determined that low \ch{Mo} content and low mixing times would likely improve $\kappa$. They used these results as guidelines to synthesize a \SI{2}{\percent} \ch{Mo} (by vol.) \ch{UO2} pellet with various mixing times, producing a pellet with 20\% better $\kappa$ than pure \ch{UO2}. Because their synthesis parameters were arbitrary and could perhaps have been chosen using domain knowledge alone, it is unclear to what extent their \gls{ml} modeling produced better results than if parameters were chosen in a typical trial-and-error scheme guided by domain knowledge. Additionally, while their model suggested that low mixing times would improve $\kappa$, the lowest tested mixing times and corresponding $\kappa$ values were not reported because it did not meet the authors' expectations. While the connection between the experimental validation and the \gls{ml} model may be weak, they were able to predict $\kappa$ for new microstructures with reasonable accuracy.
        
        \citet{zhuoEvaluatingThermalQuenching2020} predicted and tested thermal quenching temperature (temperature at which emission intensity is cut in half relative to initial) using \gls{svm} and 134 experimental training datapoints. Five compounds (\ch{Sr2ScO3F}, \ch{Cs2MgSi5O12}, \ch{Ba2P2O7}, \ch{LiBaB9O15}, and \ch{Y3Al5O12}) had predicted thermal quenching temperatures above \SI{423}{\kelvin} and exhibited thermal stability when using \ch{E^{3+}} as a substitutional atom.
        
        In earlier work, \citet{zhuoIdentifyingEfficientThermally2018} predicted and tested Debye temperature as a proxy for photoluminescent quantum yield (i.e. energy-efficiency of light bulb phosphors) using \gls{svm}, 2610 \gls{dft} training datapoints, and \gls{rfe} (\gls{fs} method) for 2071 potential phosphor hosts. The compound with highest Debye temperature and largest band gap, \ch{NaBaB9O15}, was synthesised and \ch{NaBaB9O15:Eu^{2+}} was shown to have a quantum yield of \SI{95}{\percent}. The Debye temperature \gls{rmsecv} and \gls{maecv} was \SIlist{59.9; 37.9}{\kelvin}, respectively, with most temperatures of training data between \SIrange{50}{750}{\kelvin}.
        
        \citet{luAcceleratedDiscoveryStable2018} combined various \gls{ml} techniques with \gls{dft} calculations to quickly screen \glspl{hoip} for photovoltaics based on bandgap. Six \gls{ml} regression methods (\gls{gbr}, \gls{krr}, \gls{svm}, \gls{gpr}, \gls{dt} regression, and multilayer perceptron regression) were trained using 212 reported \glspl{hoip} bandgap values. 14 selected material features were narrowed down from an initial 30 property features (including properties such as ionic radii, tolerance factor, and electronegativity) through feature engineering. The \gls{gbr} model was shown to be the most accurate, so it was then used to screen \num{5158} unexplored \glspl{hoip} (\num{346} that had been previously studied and \num{5504} that were calculated) for any promising \glspl{hoip} that are both efficient and environmentally sustainable. They successfully screened 6 orthorhombic lead-free \glspl{hoip} with proper bandgap for solar cells and room temperature thermal stability, of which two particularly stood out. Validations of these results from \gls{dft} calculations showed that the two are in excellent agreement, with the $\Delta E_g$ never being larger than \SI{0.1}{eV}. \citet{luAcceleratedDiscoveryStable2018} demonstrated a highly accurate method that can be used on a broader class of functional materials design.
        
        \citet{minMachineLearningAssisted2018} used a dataset of 300 Ni-rich \ch{LiNi_xCo_{1-x-y}Mn_{1-x-y-z}O_2} cathodes with 13 input variables (synthesis parameters, inductively coupled plasma mass spectrometry, and X-ray diffraction results) to compare the accuracy of 7 different \gls{ml} algorithms (\gls{svm}, \gls{dt}, \gls{rr}, \gls{rf}, \gls{ert} with an adaptive boosting algorithm, and \gls{ann} with multi-layer perceptron) in predicting the initial capacity, \gls{crr}, and amount of residual Li. The \gls{ert} with adaptive boosting algorithm resulted in the highest predictive accuracy, with an average coefficient of determinant, $R^2$, of \num{0.833}. Additionally, \citet{minMachineLearningAssisted2018} employed a reverse engineering model to propose optimized experimental parameters that satisfy target specifications. These optimal parameters were then fed into the trained \gls{ml} model, that makes corresponding electrochemical property predictions based on them. Experimental validations showed average differences of \SIlist{6.3;1.0;12.8}{\percent} for the capacity, \gls{crr}, and free Li, respectively. %The current model is promising and can be used to accelerate the optimization of parameters in materials synthesis.
        
        \subsubsection{Non- \glsentrytitlecasefull{svm}}
        \label{sec:101-10000:non-svm}
        
        \citet{sekoMatrixTensorbasedRecommender2018} used four descriptor-free recommender systems --- \gls{nmf}, \gls{svd}, \gls{cpd}, and Tucker decomposition --- to predict currently unknown \glspl{crc}. The Tucker decomposition recommender system had the best discovery rate which was validated by performing \gls{dft} calculations on phase stability of 27 recommended, unknown candidates, 23 of which were stable (\SI{85}{\percent} discovery rate).
        
        % ensemble \citet{gzylHalfHeuslerStructuresFullHeusler2020} already discussed
        
        % \gls{ann} \citet{yanOptimizationThermalConductivity2020} already discussed
        
        % \gls{rf} \citet{kauweMachineLearningPrediction2018} already discussed
        
        \citet{renAcceleratedDiscoveryMetallic2018} searched for metallic glasses in the \ch{Co{-}V{-}Zr} ternary system using \gls{rf} and 315 initial training datapoints, followed by a \gls{hitp} \gls{cms}, \gls{ad} scheme producing ~\num{1315} total points (including "dark", i.e. failed, experiments). Discrepancies in the initially trained model were used for retraining which improved acccuracy for the \ch{Co{-}V{-}Zr} predictions. Two additional unreported ternaries, \ch{Co{-}Ti{-}Zr} and \ch{Co{-}Fe{-}Zr}, were discovered. A "grouping" \gls{cv} approach (\gls{gcv}) was used for outside-of-dataset predictions (\cref{sec:cv}).
        
%        on previously reported observations and parameters from physiochemical theories. A synthesis method–dependent to guide high-throughput experiments to find a new system of metallic glasses in the Co-V-Zr ternary. Experimental observations are in good agreement with the predictions of the model, but there are quantitative discrepancies in the precise compositions predicted. Then they did something interesting. They used the discrepancies to retrain the ML model. The refined model has significantly improved accuracy not only for the Co-V-Zr system but also across all other available validation data. Then they used the refined model to guide the discovery of metallic glasses in two additional previously unreported ternaries. --iterative combination of ML with high throughput experimentation. Dark experiments (i.e. failed experiments). 315 starting datapoints. Special CV test for outside-of-dataset predictions ("grouping" approach). ~1000 more via a high throughput experimental method. So something like 1315 training data in total, with some downsampling schemes employed
        
        \citet{oliynykHighThroughputMachineLearningDrivenSynthesis2016} searched for Heusler-type structures using a classification \gls{rf} model with compositional descriptors and \num{1948} compounds (341 of which are Heusler) across 208 structure types as training data, achieving a sensitivity (true-positive rate) of 0.94. Of 21 synthesized compounds, 19 were predicted correctly (12/14 as Heusler and 7/7 as non-Heusler). \ch{TiRu2Ga}, a potential thermoelectric material, was also synthesized and confirmed to have Heusler structure. %could include Figure 5. Machine-learning-predicted probability of forming Heusler compounds for three series of gallides, and experimental confirmation through arc-melting and annealing at 800 °C (check marks indicate successful preparation of Heusler compound, and crosses indicate absence of Heusler compound).
        
        % \gls{dt} \citet{raccugliaMachinelearningassistedMaterialsDiscovery2016} already discussed
        
        % \citet{wenMachineLearningAssisted2019} already discussed
        
        % \gls{rfe} \citet{zhuoEvaluatingThermalQuenching2020} already discussed
        
        \citet{buciorEnergybasedDescriptorsRapidly2019} predicted hydrogen uptake in \glspl{mof} by predicting \num{50000}+ compounds via a \gls{lasso} approach with \num{1000} training \gls{gcmc} simulations and 12 binned energy features. The energy features were obtained by overlaying a 3D grid on the \gls{gcmc} simulation box, probing each grid point with a "hydrogen probe" and binning the 3D distribution into a 1D histogram with 12 bins (1 feature per bin). The predictions were screened by retrieving and running \gls{gcmc} simulations on the top \num{1000} predictions. The max \gls{gcmc} simulation in the training data was $\sim$\SI{47.5}{\gram\per\liter} \ch{H2} uptake, and 51 of the top 1000 simulations were $>$\SI{45}{\gram\per\liter}. They synthesized one promising MOF, MFU-4\textit{l}(Zn), with a predicted $\sim$\SI{54}{\gram\per\liter} \ch{H2} uptake (\SI{100}{\bar} $\rightarrow$ \SI{5}{\bar}) and experimentally characterized as having \SI{47}{\gram\per\liter} \ch{H2} uptake (\SI{100}{\bar} $\rightarrow$ \SI{5}{\bar}) which is competitive with similar experimental \glspl{mof} in the literature.
        
        % \gls{crfs} \citet{oliynykDisentanglingStructuralConfusion2017} already discussed
        
        % \gls{doe} \citet{caoHowOptimizeMaterials2018} already discussed
        
        % \gls{lr} \cite{kauweMachineLearningPrediction2018} already discussed
        
        % \cite{raccugliaMachinelearningassistedMaterialsDiscovery2016} already discussed
        
        % \cite{wenMachineLearningAssisted2019} already discussed
        
        % \gls{pr} \cite{wenMachineLearningAssisted2019} already discussed
        
        % \gls{pls} \cite{gzylHalfHeuslerStructuresFullHeusler2020} already discussed
        
        % matrix-based recommender \cite{sekoMatrixTensorbasedRecommender2018} already discussed
        
        % \gls{smote} \cite{gzylHalfHeuslerStructuresFullHeusler2020} already discussed
        
        % \gls{knn} \cite{gzylHalfHeuslerStructuresFullHeusler2020} already discussed
        
        % \cite{raccugliaMachinelearningassistedMaterialsDiscovery2016} already discussed
        
        % \cite{wenMachineLearningAssisted2019} already discussed
        
        \citet{nikolaevAutonomyMaterialsResearch2016} designed an automated method to study the synthesis and target a specified growth rate of single-walled \glspl{cnt}, called \gls{ares} which is the first to do closed-loop iterative materials experimentation. \gls{ares} was capable of designing, executing, and analyzing experiments orders of magnitude faster than current research methods. To achieve this, \gls{ares} used a \gls{rf}/\gls{ga} planner that was trained off of an initial database of \num{84} experiments that was then updated as it performed a series of approximately \num{600} experiments. \gls{ares} demonstrated an autonomous research system capable of controlling experimental variables in materials science.
        
        \citet{mannodi-kanakkithodiMachineLearningStrategy2016} trained a \gls{krr}-based \gls{ml} model using the crystal structures of 284 four-block polymers (250 training datapoints and 34 test points), including relevant property information about each: bandgap and ionic and total dielectric constant (calculated from \gls{dft}). Additionally, each polymer was fingerprinted based on their building block identities using the Pearson correlation analysis to explore the possibility of a correlation between those fingerprints and a polymer’s properties. By validating using \gls{dft} calculations and experimental values from synthesized polymers, the \gls{krr} model converted a fingerprint to property values with an average error for all three properties mentioned above of 10\% or less. A genetic algorithm then searched for materials with desired properties that can then be inputted into the \gls{krr} model, instead of traditional approaches like random search and chemical-rules based search. \citet{mannodi-kanakkithodiMachineLearningStrategy2016} demonstrated how carefully created and curated materials data can be used to train statistical learning models so that they only require a simple fingerprint of a new material to predict its properties. Furthermore, they also showed that the combination of a genetic algorithm with learning models can efficiently determine specific materials that possess certain desired properties.
        
        \citet{zhangFindingNextSuperhard2020} extracted 1062 experimentally measured load-dependent Vickers hardness data from literature and 532 unique compositions to train a supervised \gls{rf} algorithm using boosting algorithms (\gls{gbdt} and XGBoost). The \gls{rf} model’s hardness predictions were validated using two different hold-out test sets: the first with Vickers hardness measurements for 8 synthesized, unmeasured metal disilicides and the second with a customized hold-out containing several classic high hardness materials. After validation, the model screened more than \num{66000} compounds in the crystal structure database, of which 10 are predicted to be superhard at 5 N. Due to the low number of entirely new predicted materials (most had already been discovered), the hardness model was combined with a recently developed formation energy and convex hull prediction tool to find new compounds with high hardness. More than ten thermodynamically favorable compositions with hardness above \SI{40}{GPa} were discovered, proving that this model can successfully identify completely new materials with extraordinary mechanical properties.
        
    \subsection{\num{10000}+ Training Datapoints}
    \label{sec:10000+}
    
    Experimentally and computationally validated \gls{ml} articles that use more than \num{10000} training datapoints are sparse compared to the previous two training datapoint set sizes considered in this work.  This is to be expected given the difficulty of generating a reliable dataset of this magnitude, either experimental or computational. This problem is especially exacerbated in materials-related projects as many synthesis methods are lengthy and difficult to procure.  preference towards \glspl{ann} may have been expected, given the limited number of articles, no clear trend emerges.  We now present experimental \cite{gaultoisPerspectiveWebbasedMachine2016,gomez-bombarelliDesignEfficientMolecular2016,sakuraiUltranarrowBandWavelengthSelectiveThermal2019} and computational \cite{meredigCombinatorialScreeningNew2014,parkDevelopingImprovedCrystal2020} examples: \gls{ann} \cite{gomez-bombarelliDesignEfficientMolecular2016,parkDevelopingImprovedCrystal2020}, \gls{rf} \cite{gaultoisPerspectiveWebbasedMachine2016}, \gls{dt} \cite{meredigCombinatorialScreeningNew2014}, and \gls{bo} \cite{gomez-bombarelliDesignEfficientMolecular2016,sakuraiUltranarrowBandWavelengthSelectiveThermal2019}.
    
    % Experimentally validated:
    % \cite{gaultoisPerspectiveWebbasedMachine2016,buciorEnergybasedDescriptorsRapidly2019}
    % Computationally validated:
    % \cite{meredigCombinatorialScreeningNew2014,parkDevelopingImprovedCrystal2020}
    
    \subsubsection{\glsentrytitlecasefull{ann}}
        
\begin{figure}
            \centering 
            \includegraphics[width=15cm]{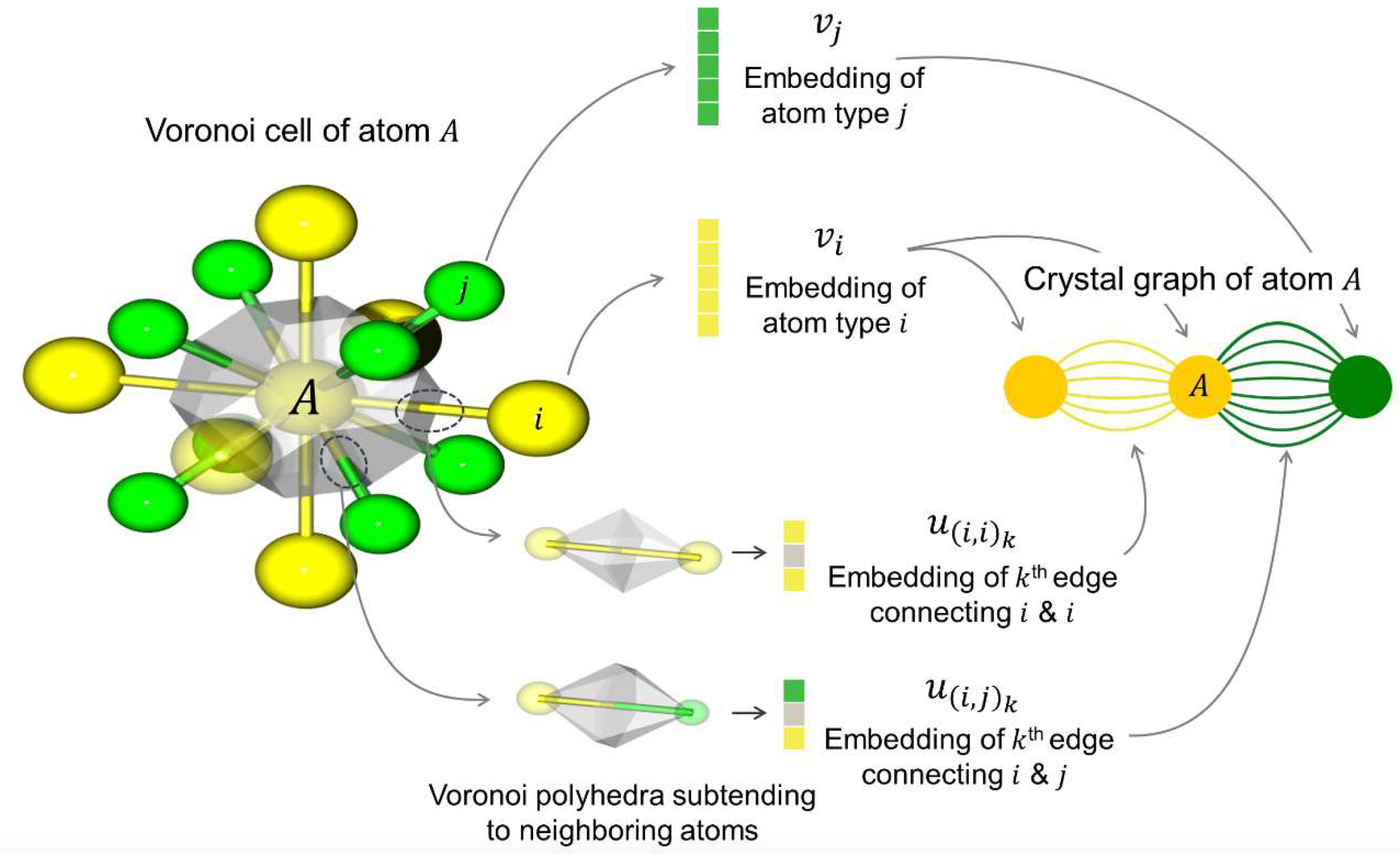}
            \caption{Visual representation of the \acrfull{icgcnn} crystal graph. On the left is an illustration of the Voronoi cell of Atom $A$, which is connected to its twelve nearest neighbors. On the right is the local environment of $A$. Each node and edge is embedded with vectors that contain information about the relationship between each constituent atom ($\mathit\mathbf{v}_i, \mathit\mathbf{v}_j$) and its neighbors ($\mathbf{u}_{(i,i)_k}, \mathbf{u}_{(i,j)_k}$). Additionally, edge vectors contain information (e.g. solid angle, area, and volume) about the Voronoi polyhedra. Reproduced with permission from Park, C. W.; Wolverton, C. Phys. Rev. Materials 2020, 4 (6), 063801. \cite{parkDevelopingImprovedCrystal2020} }
            \label{fig:park2020-icgcnn.png}
        \end{figure}

         The \gls{cgcnn} model can accurately learn material properties from graphical representations of atomic crystal structures, called “crystal graphs”. \citet{parkDevelopingImprovedCrystal2020} designed an improved framework of the \gls{cgcnn} model, called \gls{icgcnn}, which incorporated Voronoi tessellated crystal structures, 3-body explicit correlations of neighboring atoms, and an optimized chemical representation of interatomic bonds in the crystal graphs, all of which are absent in \gls{cgcnn} (\cref{fig:park2020-icgcnn.png}). First, a training/testing dataset consisting of \num{180000} \gls{dft} entries from the Open Quantum Materials Database \cite{kirklinOpenQuantumMaterials2015} was created. \gls{cgcnn} and \gls{icgcnn} were compared in their accuracy of predicting the thermodynamic stability of inorganic materials. Then, both models were used to conduct separate \gls{ml}-assisted \gls{hitp} searches to discover new stable compounds. The new framework was shown to have \SI{20}{\percent} higher accuracy than those of \gls{cgcnn} on \gls{dft} calculated thermodynamic stability and a success rate that is 2.4 times higher than \gls{cgcnn}. Using \gls{icgcnn}, they were also able to identify 97 novel stable compounds from \num{132600} screened \ch{ThCR2Si2}-type compounds through only 757 \gls{dft} calculations which corresponds to a success rate that is 130 times higher than that of an undirected \gls{hitp} search. %The significant differences in performance in that of \gls{cgcnn} and \gls{icgcnn} suggest that this new improved model can be used in the future to greatly accelerate materials discovery.
             
        \citet{gomez-bombarelliDesignEfficientMolecular2016} screened \num{40000} \gls{oled} molecules with \gls{tadf} character randomly selected from a library of 1.6 million software-generated candidates using an \gls{ann} combined with \gls{bo}. Then, the highest-ranking molecules based on \gls{eqe} predicted by the \gls{ann} were promoted to \gls{tddft} simulation. After \gls{bo}, \num{400000} molecules were screened in total. Results from the \gls{tddft} simulation found thousands of emitters predicted to be highly efficient, with about 900 being extremely promising. The top candidates, chosen by humans, were then validated using experimental synthesis. \citet{gomez-bombarelliDesignEfficientMolecular2016} was able to perform an integrated high-throughput virtual screening method targeting novel \gls{tadf} \gls{oled} emitters, which resulted in the discovery of new devices up to \SI{22}{\percent} \gls{eqe}, which can be applied to other areas of organic electronics.

        %  \gls{nlp} % technically none that fall into computationally or experimentally validated, huh..
        
    \subsubsection{\glsentrytitlecasefull{rf}}
        \citet{gaultoisPerspectiveWebbasedMachine2016} used \gls{rf} to predict promising new thermoelectric materials via a user-friendly \gls{ml}-based web engine. The engine suggested thermoelectric compositions based on a pre-screening of a dataset consisting of \num{25000} known materials from a myriad of sources, both experimental and computational. These predictions were then experimentally validated with two new compounds. They specifically focus on a set of compounds derived from the engine, \ch{RE12Co5Bi} (RE = Gd, Er), which exhibited high thermoelectric performance \cite{oliynykGd12Co2016}. The engine successfully predicted that this set of materials had low thermal and high electrical conductivities, but modest Seebeck coefficients, all of which were then additionally verified experimentally. The engine is the first example of \gls{ml} being utilized to suggest an experimentally viable new compound from true chemical white space, with no prior characterization, that can eventually replace traditional trial-and-error techniques in the search for new materials.
        
    \subsubsection{\glsentrytitlecasefull{dt}}
        %To lift the constraints of a compositional and structural search space experienced by computational screening 
        \citet{meredigCombinatorialScreeningNew2014} developed a \gls{ml} model using data from over \num{15000} \gls{dft} calculations to predict the thermodynamic stability of arbitrary compounds one million times faster than when just using \gls{dft} and without knowledge of crystal structure. The model was used to scan 1.6 million candidate compositions and predict \num{4500} new stable materials. Combining a physically motivated heuristic with a \gls{ml} model and using it on a large database of quantum mechanical calculations provides a new approach for extremely rapid computational materials screening. %This proposed method has potential to be applied to other fields to accelerate domain-specific materials discovery.
        
        \subsubsection{\glsentrytitlecasefull{bo}}
        \citet{sakuraiUltranarrowBandWavelengthSelectiveThermal2019} optimized a multilayer, ultranarrow-band wavelength-selective thermal radiator using electromagnetic simulations in sets of 200 or 400 simulations in a \gls{bo}/\gls{ad} scheme. For computational tractability, candidates were divided into groups of approximately \num{200000} each. The optimizable multilayer template consisted of 18 layers with variable total thickness (21 discrete choices) and \ch{Ge}, \ch{Si}, or \ch{SiO2} as the choices for each layer. The maximum figure of merit (a function of spectral normal intensity, spectral blackbody intensity, and min/max considered wavelengths) was typically obtained within \num{168000000} calculations, comprising $\sim$\SI{2}{\percent} of the total possible number of structures. They identified a structure with a predicted Q-factor of 273 and experimentally validated to have a Q-factor of 188 (compare with highest reported narrow-band thermal radiator Q-factor of $\sim$200 according to the authors).

\section{A Caution about \glsentrytitlecasefull{cv}}
\label{sec:cv}

\begin{figure}
    \centering
    \includegraphics[width=\textwidth]{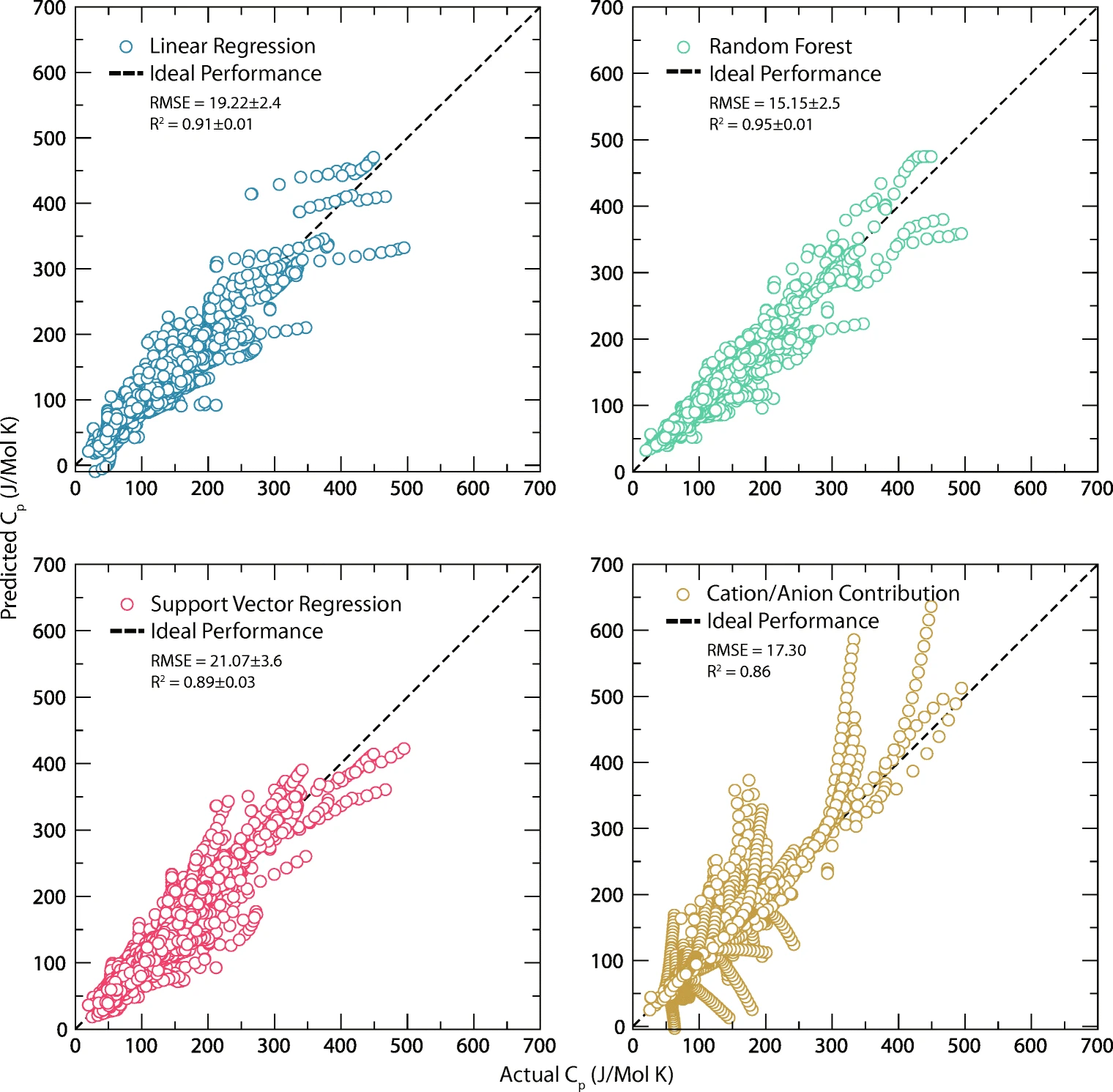}
    \caption{\Acrfull{gcv} parity plots for heat capacity predictions by \acrfull{lr} (top-left), \acrfull{rf} (top-right), \acrfull{svm} (bottom-left), and \acrfull{cac} (bottom-right) vs. actual heat capacity. \gls{gcv} was applied by sorting training data into chemical formula groups resulting in predictions that are extrapolations to new formulas rather than simple interpolation between temperatures of a certain chemical formula. This is likely the cause of parity "trails" (i.e. systemic bias for certain chemical formula groups) in \gls{lr}, \gls{rf}, and \gls{svm} methods. \gls{cac} (a legacy, non- \acrfull{ml} approach) likely exhibited optimistically low \acrfull{rmse} due to probable repeats between chemical formulae of fitted \gls{cac} coefficients (legacy work) and the \gls{gcv} data. Reproduced from Kauwe, S. K.; Graser, J.; Vazquez, A.; Sparks, T. D. Integrating Materials and Manufacturing Innovation 2018, 7 (2), 43–51\cite{kauweMachineLearningPrediction2018}; licensed under a Creative Commons Attribution (CC BY) license (\url{http://creativecommons.org/licenses/by/4.0/}). }
    \label{fig:kauwe2018-svm-lr-rf-cac}
\end{figure}

A common pitfall in materials discovery involves the use of \gls{cv}. If the goal of an approach is to predict fundamentally new materials (i.e. materials extrapolation rather than interpolation), a special "grouping" \gls{cv} scheme (termed \gls{gcv} in this work) may be used to ensure the model predictions are not overly optimistic. \citet{meredigCanMachineLearning2018} first introduced the idea of \gls{lococv} or \gls{gcv} and \citet{sparksMachineLearningStructural2020} discussed the difficulty of making predictions when many mechanisms interact to cause outstanding properties. \citet{sparksMachineLearningStructural2020} described how \gls{ml} can be used for structure-composition-property-processing relationships and review successful examples of materials discovery for structural materials (fatigue, failure), high-entropy alloys, and bulk metallic glasses. For example, in the case of \cite{renAcceleratedDiscoveryMetallic2018}, all training data for the \ch{Co{-}V{-}Zr} ternary were removed before making predictions in that group (hence \gls{gcv}). \citet{kauweMachineLearningPrediction2018} performed \gls{cv} on chemical formula groups rather than on all of the training data as a whole to make sure that cross-validated predictions were not simply interpolations between temperatures within a chemical formula group. To illustrate, the "trails" seen in the \gls{ml} parity plots of \cref{fig:kauwe2018-svm-lr-rf-cac} exhibiting systemic deviation from parity are likely present because of the \gls{gcv} scheme. By taking a non-group \gls{cv} approach, the model would likely favor temperature interpolation and mild temperature extrapolation, causing the trails to disappear at the expense of heavily overoptimistic predictive performance. We believe the question, "are my model predictions overly optimistic?", is wise to ask when pursuing true materials discovery.

%what is the name for this kind of cross-validation?

\section{An Eye Towards Extraordinary Predictions}
\label{sec:extraordinary}

Related to the need for specialized assessment of extrapolative performance (\cref{sec:cv}), making extraordinary predictions can be a difficult task. Due to ambiguity of the definition of extraordinary predictions, we provide three possible definitions:
\begin{enumerate}
    \item Experimentally or computationally validated predictions with better performance than any of the initial training dataset (also referred to as "better-than-input")
    \item Experimentally or computationally validated predictions with performance on par with top performers (e.g. falls into top \SI{1}{\percent} of the dataset as in \cite{kauweCanMachineLearning2020})
    \item Experimentally or computationally validated predictions with holistically ideal performance for a particular application including e.g. cost, toxicity, and abuse-tolerance (difficult to quantify).
\end{enumerate}

From \cref{sec:1-100}, we see that extraordinary predictions (definitions 1. and 2.) are commonplace due to a mixture of low number of training datapoints, simplicity of the model space (e.g. two continuous variables), and interpolative predictions. Likewise, from \cref{sec:101-10000} and \cref{sec:10000+}, we see that extraordinary predictions for large number of training datapoints, complex model spaces, and extrapolative (i.e. out-of-dataset) predictions are more difficult to attain. \citet{kauweCanMachineLearning2020} analyzed the ability of \gls{ml} models to predict extraordinary materials by holding out the top \SI{1}{\percent} of compounds for a given property and training on the bottom \SI{99}{\percent}. This was done for six different materials properties such as thermal expansion. They definitely show that extrapolation is possible, and furthermore, they show that a classification approach outperforms a regression approach. They reason that extrapolating extraordinary predictions is unlikely when the fundamental mechanism of the extraordinary prediction is different from the training dataset and that many examples of that mechanism need to be supplied. They also suggest that input data accuracy and consistency is a non-trivial issue. %maybe discuss how incorporation of domain knowledge can improve extrapolative performance when the same underlying mechanisms are at play

\begin{figure}
    \centering
    \includegraphics{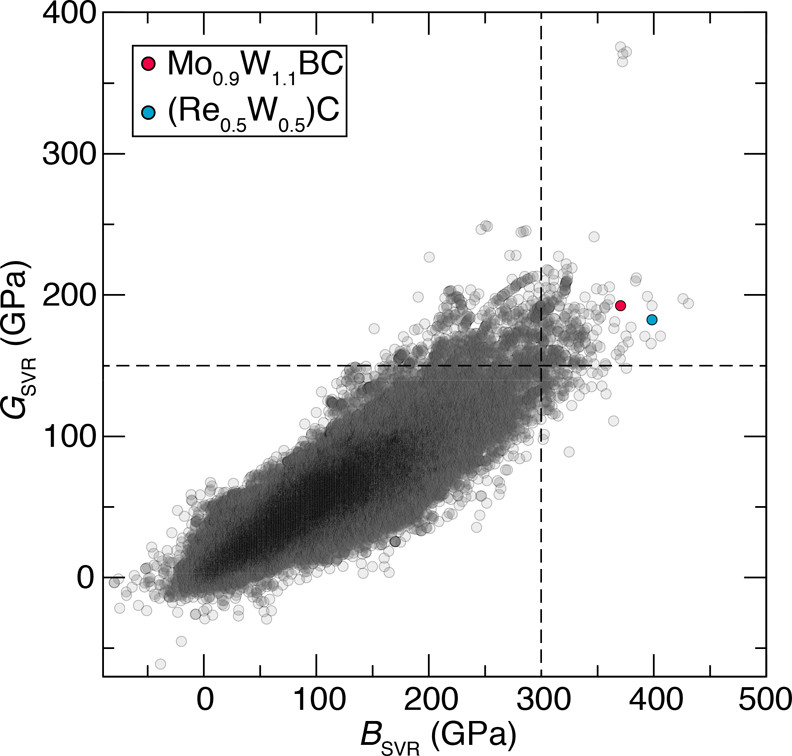}
    \caption{Predicting extraordinary ultraincompressible, superhard materials. \acrfull{svm} predictions of bulk ($B$) and shear ($G$) moduli for \num{118288} inorganic compounds. Binary candidates have already been thoroughly explored, so the top ternary (\ch{Re_{0.5}W_{0.5}C}, blue circle) and quaternary (\ch{Mo_{0.9}W_{1.1}BC}, red circle) candidate were each identified and selected for synthesis and characterization. Due to presence of graphite impurities in synthesized \ch{Re_{0.5}W_{0.5}C}, \ch{ReWC_{0.8}} was used as an alternative. Both \ch{ReWC_{0.8}} and \ch{Mo_{0.9}W_{1.1}BC} were synthesized and confirmed as being superhard at low loads and ultraincompressible, indicating a successful example of extraordinary material discovery. Reproduced with permission from Tehrani, A. M.; Oliynyk, A. O.; Parry, M.; Rizvi, Z.; Couper, S.; Lin, F.; Miyagi, L.; Sparks, T. D.; Brgoch, J. Journal of the American Chemical Society 2018, 140 (31), 9844–9853.\cite{tehraniMachineLearningDirected2018} }
    \label{fig:tehrani2018-moduli}
\end{figure}

In a successful example of extraordinary prediction (definition 2) \cite{tehraniMachineLearningDirected2018}, the top candidates from the considered ternary and quaternary inorganic solids (\cref{fig:tehrani2018-moduli}) were selected for validation and confirmed to be ultraincompressible and to be superhard at low loads. \citet{tehraniMachineLearningDirected2018} also discuss nuances of measured performance such as whether hardness at low loads is a valid metric for superhardness considerations and to what extent the predicted compounds are viable for real-life applications.

For an in-depth treatment of extraordinary material predictions, see \citet{kauweCanMachineLearning2020}.
% From our view, common characteristics of larger dataset extraordinary predictions discussed in this work involve careful feature selection with potential use of domain knowledge, 

\glsresetall %reset the abbreviations
\section{Conclusion}

\Gls{ml} techniques can be sorted into rough categories based on the size of the training data used for the model: \numrange{1}{100}, \numrange{101}{10000}, and \num{10000}+. We demonstrate the most comprehensive set of experimentally and computationally validated examples in the literature to date and to our knowledge. Based on the distribution of techniques used in the articles, it is clear that \gls{bo} and \gls{svm} are most often used for \numrange{1}{100} and \numrange{101}{10000} training dataset size ranges, respectively, whereas \num{10000}+ has too few examples with too much variation to establish a trend. The low number of \num{10000}+ validation articles relative to other size ranges illustrates the difficulty of obtaining large, high-fidelity, materials science datasets which often requires extensive curation or are simply non-existent.

We also find that \gls{ad} is successfully paired with \gls{bo}, \gls{svm}, and other validation \gls{ml} articles and that material discovery rates have been enhanced through its use. \Gls{fs} schemes, sometimes augmented by domain knowledge, play an important role in many validation articles. In other cases, experimental or computational \gls{hitp} techniques vastly increase the amount of available homogeneous data and are even paired with \gls{ad} and/or \gls{fs} schemes as described earlier.

Many materials discovery articles use and benefit from \gls{gcv} which allows for extrapolative predictive performance to be assessed more accurately. We also find that extraordinary prediction (\cref{sec:extraordinary}) is practically guaranteed for small datasets where interpolation is the primary mechanism of improved performance and much more difficult for large datasets where extrapolation is required for extraordinary material discovery.

The increase of experimentally or computationally validated articles in recent years (\nvalid{} total articles in this work) and the powerful \gls{ml}, \gls{fs}, \gls{ad}, and \gls{hitp} methods used in the articles, often in combination with each other, demonstrate that materials informatics is continuing to penetrate the materials science discipline and accelerating material discoveries for real-world applications.

% \begin{figure}
%     \centering
%     \includegraphics{valid-workflow.png}
%     \caption{Typical validation workflow}
%     \label{fig:valid-workflow}
% \end{figure}

% Transformer approach

%\include{outline}

\singlespacing

\newpage

\printglossaries

\newpage

\bibliography{mat-synth.bib}

\providecommand{\latin}[1]{#1}
\makeatletter
\providecommand{\doi}
  {\begingroup\let\do\@makeother\dospecials
  \catcode`\{=1 \catcode`\}=2 \doi@aux}
\providecommand{\doi@aux}[1]{\endgroup\texttt{#1}}
\makeatother
\providecommand*\mcitethebibliography{\thebibliography}
\csname @ifundefined\endcsname{endmcitethebibliography}
  {\let\endmcitethebibliography\endthebibliography}{}
\begin{mcitethebibliography}{83}
\providecommand*\natexlab[1]{#1}
\providecommand*\mciteSetBstSublistMode[1]{}
\providecommand*\mciteSetBstMaxWidthForm[2]{}
\providecommand*\mciteBstWouldAddEndPuncttrue
  {\def\EndOfBibitem{\unskip.}}
\providecommand*\mciteBstWouldAddEndPunctfalse
  {\let\EndOfBibitem\relax}
\providecommand*\mciteSetBstMidEndSepPunct[3]{}
\providecommand*\mciteSetBstSublistLabelBeginEnd[3]{}
\providecommand*\EndOfBibitem{}
\mciteSetBstSublistMode{f}
\mciteSetBstMaxWidthForm{subitem}{(\alph{mcitesubitemcount})}
\mciteSetBstSublistLabelBeginEnd
  {\mcitemaxwidthsubitemform\space}
  {\relax}
  {\relax}

\bibitem[Balachandran \latin{et~al.}(2018)Balachandran, Kowalski, Sehirlioglu,
  and Lookman]{balachandranExperimentalSearchHightemperature2018}
Balachandran,~P.~V.; Kowalski,~B.; Sehirlioglu,~A.; Lookman,~T. Experimental
  Search for High-Temperature Ferroelectric Perovskites Guided by Two-Step
  Machine Learning. \emph{Nature Communications} \textbf{2018}, \emph{9}\relax
\mciteBstWouldAddEndPuncttrue
\mciteSetBstMidEndSepPunct{\mcitedefaultmidpunct}
{\mcitedefaultendpunct}{\mcitedefaultseppunct}\relax
\EndOfBibitem
\bibitem[Bucior \latin{et~al.}(2019)Bucior, Bobbitt, Islamoglu, Goswami,
  Gopalan, Yildirim, Farha, Bagheri, and
  Snurr]{buciorEnergybasedDescriptorsRapidly2019}
Bucior,~B.~J.; Bobbitt,~N.~S.; Islamoglu,~T.; Goswami,~S.; Gopalan,~A.;
  Yildirim,~T.; Farha,~O.~K.; Bagheri,~N.; Snurr,~R.~Q. Energy-Based
  Descriptors to Rapidly Predict Hydrogen Storage in Metal\textendash Organic
  Frameworks. \emph{Molecular Systems Design \& Engineering} \textbf{2019},
  \emph{4}, 162--174\relax
\mciteBstWouldAddEndPuncttrue
\mciteSetBstMidEndSepPunct{\mcitedefaultmidpunct}
{\mcitedefaultendpunct}{\mcitedefaultseppunct}\relax
\EndOfBibitem
\bibitem[Cao \latin{et~al.}(2018)Cao, Adutwum, Oliynyk, Luber, Olsen, Mar, and
  Buriak]{caoHowOptimizeMaterials2018}
Cao,~B.; Adutwum,~L.~A.; Oliynyk,~A.~O.; Luber,~E.~J.; Olsen,~B.~C.; Mar,~A.;
  Buriak,~J.~M. How to Optimize Materials and Devices via Design of Experiments
  and Machine Learning: {{Demonstration}} Using Organic Photovoltaics.
  \emph{ACS Nano} \textbf{2018}, \emph{12}, 7434--7444\relax
\mciteBstWouldAddEndPuncttrue
\mciteSetBstMidEndSepPunct{\mcitedefaultmidpunct}
{\mcitedefaultendpunct}{\mcitedefaultseppunct}\relax
\EndOfBibitem
\bibitem[Chen \latin{et~al.}(2020)Chen, Tian, Zhou, Fang, Ding, Sun, and
  Xue]{chenMachineLearningAssisted2020}
Chen,~Y.; Tian,~Y.; Zhou,~Y.; Fang,~D.; Ding,~X.; Sun,~J.; Xue,~D. Machine
  Learning Assisted Multi-Objective Optimization for Materials Processing
  Parameters: {{A}} Case Study in {{Mg}} Alloy. \emph{Journal of Alloys and
  Compounds} \textbf{2020}, \emph{844}, 156159\relax
\mciteBstWouldAddEndPuncttrue
\mciteSetBstMidEndSepPunct{\mcitedefaultmidpunct}
{\mcitedefaultendpunct}{\mcitedefaultseppunct}\relax
\EndOfBibitem
\bibitem[Gaultois \latin{et~al.}(2016)Gaultois, Oliynyk, Mar, Sparks,
  Mulholland, and Meredig]{gaultoisPerspectiveWebbasedMachine2016}
Gaultois,~M.~W.; Oliynyk,~A.~O.; Mar,~A.; Sparks,~T.~D.; Mulholland,~G.~J.;
  Meredig,~B. Perspective: {{Web}}-Based Machine Learning Models for Real-Time
  Screening of Thermoelectric Materials Properties. \emph{APL Materials}
  \textbf{2016}, \emph{4}\relax
\mciteBstWouldAddEndPuncttrue
\mciteSetBstMidEndSepPunct{\mcitedefaultmidpunct}
{\mcitedefaultendpunct}{\mcitedefaultseppunct}\relax
\EndOfBibitem
\bibitem[{G{\'o}mez-Bombarelli} \latin{et~al.}(2016){G{\'o}mez-Bombarelli},
  {Aguilera-Iparraguirre}, Hirzel, Duvenaud, Maclaurin, {Blood-Forsythe}, Chae,
  Einzinger, Ha, Wu, Markopoulos, Jeon, Kang, Miyazaki, Numata, Kim, Huang,
  Hong, Baldo, Adams, and
  {Aspuru-Guzik}]{gomez-bombarelliDesignEfficientMolecular2016}
{G{\'o}mez-Bombarelli},~R. \latin{et~al.}  Design of Efficient Molecular
  Organic Light-Emitting Diodes by a High-Throughput Virtual Screening and
  Experimental Approach. \emph{Nature Materials} \textbf{2016}, \emph{15},
  1120--1127\relax
\mciteBstWouldAddEndPuncttrue
\mciteSetBstMidEndSepPunct{\mcitedefaultmidpunct}
{\mcitedefaultendpunct}{\mcitedefaultseppunct}\relax
\EndOfBibitem
\bibitem[Gzyl \latin{et~al.}(2020)Gzyl, Oliynyk, and
  Mar]{gzylHalfHeuslerStructuresFullHeusler2020}
Gzyl,~A.~S.; Oliynyk,~A.~O.; Mar,~A. Half-{{Heusler Structures}} with
  {{Full}}-{{Heusler Counterparts}}: {{Machine}}-{{Learning Predictions}} and
  {{Experimental Validation}}. \emph{Crystal Growth \& Design} \textbf{2020},
  \relax
\mciteBstWouldAddEndPunctfalse
\mciteSetBstMidEndSepPunct{\mcitedefaultmidpunct}
{}{\mcitedefaultseppunct}\relax
\EndOfBibitem
\bibitem[Gzyl \latin{et~al.}(2019)Gzyl, Oliynyk, Adutwum, and
  Mar]{gzylSolvingColoringProblem2019}
Gzyl,~A.~S.; Oliynyk,~A.~O.; Adutwum,~L.~A.; Mar,~A. Solving the {{Coloring
  Problem}} in {{Half}}-{{Heusler Structures}}: {{Machine}}-{{Learning
  Predictions}} and {{Experimental Validation}}. \emph{Inorganic Chemistry}
  \textbf{2019}, \emph{58}, 9280--9289\relax
\mciteBstWouldAddEndPuncttrue
\mciteSetBstMidEndSepPunct{\mcitedefaultmidpunct}
{\mcitedefaultendpunct}{\mcitedefaultseppunct}\relax
\EndOfBibitem
\bibitem[Homma \latin{et~al.}(2020)Homma, Liu, Sumita, Tamura, Fushimi, Iwata,
  Tsuda, and Kaneta]{hommaOptimizationHeterogeneousTernary2020}
Homma,~K.; Liu,~Y.; Sumita,~M.; Tamura,~R.; Fushimi,~N.; Iwata,~J.; Tsuda,~K.;
  Kaneta,~C. Optimization of a {{Heterogeneous Ternary
  Li3PO4}}-{{Li3BO3}}-{{Li2SO4Mixture}} for {{Li}}-{{Ion Conductivity}} by
  {{Machine Learning}}. \emph{Journal of Physical Chemistry C} \textbf{2020},
  \emph{124}, 12865--12870\relax
\mciteBstWouldAddEndPuncttrue
\mciteSetBstMidEndSepPunct{\mcitedefaultmidpunct}
{\mcitedefaultendpunct}{\mcitedefaultseppunct}\relax
\EndOfBibitem
\bibitem[Hou \latin{et~al.}(2019)Hou, Takagiwa, Shinohara, Xu, and
  Tsuda]{houMachineLearningAssistedDevelopmentTheoretical2019}
Hou,~Z.; Takagiwa,~Y.; Shinohara,~Y.; Xu,~Y.; Tsuda,~K.
  Machine-{{Learning}}-{{Assisted Development}} and {{Theoretical
  Consideration}} for the {{Al}} 2 {{Fe}} 3 {{Si}} 3 {{Thermoelectric
  Material}}. \emph{ACS Applied Materials and Interfaces} \textbf{2019},
  \emph{11}, 11545--11554\relax
\mciteBstWouldAddEndPuncttrue
\mciteSetBstMidEndSepPunct{\mcitedefaultmidpunct}
{\mcitedefaultendpunct}{\mcitedefaultseppunct}\relax
\EndOfBibitem
\bibitem[Iwasaki \latin{et~al.}(2019)Iwasaki, Sawada, Stanev, Ishida, Kirihara,
  Omori, Someya, Takeuchi, Saitoh, and
  Yorozu]{iwasakiIdentificationAdvancedSpindriven2019}
Iwasaki,~Y.; Sawada,~R.; Stanev,~V.; Ishida,~M.; Kirihara,~A.; Omori,~Y.;
  Someya,~H.; Takeuchi,~I.; Saitoh,~E.; Yorozu,~S. Identification of Advanced
  Spin-Driven Thermoelectric Materials via Interpretable Machine Learning.
  \emph{npj Computational Materials} \textbf{2019}, \emph{5}, 6--11\relax
\mciteBstWouldAddEndPuncttrue
\mciteSetBstMidEndSepPunct{\mcitedefaultmidpunct}
{\mcitedefaultendpunct}{\mcitedefaultseppunct}\relax
\EndOfBibitem
\bibitem[Kauwe \latin{et~al.}(2018)Kauwe, Graser, Vazquez, and
  Sparks]{kauweMachineLearningPrediction2018}
Kauwe,~S.~K.; Graser,~J.; Vazquez,~A.; Sparks,~T.~D. Machine {{Learning
  Prediction}} of {{Heat Capacity}} for {{Solid Inorganics}}. \emph{Integrating
  Materials and Manufacturing Innovation} \textbf{2018}, \emph{7}, 43--51\relax
\mciteBstWouldAddEndPuncttrue
\mciteSetBstMidEndSepPunct{\mcitedefaultmidpunct}
{\mcitedefaultendpunct}{\mcitedefaultseppunct}\relax
\EndOfBibitem
\bibitem[Kim \latin{et~al.}(2018)Kim, Kang, Yoo, Kwon, Nam, Lee, Kim, Choi,
  Jung, Kim, Son, Son, Lee, Kim, Shin, and
  Hwang]{kimDeeplearningbasedInverseDesign2018}
Kim,~K. \latin{et~al.}  Deep-Learning-Based Inverse Design Model for
  Intelligent Discovery of Organic Molecules. \emph{npj Computational
  Materials} \textbf{2018}, \emph{4}, 67\relax
\mciteBstWouldAddEndPuncttrue
\mciteSetBstMidEndSepPunct{\mcitedefaultmidpunct}
{\mcitedefaultendpunct}{\mcitedefaultseppunct}\relax
\EndOfBibitem
\bibitem[Li \latin{et~al.}(2018)Li, Hou, Gao, Zeng, Ao, Zhou, Da, Liu, Sun, and
  Zhang]{liEfficientOptimizationPerformance2018}
Li,~X.; Hou,~Z.; Gao,~S.; Zeng,~Y.; Ao,~J.; Zhou,~Z.; Da,~B.; Liu,~W.; Sun,~Y.;
  Zhang,~Y. Efficient {{Optimization}} of the {{Performance}} of
  {{Mn2}}+-{{Doped Kesterite Solar Cell}}: {{Machine Learning Aided Synthesis}}
  of {{High Efficient Cu2}}({{Mn}},{{Zn}}){{Sn}}({{S}},{{Se}})4 {{Solar
  Cells}}. \emph{Solar RRL} \textbf{2018}, \emph{2}\relax
\mciteBstWouldAddEndPuncttrue
\mciteSetBstMidEndSepPunct{\mcitedefaultmidpunct}
{\mcitedefaultendpunct}{\mcitedefaultseppunct}\relax
\EndOfBibitem
\bibitem[Menon \latin{et~al.}(2019)Menon, Childs, Pocz{\'o}s, Washburn, and
  Kurtis]{menonMolecularEngineeringSuperplasticizers2019}
Menon,~A.; Childs,~C.~M.; Pocz{\'o}s,~B.; Washburn,~N.~R.; Kurtis,~K.~E.
  Molecular {{Engineering}} of {{Superplasticizers}} for
  {{Metakaolin}}-{{Portland Cement Blends}} with {{Hierarchical Machine
  Learning}}. \emph{Advanced Theory and Simulations} \textbf{2019}, \emph{2},
  1800164\relax
\mciteBstWouldAddEndPuncttrue
\mciteSetBstMidEndSepPunct{\mcitedefaultmidpunct}
{\mcitedefaultendpunct}{\mcitedefaultseppunct}\relax
\EndOfBibitem
\bibitem[Min \latin{et~al.}(2018)Min, Choi, Park, and
  Cho]{minMachineLearningAssisted2018}
Min,~K.; Choi,~B.; Park,~K.; Cho,~E. Machine Learning Assisted Optimization of
  Electrochemical Properties for {{Ni}}-Rich Cathode Materials.
  \emph{Scientific Reports} \textbf{2018}, \emph{8}, 15778\relax
\mciteBstWouldAddEndPuncttrue
\mciteSetBstMidEndSepPunct{\mcitedefaultmidpunct}
{\mcitedefaultendpunct}{\mcitedefaultseppunct}\relax
\EndOfBibitem
\bibitem[Nikolaev \latin{et~al.}(2016)Nikolaev, Hooper, Webber, Rao, Decker,
  Krein, Poleski, Barto, and Maruyama]{nikolaevAutonomyMaterialsResearch2016}
Nikolaev,~P.; Hooper,~D.; Webber,~F.; Rao,~R.; Decker,~K.; Krein,~M.;
  Poleski,~J.; Barto,~R.; Maruyama,~B. Autonomy in Materials Research: A Case
  Study in Carbon Nanotube Growth. \emph{npj Computational Materials}
  \textbf{2016}, \emph{2}, 16031\relax
\mciteBstWouldAddEndPuncttrue
\mciteSetBstMidEndSepPunct{\mcitedefaultmidpunct}
{\mcitedefaultendpunct}{\mcitedefaultseppunct}\relax
\EndOfBibitem
\bibitem[Oliynyk \latin{et~al.}(2016)Oliynyk, Adutwum, Harynuk, and
  Mar]{oliynykClassifyingCrystalStructures2016}
Oliynyk,~A.~O.; Adutwum,~L.~A.; Harynuk,~J.~J.; Mar,~A. Classifying {{Crystal
  Structures}} of {{Binary Compounds AB}} through {{Cluster Resolution Feature
  Selection}} and {{Support Vector Machine Analysis}}. \emph{Chemistry of
  Materials} \textbf{2016}, \emph{28}, 6672--6681\relax
\mciteBstWouldAddEndPuncttrue
\mciteSetBstMidEndSepPunct{\mcitedefaultmidpunct}
{\mcitedefaultendpunct}{\mcitedefaultseppunct}\relax
\EndOfBibitem
\bibitem[Oliynyk \latin{et~al.}(2017)Oliynyk, Adutwum, Rudyk, Pisavadia, Lotfi,
  Hlukhyy, Harynuk, Mar, and
  Brgoch]{oliynykDisentanglingStructuralConfusion2017}
Oliynyk,~A.~O.; Adutwum,~L.~A.; Rudyk,~B.~W.; Pisavadia,~H.; Lotfi,~S.;
  Hlukhyy,~V.; Harynuk,~J.~J.; Mar,~A.; Brgoch,~J. Disentangling {{Structural
  Confusion}} through {{Machine Learning}}: {{Structure Prediction}} and
  {{Polymorphism}} of {{Equiatomic Ternary Phases ABC}}. \emph{Journal of the
  American Chemical Society} \textbf{2017}, \emph{139}, 17870--17881\relax
\mciteBstWouldAddEndPuncttrue
\mciteSetBstMidEndSepPunct{\mcitedefaultmidpunct}
{\mcitedefaultendpunct}{\mcitedefaultseppunct}\relax
\EndOfBibitem
\bibitem[Oliynyk \latin{et~al.}(2016)Oliynyk, Antono, Sparks, Ghadbeigi,
  Gaultois, Meredig, and
  Mar]{oliynykHighThroughputMachineLearningDrivenSynthesis2016}
Oliynyk,~A.~O.; Antono,~E.; Sparks,~T.~D.; Ghadbeigi,~L.; Gaultois,~M.~W.;
  Meredig,~B.; Mar,~A. High-{{Throughput Machine}}-{{Learning}}-{{Driven
  Synthesis}} of {{Full}}-{{Heusler Compounds}}. \emph{Chemistry of Materials}
  \textbf{2016}, \emph{28}\relax
\mciteBstWouldAddEndPuncttrue
\mciteSetBstMidEndSepPunct{\mcitedefaultmidpunct}
{\mcitedefaultendpunct}{\mcitedefaultseppunct}\relax
\EndOfBibitem
\bibitem[Raccuglia \latin{et~al.}(2016)Raccuglia, Elbert, Adler, Falk, Wenny,
  Mollo, Zeller, Friedler, Schrier, and
  Norquist]{raccugliaMachinelearningassistedMaterialsDiscovery2016}
Raccuglia,~P.; Elbert,~K.~C.; Adler,~P.~D.; Falk,~C.; Wenny,~M.~B.; Mollo,~A.;
  Zeller,~M.; Friedler,~S.~A.; Schrier,~J.; Norquist,~A.~J.
  Machine-Learning-Assisted Materials Discovery Using Failed Experiments.
  \emph{Nature} \textbf{2016}, \emph{533}, 73--76\relax
\mciteBstWouldAddEndPuncttrue
\mciteSetBstMidEndSepPunct{\mcitedefaultmidpunct}
{\mcitedefaultendpunct}{\mcitedefaultseppunct}\relax
\EndOfBibitem
\bibitem[Ren \latin{et~al.}(2018)Ren, Ward, Williams, Laws, Wolverton,
  {Hattrick-Simpers}, and Mehta]{renAcceleratedDiscoveryMetallic2018}
Ren,~F.; Ward,~L.; Williams,~T.; Laws,~K.~J.; Wolverton,~C.;
  {Hattrick-Simpers},~J.; Mehta,~A. Accelerated Discovery of Metallic Glasses
  through Iteration of Machine Learning and High-Throughput Experiments.
  \emph{Science Advances} \textbf{2018}, \emph{4}\relax
\mciteBstWouldAddEndPuncttrue
\mciteSetBstMidEndSepPunct{\mcitedefaultmidpunct}
{\mcitedefaultendpunct}{\mcitedefaultseppunct}\relax
\EndOfBibitem
\bibitem[Rickman \latin{et~al.}(2019)Rickman, Chan, Harmer, Smeltzer, Marvel,
  Roy, and Balasubramanian]{rickmanMaterialsInformaticsScreening2019}
Rickman,~J.~M.; Chan,~H.~M.; Harmer,~M.~P.; Smeltzer,~J.~A.; Marvel,~C.~J.;
  Roy,~A.; Balasubramanian,~G. Materials Informatics for the Screening of
  Multi-Principal Elements and High-Entropy Alloys. \emph{Nature
  Communications} \textbf{2019}, \emph{10}, 1--10\relax
\mciteBstWouldAddEndPuncttrue
\mciteSetBstMidEndSepPunct{\mcitedefaultmidpunct}
{\mcitedefaultendpunct}{\mcitedefaultseppunct}\relax
\EndOfBibitem
\bibitem[Sakurai \latin{et~al.}(2019)Sakurai, Yada, Simomura, Ju, Kashiwagi,
  Okada, Nagao, Tsuda, and
  Shiomi]{sakuraiUltranarrowBandWavelengthSelectiveThermal2019}
Sakurai,~A.; Yada,~K.; Simomura,~T.; Ju,~S.; Kashiwagi,~M.; Okada,~H.;
  Nagao,~T.; Tsuda,~K.; Shiomi,~J. Ultranarrow-{{Band Wavelength}}-{{Selective
  Thermal Emission}} with {{Aperiodic Multilayered Metamaterials Designed}} by
  {{Bayesian Optimization}}. \emph{ACS Central Science} \textbf{2019},
  \emph{5}, 319--326\relax
\mciteBstWouldAddEndPuncttrue
\mciteSetBstMidEndSepPunct{\mcitedefaultmidpunct}
{\mcitedefaultendpunct}{\mcitedefaultseppunct}\relax
\EndOfBibitem
\bibitem[Shamp \latin{et~al.}(2016)Shamp, Terpstra, Bi, Falls, Avery, and
  Zurek]{shampDecompositionProductsPhosphine2016}
Shamp,~A.; Terpstra,~T.; Bi,~T.; Falls,~Z.; Avery,~P.; Zurek,~E. Decomposition
  {{Products}} of {{Phosphine}} under {{Pressure}}: {{PH2 Stable}} and
  {{Superconducting}}? \emph{Journal of the American Chemical Society}
  \textbf{2016}, \emph{138}, 1884--1892\relax
\mciteBstWouldAddEndPuncttrue
\mciteSetBstMidEndSepPunct{\mcitedefaultmidpunct}
{\mcitedefaultendpunct}{\mcitedefaultseppunct}\relax
\EndOfBibitem
\bibitem[Tehrani \latin{et~al.}(2018)Tehrani, Oliynyk, Parry, Rizvi, Couper,
  Lin, Miyagi, Sparks, and Brgoch]{tehraniMachineLearningDirected2018}
Tehrani,~A.~M.; Oliynyk,~A.~O.; Parry,~M.; Rizvi,~Z.; Couper,~S.; Lin,~F.;
  Miyagi,~L.; Sparks,~T.~D.; Brgoch,~J. Machine {{Learning Directed Search}}
  for {{Ultraincompressible}}, {{Superhard Materials}}. \emph{Journal of the
  American Chemical Society} \textbf{2018}, \emph{140}, 9844--9853\relax
\mciteBstWouldAddEndPuncttrue
\mciteSetBstMidEndSepPunct{\mcitedefaultmidpunct}
{\mcitedefaultendpunct}{\mcitedefaultseppunct}\relax
\EndOfBibitem
\bibitem[Wahab \latin{et~al.}(2020)Wahab, Jain, Tyrrell, Seas, Kotthoff, and
  Johnson]{wahabMachinelearningassistedFabricationBayesian2020}
Wahab,~H.; Jain,~V.; Tyrrell,~A.~S.; Seas,~M.~A.; Kotthoff,~L.; Johnson,~P.~A.
  Machine-Learning-Assisted Fabrication: {{Bayesian}} Optimization of
  Laser-Induced Graphene Patterning Using in-Situ {{Raman}} Analysis.
  \emph{Carbon} \textbf{2020}, \emph{167}, 609--619\relax
\mciteBstWouldAddEndPuncttrue
\mciteSetBstMidEndSepPunct{\mcitedefaultmidpunct}
{\mcitedefaultendpunct}{\mcitedefaultseppunct}\relax
\EndOfBibitem
\bibitem[Wakabayashi \latin{et~al.}(2019)Wakabayashi, Otsuka, Krockenberger,
  Sawada, Taniyasu, and
  Yamamoto]{wakabayashiMachinelearningassistedThinfilmGrowth2019}
Wakabayashi,~Y.~K.; Otsuka,~T.; Krockenberger,~Y.; Sawada,~H.; Taniyasu,~Y.;
  Yamamoto,~H. Machine-Learning-Assisted Thin-Film Growth: {{Bayesian}}
  Optimization in Molecular Beam Epitaxy of {{SrRuO3}} Thin Films. \emph{APL
  Materials} \textbf{2019}, \emph{7}\relax
\mciteBstWouldAddEndPuncttrue
\mciteSetBstMidEndSepPunct{\mcitedefaultmidpunct}
{\mcitedefaultendpunct}{\mcitedefaultseppunct}\relax
\EndOfBibitem
\bibitem[Weng \latin{et~al.}(2020)Weng, Song, Zhu, Yan, Sun, Grice, Yan, and
  Yin]{wengSimpleDescriptorDerived2020}
Weng,~B.; Song,~Z.; Zhu,~R.; Yan,~Q.; Sun,~Q.; Grice,~C.~G.; Yan,~Y.;
  Yin,~W.~J. Simple Descriptor Derived from Symbolic Regression Accelerating
  the Discovery of New Perovskite Catalysts. \emph{Nature Communications}
  \textbf{2020}, \emph{11}, 1--8\relax
\mciteBstWouldAddEndPuncttrue
\mciteSetBstMidEndSepPunct{\mcitedefaultmidpunct}
{\mcitedefaultendpunct}{\mcitedefaultseppunct}\relax
\EndOfBibitem
\bibitem[Wen \latin{et~al.}(2019)Wen, Zhang, Wang, Xue, Bai, Antonov, Dai,
  Lookman, and Su]{wenMachineLearningAssisted2019}
Wen,~C.; Zhang,~Y.; Wang,~C.; Xue,~D.; Bai,~Y.; Antonov,~S.; Dai,~L.;
  Lookman,~T.; Su,~Y. Machine Learning Assisted Design of High Entropy Alloys
  with Desired Property. \emph{Acta Materialia} \textbf{2019}, \emph{170},
  109--117\relax
\mciteBstWouldAddEndPuncttrue
\mciteSetBstMidEndSepPunct{\mcitedefaultmidpunct}
{\mcitedefaultendpunct}{\mcitedefaultseppunct}\relax
\EndOfBibitem
\bibitem[Wu \latin{et~al.}(2019)Wu, Kondo, Kakimoto, Yang, Yamada, Kuwajima,
  Lambard, Hongo, Xu, Shiomi, Schick, Morikawa, and
  Yoshida]{wuMachinelearningassistedDiscoveryPolymers2019}
Wu,~S.; Kondo,~Y.; Kakimoto,~M.-a.; Yang,~B.; Yamada,~H.; Kuwajima,~I.;
  Lambard,~G.; Hongo,~K.; Xu,~Y.; Shiomi,~J.; Schick,~C.; Morikawa,~J.;
  Yoshida,~R. Machine-Learning-Assisted Discovery of Polymers with High Thermal
  Conductivity Using a Molecular Design Algorithm. \emph{npj Computational
  Materials} \textbf{2019}, \emph{5}, 66\relax
\mciteBstWouldAddEndPuncttrue
\mciteSetBstMidEndSepPunct{\mcitedefaultmidpunct}
{\mcitedefaultendpunct}{\mcitedefaultseppunct}\relax
\EndOfBibitem
\bibitem[Xue \latin{et~al.}(2016)Xue, Balachandran, Yuan, Hu, Qian, Dougherty,
  and Lookman]{xueAcceleratedSearchBaTiO3based2016}
Xue,~D.; Balachandran,~P.~V.; Yuan,~R.; Hu,~T.; Qian,~X.; Dougherty,~E.~R.;
  Lookman,~T. Accelerated Search for {{BaTiO3}}-Based Piezoelectrics with
  Vertical Morphotropic Phase Boundary Using {{Bayesian}} Learning.
  \emph{Proceedings of the National Academy of Sciences of the United States of
  America} \textbf{2016}, \emph{113}, 13301--13306\relax
\mciteBstWouldAddEndPuncttrue
\mciteSetBstMidEndSepPunct{\mcitedefaultmidpunct}
{\mcitedefaultendpunct}{\mcitedefaultseppunct}\relax
\EndOfBibitem
\bibitem[Xue \latin{et~al.}(2017)Xue, Xue, Yuan, Zhou, Balachandran, Ding, Sun,
  and Lookman]{xueInformaticsApproachTransformation2017}
Xue,~D.; Xue,~D.; Yuan,~R.; Zhou,~Y.; Balachandran,~P.~V.; Ding,~X.; Sun,~J.;
  Lookman,~T. An Informatics Approach to Transformation Temperatures of
  {{NiTi}}-Based Shape Memory Alloys. \emph{Acta Materialia} \textbf{2017},
  \emph{125}, 532--541\relax
\mciteBstWouldAddEndPuncttrue
\mciteSetBstMidEndSepPunct{\mcitedefaultmidpunct}
{\mcitedefaultendpunct}{\mcitedefaultseppunct}\relax
\EndOfBibitem
\bibitem[Yuan \latin{et~al.}(2018)Yuan, Liu, Balachandran, Xue, Zhou, Ding,
  Sun, Xue, and Lookman]{yuanAcceleratedDiscoveryLarge2018}
Yuan,~R.; Liu,~Z.; Balachandran,~P.~V.; Xue,~D.~D.; Zhou,~Y.; Ding,~X.;
  Sun,~J.; Xue,~D.~D.; Lookman,~T. Accelerated {{Discovery}} of {{Large
  Electrostrains}} in {{BaTiO3}}-{{Based Piezoelectrics Using Active
  Learning}}. \emph{Advanced Materials} \textbf{2018}, \emph{30}\relax
\mciteBstWouldAddEndPuncttrue
\mciteSetBstMidEndSepPunct{\mcitedefaultmidpunct}
{\mcitedefaultendpunct}{\mcitedefaultseppunct}\relax
\EndOfBibitem
\bibitem[Zhang \latin{et~al.}(2020)Zhang, Mansouri~Tehrani, Oliynyk, Day, and
  Brgoch]{zhangFindingNextSuperhard2020}
Zhang,~Z.; Mansouri~Tehrani,~A.; Oliynyk,~A.~O.; Day,~B.; Brgoch,~J. Finding
  the {{Next Superhard Material}} through {{Ensemble Learning}}. \emph{Advanced
  Materials} \textbf{2020}, 2005112\relax
\mciteBstWouldAddEndPuncttrue
\mciteSetBstMidEndSepPunct{\mcitedefaultmidpunct}
{\mcitedefaultendpunct}{\mcitedefaultseppunct}\relax
\EndOfBibitem
\bibitem[Zhang \latin{et~al.}(2018)Zhang, Oliynyk, Duarte, Iyer, Ghadbeigi,
  Kauwe, Sparks, and Mar]{zhangNotJustPar2018}
Zhang,~D.; Oliynyk,~A.~O.; Duarte,~G.~M.; Iyer,~A.~K.; Ghadbeigi,~L.;
  Kauwe,~S.~K.; Sparks,~T.~D.; Mar,~A. Not {{Just Par}} for the {{Course}}: 73
  {{Quaternary Germanides RE4 M2 XGe4}} ({{RE}} = {{La}}-{{Nd}}, {{Sm}},
  {{Gd}}-{{Tm}}, {{Lu}}; {{M}} = {{Mn}}-{{Ni}}; {{X}} = {{Ag}}, {{Cd}}) and the
  {{Search}} for {{Intermetallics}} with {{Low Thermal Conductivity}}.
  \emph{Inorganic Chemistry} \textbf{2018}, \emph{57}, 14249--14259\relax
\mciteBstWouldAddEndPuncttrue
\mciteSetBstMidEndSepPunct{\mcitedefaultmidpunct}
{\mcitedefaultendpunct}{\mcitedefaultseppunct}\relax
\EndOfBibitem
\bibitem[Zhuo \latin{et~al.}(2020)Zhuo, Hariyani, Armijo, Abolade~Lawson, and
  Brgoch]{zhuoEvaluatingThermalQuenching2020}
Zhuo,~Y.; Hariyani,~S.; Armijo,~E.; Abolade~Lawson,~Z.; Brgoch,~J. Evaluating
  {{Thermal Quenching Temperature}} in {{Eu3}}+-{{Substituted Oxide Phosphors}}
  via {{Machine Learning}}. \emph{ACS Applied Materials and Interfaces}
  \textbf{2020}, \emph{12}, 5244--5250\relax
\mciteBstWouldAddEndPuncttrue
\mciteSetBstMidEndSepPunct{\mcitedefaultmidpunct}
{\mcitedefaultendpunct}{\mcitedefaultseppunct}\relax
\EndOfBibitem
\bibitem[Zhuo \latin{et~al.}(2018)Zhuo, Mansouri~Tehrani, Oliynyk, Duke, and
  Brgoch]{zhuoIdentifyingEfficientThermally2018}
Zhuo,~Y.; Mansouri~Tehrani,~A.; Oliynyk,~A.~O.; Duke,~A.~C.; Brgoch,~J.
  Identifying an Efficient, Thermally Robust Inorganic Phosphor Host via
  Machine Learning. \emph{Nature Communications} \textbf{2018}, \emph{9}\relax
\mciteBstWouldAddEndPuncttrue
\mciteSetBstMidEndSepPunct{\mcitedefaultmidpunct}
{\mcitedefaultendpunct}{\mcitedefaultseppunct}\relax
\EndOfBibitem
\bibitem[Balachandran \latin{et~al.}(2016)Balachandran, Xue, Theiler, Hogden,
  and Lookman]{balachandranAdaptiveStrategiesMaterials2016}
Balachandran,~P.~V.; Xue,~D.; Theiler,~J.; Hogden,~J.; Lookman,~T. Adaptive
  {{Strategies}} for {{Materials Design}} Using {{Uncertainties}}.
  \emph{Scientific Reports} \textbf{2016}, \emph{6}, 19660\relax
\mciteBstWouldAddEndPuncttrue
\mciteSetBstMidEndSepPunct{\mcitedefaultmidpunct}
{\mcitedefaultendpunct}{\mcitedefaultseppunct}\relax
\EndOfBibitem
\bibitem[Balachandran(2020)]{balachandranDatadrivenDesignB202020}
Balachandran,~P.~V. Data-Driven Design of {{B20}} Alloys with Targeted Magnetic
  Properties Guided by Machine Learning and Density Functional Theory.
  \emph{Journal of Materials Research} \textbf{2020}, \emph{35}, 890--897\relax
\mciteBstWouldAddEndPuncttrue
\mciteSetBstMidEndSepPunct{\mcitedefaultmidpunct}
{\mcitedefaultendpunct}{\mcitedefaultseppunct}\relax
\EndOfBibitem
\bibitem[Balachandran \latin{et~al.}(2017)Balachandran, Young, Lookman, and
  Rondinelli]{balachandranLearningDataDesign2017}
Balachandran,~P.~V.; Young,~J.; Lookman,~T.; Rondinelli,~J.~M. Learning from
  Data to Design Functional Materials without Inversion Symmetry. \emph{Nature
  Communications} \textbf{2017}, \emph{8}\relax
\mciteBstWouldAddEndPuncttrue
\mciteSetBstMidEndSepPunct{\mcitedefaultmidpunct}
{\mcitedefaultendpunct}{\mcitedefaultseppunct}\relax
\EndOfBibitem
\bibitem[Balachandran \latin{et~al.}(2017)Balachandran, Shearman, Theiler, and
  Lookman]{balachandranPredictingDisplacementsOctahedral2017}
Balachandran,~P.~V.; Shearman,~T.; Theiler,~J.; Lookman,~T. Predicting
  Displacements of Octahedral Cations in Ferroelectric Perovskites Using
  Machine Learning. \emph{Acta Crystallographica Section B: Structural Science,
  Crystal Engineering and Materials} \textbf{2017}, \emph{73}, 962--967\relax
\mciteBstWouldAddEndPuncttrue
\mciteSetBstMidEndSepPunct{\mcitedefaultmidpunct}
{\mcitedefaultendpunct}{\mcitedefaultseppunct}\relax
\EndOfBibitem
\bibitem[Ju \latin{et~al.}(2017)Ju, Shiga, Feng, Hou, Tsuda, and
  Shiomi]{juDesigningNanostructuresPhonon2017}
Ju,~S.; Shiga,~T.; Feng,~L.; Hou,~Z.; Tsuda,~K.; Shiomi,~J. Designing
  {{Nanostructures}} for {{Phonon Transport}} via {{Bayesian Optimization}}.
  \emph{Physical Review X} \textbf{2017}, \emph{7}, 021024\relax
\mciteBstWouldAddEndPuncttrue
\mciteSetBstMidEndSepPunct{\mcitedefaultmidpunct}
{\mcitedefaultendpunct}{\mcitedefaultseppunct}\relax
\EndOfBibitem
\bibitem[Lu \latin{et~al.}(2018)Lu, Zhou, Ouyang, Guo, Li, and
  Wang]{luAcceleratedDiscoveryStable2018}
Lu,~S.; Zhou,~Q.; Ouyang,~Y.; Guo,~Y.; Li,~Q.; Wang,~J. Accelerated Discovery
  of Stable Lead-Free Hybrid Organic-Inorganic Perovskites via Machine
  Learning. \emph{Nature Communications} \textbf{2018}, \emph{9}, 3405\relax
\mciteBstWouldAddEndPuncttrue
\mciteSetBstMidEndSepPunct{\mcitedefaultmidpunct}
{\mcitedefaultendpunct}{\mcitedefaultseppunct}\relax
\EndOfBibitem
\bibitem[{Mannodi-Kanakkithodi} \latin{et~al.}(2016){Mannodi-Kanakkithodi},
  Pilania, Huan, Lookman, and
  Ramprasad]{mannodi-kanakkithodiMachineLearningStrategy2016}
{Mannodi-Kanakkithodi},~A.; Pilania,~G.; Huan,~T.~D.; Lookman,~T.;
  Ramprasad,~R. Machine {{Learning Strategy}} for {{Accelerated Design}} of
  {{Polymer Dielectrics}}. \emph{Scientific Reports} \textbf{2016}, \emph{6},
  20952\relax
\mciteBstWouldAddEndPuncttrue
\mciteSetBstMidEndSepPunct{\mcitedefaultmidpunct}
{\mcitedefaultendpunct}{\mcitedefaultseppunct}\relax
\EndOfBibitem
\bibitem[Meredig \latin{et~al.}(2014)Meredig, Agrawal, Kirklin, Saal, Doak,
  Thompson, Zhang, Choudhary, and
  Wolverton]{meredigCombinatorialScreeningNew2014}
Meredig,~B.; Agrawal,~A.; Kirklin,~S.; Saal,~J.~E.; Doak,~J.~W.; Thompson,~A.;
  Zhang,~K.; Choudhary,~A.; Wolverton,~C. Combinatorial Screening for New
  Materials in Unconstrained Composition Space with Machine Learning.
  \emph{Physical Review B} \textbf{2014}, \emph{89}, 094104\relax
\mciteBstWouldAddEndPuncttrue
\mciteSetBstMidEndSepPunct{\mcitedefaultmidpunct}
{\mcitedefaultendpunct}{\mcitedefaultseppunct}\relax
\EndOfBibitem
\bibitem[Park and Wolverton(2020)Park, and
  Wolverton]{parkDevelopingImprovedCrystal2020}
Park,~C.~W.; Wolverton,~C. Developing an Improved Crystal Graph Convolutional
  Neural Network Framework for Accelerated Materials Discovery. \emph{Physical
  Review Materials} \textbf{2020}, \emph{4}, 063801\relax
\mciteBstWouldAddEndPuncttrue
\mciteSetBstMidEndSepPunct{\mcitedefaultmidpunct}
{\mcitedefaultendpunct}{\mcitedefaultseppunct}\relax
\EndOfBibitem
\bibitem[Seko \latin{et~al.}(2018)Seko, Hayashi, Kashima, and
  Tanaka]{sekoMatrixTensorbasedRecommender2018}
Seko,~A.; Hayashi,~H.; Kashima,~H.; Tanaka,~I. Matrix- and Tensor-Based
  Recommender Systems for the Discovery of Currently Unknown Inorganic
  Compounds. \emph{Physical Review Materials} \textbf{2018}, \emph{2},
  013805\relax
\mciteBstWouldAddEndPuncttrue
\mciteSetBstMidEndSepPunct{\mcitedefaultmidpunct}
{\mcitedefaultendpunct}{\mcitedefaultseppunct}\relax
\EndOfBibitem
\bibitem[Sendek \latin{et~al.}(2017)Sendek, Yang, Cubuk, Duerloo, Cui, and
  Reed]{sendekHolisticComputationalStructure2017}
Sendek,~A.~D.; Yang,~Q.; Cubuk,~E.~D.; Duerloo,~K. A.~N.; Cui,~Y.; Reed,~E.~J.
  Holistic Computational Structure Screening of More than 12 000 Candidates for
  Solid Lithium-Ion Conductor Materials. \emph{Energy and Environmental
  Science} \textbf{2017}, \emph{10}, 306--320\relax
\mciteBstWouldAddEndPuncttrue
\mciteSetBstMidEndSepPunct{\mcitedefaultmidpunct}
{\mcitedefaultendpunct}{\mcitedefaultseppunct}\relax
\EndOfBibitem
\bibitem[Talapatra \latin{et~al.}(2018)Talapatra, Boluki, Duong, Qian,
  Dougherty, and Arr{\'o}yave]{talapatraAutonomousEfficientExperiment2018}
Talapatra,~A.; Boluki,~S.; Duong,~T.; Qian,~X.; Dougherty,~E.; Arr{\'o}yave,~R.
  Autonomous Efficient Experiment Design for Materials Discovery with
  {{Bayesian}} Model Averaging. \emph{Physical Review Materials} \textbf{2018},
  \emph{2}\relax
\mciteBstWouldAddEndPuncttrue
\mciteSetBstMidEndSepPunct{\mcitedefaultmidpunct}
{\mcitedefaultendpunct}{\mcitedefaultseppunct}\relax
\EndOfBibitem
\bibitem[Meredig(2019)]{meredigFiveHighImpactResearch2019}
Meredig,~B. Five {{High}}-{{Impact Research Areas}} in {{Machine Learning}} for
  {{Materials Science}}. \emph{Chemistry of Materials} \textbf{2019},
  \emph{31}, 9579--9581\relax
\mciteBstWouldAddEndPuncttrue
\mciteSetBstMidEndSepPunct{\mcitedefaultmidpunct}
{\mcitedefaultendpunct}{\mcitedefaultseppunct}\relax
\EndOfBibitem
\bibitem[Murdock \latin{et~al.}(2020)Murdock, Kauwe, Wang, and
  Sparks]{murdockDomainKnowledgeNecessary2020}
Murdock,~R.~J.; Kauwe,~S.~K.; Wang,~A. Y.-T.; Sparks,~T.~D. Is Domain Knowledge
  Necessary for Machine Learning Materials Properties? \emph{ChemRxiv}
  \textbf{2020}, 8\relax
\mciteBstWouldAddEndPuncttrue
\mciteSetBstMidEndSepPunct{\mcitedefaultmidpunct}
{\mcitedefaultendpunct}{\mcitedefaultseppunct}\relax
\EndOfBibitem
\bibitem[Wang \latin{et~al.}(2020)Wang, Murdock, Kauwe, Oliynyk, Gurlo, Brgoch,
  Persson, and Sparks]{wangMachineLearningMaterials2020}
Wang,~A. Y.-T.; Murdock,~R.~J.; Kauwe,~S.~K.; Oliynyk,~A.~O.; Gurlo,~A.;
  Brgoch,~J.; Persson,~K.~A.; Sparks,~T.~D. Machine {{Learning}} for
  {{Materials Scientists}}: {{An Introductory Guide}} toward {{Best
  Practices}}. \emph{Chem. Mater.} \textbf{2020}, 12\relax
\mciteBstWouldAddEndPuncttrue
\mciteSetBstMidEndSepPunct{\mcitedefaultmidpunct}
{\mcitedefaultendpunct}{\mcitedefaultseppunct}\relax
\EndOfBibitem
\bibitem[Paszke \latin{et~al.}(2019)Paszke, Gross, Massa, Lerer, Bradbury,
  Chanan, Killeen, Lin, Gimelshein, Antiga, Desmaison, Kopf, Yang, DeVito,
  Raison, Tejani, Chilamkurthy, Steiner, Fang, Bai, and
  Chintala]{paszkePyTorchImperativeStyle2019}
Paszke,~A. \latin{et~al.}  In \emph{Advances in Neural Information Processing
  Systems 32}; Wallach,~H., Larochelle,~H., Beygelzimer,~A.,
  {dAlch{\'e}-Buc},~F., Fox,~E., Garnett,~R., Eds.; {Curran Associates, Inc.},
  2019; pp 8024--8035\relax
\mciteBstWouldAddEndPuncttrue
\mciteSetBstMidEndSepPunct{\mcitedefaultmidpunct}
{\mcitedefaultendpunct}{\mcitedefaultseppunct}\relax
\EndOfBibitem
\bibitem[Pedregosa \latin{et~al.}(2011)Pedregosa, Varoquaux, Gramfort, Michel,
  Thirion, Grisel, Blondel, Prettenhofer, Weiss, Dubourg, Vanderplas, Passos,
  Cournapeau, Brucher, Perrot, and
  Duchesnay]{pedregosaScikitlearnMachineLearning2011}
Pedregosa,~F. \latin{et~al.}  Scikit-Learn: {{Machine}} Learning in {{Python}}.
  \emph{Journal of Machine Learning Research} \textbf{2011}, \emph{12},
  2825--2830\relax
\mciteBstWouldAddEndPuncttrue
\mciteSetBstMidEndSepPunct{\mcitedefaultmidpunct}
{\mcitedefaultendpunct}{\mcitedefaultseppunct}\relax
\EndOfBibitem
\bibitem[Ueno \latin{et~al.}(2016)Ueno, Rhone, Hou, Mizoguchi, and
  Tsuda]{uenoCOMBOEfficientBayesian2016}
Ueno,~T.; Rhone,~T.~D.; Hou,~Z.; Mizoguchi,~T.; Tsuda,~K. {{COMBO}}: {{An}}
  Efficient {{Bayesian}} Optimization Library for Materials Science.
  \emph{Materials Discovery} \textbf{2016}, \emph{4}, 18--21\relax
\mciteBstWouldAddEndPuncttrue
\mciteSetBstMidEndSepPunct{\mcitedefaultmidpunct}
{\mcitedefaultendpunct}{\mcitedefaultseppunct}\relax
\EndOfBibitem
\bibitem[Ong(2013)]{ongPythonMaterialsGenomics2013}
Ong,~S.~P. Python {{Materials Genomics}} (Pymatgen): {{A}} Robust, Open-Source
  Python Library for Materials Analysis. \emph{Computational Materials Science}
  \textbf{2013}, 6\relax
\mciteBstWouldAddEndPuncttrue
\mciteSetBstMidEndSepPunct{\mcitedefaultmidpunct}
{\mcitedefaultendpunct}{\mcitedefaultseppunct}\relax
\EndOfBibitem
\bibitem[Ward \latin{et~al.}(2016)Ward, Agrawal, Choudhary, and
  Wolverton]{wardGeneralpurposeMachineLearning2016}
Ward,~L.; Agrawal,~A.; Choudhary,~A.; Wolverton,~C. A General-Purpose Machine
  Learning Framework for Predicting Properties of Inorganic Materials.
  \emph{npj Computational Materials} \textbf{2016}, \relax
\mciteBstWouldAddEndPunctfalse
\mciteSetBstMidEndSepPunct{\mcitedefaultmidpunct}
{}{\mcitedefaultseppunct}\relax
\EndOfBibitem
\bibitem[Choudhary \latin{et~al.}(2020)Choudhary, Garrity, Reid, DeCost,
  Biacchi, Hight~Walker, Trautt, {Hattrick-Simpers}, Kusne, Centrone, Davydov,
  Jiang, Pachter, Cheon, Reed, Agrawal, Qian, Sharma, Zhuang, Kalinin, Sumpter,
  Pilania, Acar, Mandal, Haule, Vanderbilt, Rabe, and
  Tavazza]{choudharyJointAutomatedRepository2020}
Choudhary,~K. \latin{et~al.}  The Joint Automated Repository for Various
  Integrated Simulations ({{JARVIS}}) for Data-Driven Materials Design.
  \emph{npj Computational Materials} \textbf{2020}, \emph{6}, 173\relax
\mciteBstWouldAddEndPuncttrue
\mciteSetBstMidEndSepPunct{\mcitedefaultmidpunct}
{\mcitedefaultendpunct}{\mcitedefaultseppunct}\relax
\EndOfBibitem
\bibitem[The~MathWorks(2020)]{themathworksStatisticsMachineLearning2020}
The~MathWorks,~I. Statistics and {{Machine Learning Toolbox}}. 2020\relax
\mciteBstWouldAddEndPuncttrue
\mciteSetBstMidEndSepPunct{\mcitedefaultmidpunct}
{\mcitedefaultendpunct}{\mcitedefaultseppunct}\relax
\EndOfBibitem
\bibitem[The~MathWorks(2020)]{themathworksDeepLearningToolbox2020}
The~MathWorks,~I. Deep {{Learning Toolbox}}. 2020\relax
\mciteBstWouldAddEndPuncttrue
\mciteSetBstMidEndSepPunct{\mcitedefaultmidpunct}
{\mcitedefaultendpunct}{\mcitedefaultseppunct}\relax
\EndOfBibitem
\bibitem[Kuhn(2008)]{kuhnBuildingPredictiveModels2008}
Kuhn,~M. Building Predictive Models in r Using the Caret Package. \emph{Journal
  of Statistical Software, Articles} \textbf{2008}, \emph{28}, 1--26\relax
\mciteBstWouldAddEndPuncttrue
\mciteSetBstMidEndSepPunct{\mcitedefaultmidpunct}
{\mcitedefaultendpunct}{\mcitedefaultseppunct}\relax
\EndOfBibitem
\bibitem[Meyer \latin{et~al.}(2020)Meyer, Dimitriadou, Hornik, Weingessel,
  Leisch, Chang, and Lin]{meyerE1071MiscFunctions2020}
Meyer,~D.; Dimitriadou,~E.; Hornik,~K.; Weingessel,~A.; Leisch,~F.;
  Chang,~C.-C.; Lin,~C.-C. E1071: {{Misc}} Functions of the Department of
  Statistics, Probability Theory Group. 2020\relax
\mciteBstWouldAddEndPuncttrue
\mciteSetBstMidEndSepPunct{\mcitedefaultmidpunct}
{\mcitedefaultendpunct}{\mcitedefaultseppunct}\relax
\EndOfBibitem
\bibitem[Venables and Ripley(2002)Venables, and
  Ripley]{venablesModernAppliedStatistics2002}
Venables,~W.~N.; Ripley,~B.~D. \emph{Modern Applied Statistics with s}, 4th
  ed.; {Springer}: {New York}, 2002\relax
\mciteBstWouldAddEndPuncttrue
\mciteSetBstMidEndSepPunct{\mcitedefaultmidpunct}
{\mcitedefaultendpunct}{\mcitedefaultseppunct}\relax
\EndOfBibitem
\bibitem[Butler \latin{et~al.}(2018)Butler, Davies, Cartwright, Isayev, and
  Walsh]{butlerMachineLearningMolecular2018}
Butler,~K.~T.; Davies,~D.~W.; Cartwright,~H.; Isayev,~O.; Walsh,~A. Machine
  Learning for Molecular and Materials Science. \emph{Nature} \textbf{2018},
  \emph{559}, 547--555\relax
\mciteBstWouldAddEndPuncttrue
\mciteSetBstMidEndSepPunct{\mcitedefaultmidpunct}
{\mcitedefaultendpunct}{\mcitedefaultseppunct}\relax
\EndOfBibitem
\bibitem[Gaultois \latin{et~al.}(2013)Gaultois, Sparks, Borg, Seshadri,
  Bonificio, and Clarke]{gaultoisDatadrivenReviewThermoelectric2013}
Gaultois,~M.~W.; Sparks,~T.~D.; Borg,~C.~K.; Seshadri,~R.; Bonificio,~W.~D.;
  Clarke,~D.~R. Data-Driven Review of Thermoelectric Materials: {{Performance}}
  and Resource Onsiderations. \emph{Chemistry of Materials} \textbf{2013},
  \emph{25}, 2911--2920\relax
\mciteBstWouldAddEndPuncttrue
\mciteSetBstMidEndSepPunct{\mcitedefaultmidpunct}
{\mcitedefaultendpunct}{\mcitedefaultseppunct}\relax
\EndOfBibitem
\bibitem[Hoar \latin{et~al.}(2020)Hoar, Lu, and
  Liu]{hoarMachineLearningEnabledExplorationMorphology2020}
Hoar,~B.~B.; Lu,~S.; Liu,~C. Machine-{{Learning}}-{{Enabled Exploration}} of
  {{Morphology Influence}} on {{Wire}}-{{Array Electrodes}} for
  {{Electrochemical Nitrogen Fixation}}. \emph{Journal of Physical Chemistry
  Letters} \textbf{2020}, \emph{11}, 4625--4630\relax
\mciteBstWouldAddEndPuncttrue
\mciteSetBstMidEndSepPunct{\mcitedefaultmidpunct}
{\mcitedefaultendpunct}{\mcitedefaultseppunct}\relax
\EndOfBibitem
\bibitem[Yan \latin{et~al.}(2020)Yan, Gao, Liu, Zhang, and
  Cheng]{yanOptimizationThermalConductivity2020}
Yan,~B.; Gao,~R.; Liu,~P.; Zhang,~P.; Cheng,~L. Optimization of Thermal
  Conductivity of {{UO2}}\textendash{{Mo}} Composite with Continuous {{Mo}}
  Channel Based on Finite Element Method and Machine Learning.
  \emph{International Journal of Heat and Mass Transfer} \textbf{2020},
  \emph{159}, 120067\relax
\mciteBstWouldAddEndPuncttrue
\mciteSetBstMidEndSepPunct{\mcitedefaultmidpunct}
{\mcitedefaultendpunct}{\mcitedefaultseppunct}\relax
\EndOfBibitem
\bibitem[Jain \latin{et~al.}(2013)Jain, Ong, Hautier, Chen, Richards, Dacek,
  Cholia, Gunter, Skinner, Ceder, and
  a.~Persson]{jainMaterialsProjectMaterials2013}
Jain,~A.; Ong,~S.~P.; Hautier,~G.; Chen,~W.; Richards,~W.~D.; Dacek,~S.;
  Cholia,~S.; Gunter,~D.; Skinner,~D.; Ceder,~G.; a.~Persson,~K. The
  {{Materials Project}}: {{A}} Materials Genome Approach to Accelerating
  Materials Innovation. \emph{APL Materials} \textbf{2013}, \emph{1},
  011002\relax
\mciteBstWouldAddEndPuncttrue
\mciteSetBstMidEndSepPunct{\mcitedefaultmidpunct}
{\mcitedefaultendpunct}{\mcitedefaultseppunct}\relax
\EndOfBibitem
\bibitem[Kirklin \latin{et~al.}(2015)Kirklin, Saal, Meredig, Thompson, Doak,
  Aykol, R{\"u}hl, and Wolverton]{kirklinOpenQuantumMaterials2015}
Kirklin,~S.; Saal,~J.~E.; Meredig,~B.; Thompson,~A.; Doak,~J.~W.; Aykol,~M.;
  R{\"u}hl,~S.; Wolverton,~C. The {{Open Quantum Materials Database}}
  ({{OQMD}}): Assessing the Accuracy of {{DFT}} Formation Energies. \emph{npj
  Computational Materials} \textbf{2015}, \emph{1}, 15010\relax
\mciteBstWouldAddEndPuncttrue
\mciteSetBstMidEndSepPunct{\mcitedefaultmidpunct}
{\mcitedefaultendpunct}{\mcitedefaultseppunct}\relax
\EndOfBibitem
\bibitem[Villars and Cenzual(2014)Villars, and
  Cenzual]{villarsPearsonCrystalData2014}
Villars,~P.; Cenzual,~K. Pearson's {{Crystal Data}}: {{Crystal Structure
  Database}} for {{Inorganic Compounds}}. 2014\relax
\mciteBstWouldAddEndPuncttrue
\mciteSetBstMidEndSepPunct{\mcitedefaultmidpunct}
{\mcitedefaultendpunct}{\mcitedefaultseppunct}\relax
\EndOfBibitem
\bibitem[Ward \latin{et~al.}(2018)Ward, Dunn, Faghaninia, Zimmermann, Bajaj,
  Wang, Montoya, Chen, Bystrom, Dylla, Chard, Asta, Persson, Snyder, Foster,
  and Jain]{wardMatminerOpenSource2018}
Ward,~L. \latin{et~al.}  Matminer: {{An}} Open Source Toolkit for Materials
  Data Mining. \emph{Computational Materials Science} \textbf{2018},
  \emph{152}, 60--69\relax
\mciteBstWouldAddEndPuncttrue
\mciteSetBstMidEndSepPunct{\mcitedefaultmidpunct}
{\mcitedefaultendpunct}{\mcitedefaultseppunct}\relax
\EndOfBibitem
\bibitem[Oliynyk and Buriak(2019)Oliynyk, and
  Buriak]{oliynykVirtualIssueMachineLearning2019}
Oliynyk,~A.~O.; Buriak,~J.~M. Virtual {{Issue}} on {{Machine}}-{{Learning
  Discoveries}} in {{Materials Science}}. \emph{Chemistry of Materials}
  \textbf{2019}, \emph{31}, 8243--8247\relax
\mciteBstWouldAddEndPuncttrue
\mciteSetBstMidEndSepPunct{\mcitedefaultmidpunct}
{\mcitedefaultendpunct}{\mcitedefaultseppunct}\relax
\EndOfBibitem
\bibitem[Saal \latin{et~al.}(2020)Saal, Oliynyk, and
  Meredig]{saalMachineLearningMaterials2020}
Saal,~J.~E.; Oliynyk,~A.~O.; Meredig,~B. Machine {{Learning}} in {{Materials
  Discovery}}: {{Confirmed Predictions}} and {{Their Underlying Approaches}}.
  \emph{Annual Review of Materials Research} \textbf{2020}, \emph{50},
  49--69\relax
\mciteBstWouldAddEndPuncttrue
\mciteSetBstMidEndSepPunct{\mcitedefaultmidpunct}
{\mcitedefaultendpunct}{\mcitedefaultseppunct}\relax
\EndOfBibitem
\bibitem[Bera \latin{et~al.}(2014)Bera, Jacob, Opahle, Gunda, Chmielowski,
  Dennler, and Madsen]{beraIntegratedComputationalMaterials2014}
Bera,~C.; Jacob,~S.; Opahle,~I.; Gunda,~N. S.~H.; Chmielowski,~R.; Dennler,~G.;
  Madsen,~G. K.~H. Integrated Computational Materials Discovery of Silver Doped
  Tin Sulfide as a Thermoelectric Material. \emph{Phys. Chem. Chem. Phys.}
  \textbf{2014}, \emph{16}, 19894--19899\relax
\mciteBstWouldAddEndPuncttrue
\mciteSetBstMidEndSepPunct{\mcitedefaultmidpunct}
{\mcitedefaultendpunct}{\mcitedefaultseppunct}\relax
\EndOfBibitem
\bibitem[Ghosh and Harimkar(2012)Ghosh, and
  Harimkar]{ghoshConsolidationSynthesisMAX2012}
Ghosh,~N.; Harimkar,~S. In \emph{Advances in Science and Technology of Mn+1axn
  Phases}; Low,~I., Ed.; {Woodhead Publishing}, 2012; pp 47--80\relax
\mciteBstWouldAddEndPuncttrue
\mciteSetBstMidEndSepPunct{\mcitedefaultmidpunct}
{\mcitedefaultendpunct}{\mcitedefaultseppunct}\relax
\EndOfBibitem
\bibitem[Tummers \latin{et~al.}(2015)Tummers, {van der Laan}, and
  Huyser]{tummersDataThiefIIISoftware2015}
Tummers,~B.; {van der Laan},~J.; Huyser,~K. {{DataThief III}} Software.
  2015\relax
\mciteBstWouldAddEndPuncttrue
\mciteSetBstMidEndSepPunct{\mcitedefaultmidpunct}
{\mcitedefaultendpunct}{\mcitedefaultseppunct}\relax
\EndOfBibitem
\bibitem[MathWorksHelpCenter2020(2020)]{MathWorksHelpCenter2020}
\emph{{{MathWorks Help Center Documentation}}: {{Dummy Variables}}}; 2020\relax
\mciteBstWouldAddEndPuncttrue
\mciteSetBstMidEndSepPunct{\mcitedefaultmidpunct}
{\mcitedefaultendpunct}{\mcitedefaultseppunct}\relax
\EndOfBibitem
\bibitem[Jones and Schonlau(1998)Jones, and
  Schonlau]{jonesEfficientGlobalOptimization1998}
Jones,~D.~R.; Schonlau,~M. Efficient {{Global Optimization}} of {{Expensive
  Black}}-{{Box Functions}}. \emph{Journal of Global Optimization}
  \textbf{1998}, 38\relax
\mciteBstWouldAddEndPuncttrue
\mciteSetBstMidEndSepPunct{\mcitedefaultmidpunct}
{\mcitedefaultendpunct}{\mcitedefaultseppunct}\relax
\EndOfBibitem
\bibitem[Oliynyk \latin{et~al.}(2016)Oliynyk, Sparks, Gaultois, Ghadbeigi, and
  Mar]{oliynykGd12Co2016}
Oliynyk,~A.~O.; Sparks,~T.~D.; Gaultois,~M.~W.; Ghadbeigi,~L.; Mar,~A. Gd
  {\textsubscript{12}} {{Co}} {\textsubscript{5.3}} {{Bi}} and {{Gd}}
  {\textsubscript{12}} {{Co}} {\textsubscript{5}} {{Bi}}, {{Crystalline
  Doppelg\"anger}} with {{Low Thermal Conductivities}}. \emph{Inorganic
  Chemistry} \textbf{2016}, \emph{55}, 6625--6633\relax
\mciteBstWouldAddEndPuncttrue
\mciteSetBstMidEndSepPunct{\mcitedefaultmidpunct}
{\mcitedefaultendpunct}{\mcitedefaultseppunct}\relax
\EndOfBibitem
\bibitem[Meredig \latin{et~al.}(2018)Meredig, Antono, Church, Hutchinson, Ling,
  Paradiso, Blaiszik, Foster, Gibbons, {Hattrick-Simpers}, Mehta, and
  Ward]{meredigCanMachineLearning2018}
Meredig,~B.; Antono,~E.; Church,~C.; Hutchinson,~M.; Ling,~J.; Paradiso,~S.;
  Blaiszik,~B.; Foster,~I.; Gibbons,~B.; {Hattrick-Simpers},~J.; Mehta,~A.;
  Ward,~L. Can Machine Learning Identify the next High-Temperature
  Superconductor? {{Examining}} Extrapolation Performance for Materials
  Discovery. \emph{Molecular Systems Design \& Engineering} \textbf{2018},
  \emph{3}, 819--825\relax
\mciteBstWouldAddEndPuncttrue
\mciteSetBstMidEndSepPunct{\mcitedefaultmidpunct}
{\mcitedefaultendpunct}{\mcitedefaultseppunct}\relax
\EndOfBibitem
\bibitem[Sparks \latin{et~al.}(2020)Sparks, Kauwe, Parry, Tehrani, and
  Brgoch]{sparksMachineLearningStructural2020}
Sparks,~T.~D.; Kauwe,~S.~K.; Parry,~M.~E.; Tehrani,~A.~M.; Brgoch,~J. Machine
  {{Learning}} for {{Structural Materials}}. \emph{Annual Review of Materials
  Research} \textbf{2020}, \emph{50}, 27--48\relax
\mciteBstWouldAddEndPuncttrue
\mciteSetBstMidEndSepPunct{\mcitedefaultmidpunct}
{\mcitedefaultendpunct}{\mcitedefaultseppunct}\relax
\EndOfBibitem
\bibitem[Kauwe \latin{et~al.}(2020)Kauwe, Graser, Murdock, and
  Sparks]{kauweCanMachineLearning2020}
Kauwe,~S.~K.; Graser,~J.; Murdock,~R.; Sparks,~T.~D. Can Machine Learning Find
  Extraordinary Materials? \emph{Computational Materials Science}
  \textbf{2020}, \emph{174}\relax
\mciteBstWouldAddEndPuncttrue
\mciteSetBstMidEndSepPunct{\mcitedefaultmidpunct}
{\mcitedefaultendpunct}{\mcitedefaultseppunct}\relax
\EndOfBibitem
\end{mcitethebibliography}

\end{document}